\documentclass[aps,prd,showkeys,preprint,amsmath,amssymb,nofootinbib]{revtex4-1}
\usepackage{comment}
\usepackage{color}
\usepackage{amsmath}
\usepackage{amssymb}
\usepackage{amsthm}
\usepackage{mathrsfs}
\usepackage{graphicx}
\usepackage{marvosym}
\usepackage{amsmath,bm}
\usepackage{array}
\usepackage{latexsym}
\usepackage{enumerate}
\usepackage{slashed}
\usepackage{hyperref}
\usepackage{color}
\usepackage[margin=20pt,small]{caption}
\hypersetup{pdftex,colorlinks=true,linkcolor=blue,citecolor=blue,menucolor=black,urlcolor=blue,filecolor=blue}

\newcommand{\order}{{\mathcal O}}

\allowdisplaybreaks[3]

\newcommand{\nn}{\nonumber}
\usepackage[normalem]{ulem} 

\newcommand{\gam}{ \gamma }
\newcommand{\spinGamma}{{\sigma}}

\newcommand{ \cS }{ {\mathcal S} }
\newcommand{ \cP }{ {\mathcal P} }
\newcommand{ \cV }{ {\mathcal V} }
\newcommand{ \cA }{ {\mathcal A} }
\newcommand{ \cD }{ \mathcal D }
\newcommand{\M}{ {\mathcal M} }

\newcommand{\bp}{{\bf{p}}}
\newcommand{\bq}{{\bf{q}}}
\newcommand{\bv}{{\bf{v}}}
\newcommand{\bk}{{\bf{k}}}

\begin{document}

\title{Effective quantum kinetic theory for spin transport of fermions with collsional effects}
\author{Di-Lun Yang$^{1}$, Koichi Hattori$^{2}$, Yoshimasa Hidaka$^{3,4}$}
\affiliation{
	$^1$Faculty of Science and Technology,  Keio University, Yokohama 223-8522, Japan.\\
	$^2$Yukawa Institute for Theoretical Physics, Kyoto University, Kyoto 606-8502, Japan.\\
	$^3$RIKEN Nishina Center,
	RIKEN, Wako, Saitama 351-0198, Japan.\\
	$^4$RIKEN iTHEMS, RIKEN, Wako, Saitama 351-0198, Japan.
} 

\preprint{RIKEN-QHP-441}
\preprint{YITP-20-17}
\date{\today}

\begin{abstract}
We systematically derive the collision term for the axial kinetic theory, a quantum kinetic theory delineating the coupled dynamics of the vector/axial charges and spin transport carried by the massive spin-1/2 fermions traversing a medium.
We employ the Wigner-function approach and propose a consistent power-counting scheme where the axial-charge distribution function, a non-conserved quantity for massive particles, is accounted as the first-order quantity in the $  \hbar $ expansion, while the vector-charge distribution function the zeroth-order quantity.
This specific power-counting scheme allows us to organize a reduced $ \hbar$ expansion for the collision term 
and to formally identity the spin-diffusion effect  
and the spin-polarization effect at the same order. 
We confirm that the obtained collisional axial kinetic theory smoothly 
reduces to the chiral kinetic theory in the massless limit, 
serving as a consistency check. In the absence of electromagnetic fields, we further present the simplified axial kinetic equations suitable for tracking dynamical spin polarization of heavy and light fermions, respectively.
As an application to the weakly coupled quark-gluon plasma at high temperature, 
we compute the spin-diffusion term  
for massive quarks within the leading-log approximation. 
The formal expression for the first-order terms provides a path toward 
evaluation of the spin polarization effect in quantum chromodynamics. 
\end{abstract}

\maketitle

\section{Introduction}
The observations of global polarization of $\Lambda$ hyperons in heavy ion collisions (HIC) have triggered intensive studies for the spin polarization of relativistic fermions \cite{STAR:2017ckg,Adam:2018ivw}. In particular, the experimental measurements have been motivated by theoretical proposals upon the polarization led by non-head-on scattering of hard partons \cite{Liang:2004ph} and by the thermal vorticity from the statistical model in equilibrium \cite{Becattini:2013vja}. Although the simulations based on the modified Cooper-Frye formula for spin polarization \cite{Becattini2013a,Fang:2016vpj} have shown remarkable agreements with global polarization of $\Lambda$ \cite{Karpenko:2016jyx,Li:2017slc}, more recent observations for local polarization caused the new tension between experiments and theories \cite{Becattini:2017gcx,Wei:2018zfb}. Further phenomenological studies have alluded to an essential role for non-equilibrium corrections \cite{Wu:2019eyi,Florkowski:2019voj}, where the spin polarization is no longer just dictated by thermal vorticity and equilibrium distribution functions for $\Lambda$. Also, the feed-down effects are analyzed in Refs.~\cite{Becattini:2019ntv, Xia:2019fjf}, which show only minor corrections upon the local polarization. It is hence imperative to understand the dynamics of the spin polarization for not only hadrons but also quarks (and even gluons) in quark gluon plasmas (QGP). In general, the current studies in theory for the dynamical spin polarization in heavy ion physics may be divided into two directions: One is to construct the so-called hydrodynamics of spin as a macroscopic effective theory incorporating spin as a hydrodynamic variable obeying angular momentum conservation \cite{Florkowski:2017ruc,Florkowski:2018fap,Hattori:2019lfp}. Alternatively, non-equilibrium spin transport can be studied through quantum kinetic theory as a microscopic theory having a direct connection to the underlying quantum field theory. 

In fact, the construction for quantum kinetic theory of massless fermions, known as chiral kinetic theory (CKT), was initiated by Refs.~\cite{Son:2012wh,Stephanov:2012ki} from Berry connection and by Refs.~\cite{Chen:2012ca,Son:2012zy} from the Wigner-function approach based on quantum field theory with the motivation to explore non-equilibrium transport in chiral matter beyond the renown anomalous phenomena in equilibrium such as the chiral magnetic/vortical effects \cite{Vilenkin:1979ui, Kharzeev:2007jp,Fukushima:2008xe}. There have been also plenty of followup studies for extension and applications \cite{Manuel:2013zaa,Chen:2014cla,Chen:2015gta,Gorbar:2016ygi,Kharzeev:2016sut,Hidaka:2016yjf,Hidaka:2017auj,Hidaka:2018ekt,
Mueller:2017arw,Mueller:2017lzw,Huang:2017tsq, Huang:2018wdl,Carignano:2018gqt,Dayi:2018xdy,Liu:2018xip,Sun:2018idn,Rybalka:2018uzh,Lin:2019ytz,Yamamoto:2020zrs}. In particluar, the issue for Lorentz covariance associated with the side-jump phenomenon was addressed in Refs.~\cite{Chen:2014cla,Chen:2015gta} and further refined in Refs.~\cite{Hidaka:2016yjf} from the Wigner-function aprroach with systematic inclusion of background electromagnetic fields and collisions in CKT. The side-jump effect is also shown to contribute to anti-symmetric part of the canonical energy momentum tensor responsible for angular-momentum transfer between spin and orbital angular momentum in chiral fluids \cite{Yang:2018lew} (see also Ref.~\cite{Fukushima:2018osn} for a related study), which further manifests the origin of side jumps in connection to spin-orbit interaction. Overall, the CKT can be regarded as a modified Boltzmann(Vlasov) equation involving quantum corrections such as the chiral anomaly, magnetic-moment coupling, and spin-orbit interaction.
Moreover, the CKT has also been recently applied to the study of the $\Lambda$ polarization in heavy ion collisions \cite{Sun:2017xhx,Liu:2019krs}.   

However, in order to consistently investigate the spin transport of $\Lambda$ or strange quarks as the seed for $\Lambda$ in QGP before hadronization, it is inevitable to consider the mass corrections. Unlike massless fermions of which the spin orientation is enslaved to the momentum direction, 
the spin is now a new dynamical degree of freedom coupled with the charge transport for massive fermions. 
There have been recent studies on the construction of the ``free-streaming'' quantum kinetic theory, 
without the collisional effects, 
delineating the entangled dynamics between the vector/axial charges and the spin polarization for massive fermions \cite{Mueller:2019gjj,Weickgenannt:2019dks,Gao:2019znl,Hattori:2019ahi,Wang:2019moi,Liu:2020ymh}. 
In particular, the axial kinetic theory, 
derived from the Wigner-function approach up to $\mathcal{O}(\hbar)$ in Ref.~\cite{Hattori:2019ahi}, 
successfully cover both the massless and massive cases, 
which is made possible by maintaining the general spin frame vector 
allowed in the solution of the constraint equations. 
The set of quantum kinetic theories consists of a scalar kinetic equation (SKE) for the vector-charge 
and an axial-vector kinetic equation (AKE) for the spin degrees of freedom.

The next key ingredient for describing the spin diffusion/polarization 
is the collisional effects, which have not been incorporated in the above studies. 
On the other hand, in Ref.~\cite{Li:2019qkf}, the spin-diffusion term at $\mathcal{O}(\hbar^0)$ in collisions was computed in perturbative QCD (see also Refs.~\cite{Zhang:2019xya,Kapusta:2019sad} for other studies of collisions). 
Nonetheless, the inclusion of $\mathcal{O}(\hbar^1)$ corrections to the collisional effects, 
which is responsible for generating the spin polarization, has never been achieved. Therefore, we investigate the field-theoretical framework
to systematically include the collisional effects to the axial kinetic theory.  
As will be explained in this paper, one will confront with tough technical problems. 
The full master equations directly obtained from the Kadanoff-Byam equations 
are quite involved with the quantum corrections up to $\mathcal{O}(\hbar^1)$. 
On the other hand, similar to the previous construction for the free-streaming part~\cite{Hattori:2019ahi}, 
we also have to ensure the smooth connection between the collisional AKE and the CKT in the massless limit. 
In spite of the potential importance of finite-mass effects in the collisions terms for the spin rotations and kinematics, 
the basic framework for the collisional effects has not been touched thus far. 
We, therefore, intend to establish the collisional SKE and AKE for arbitrary mass 
within a plausible approximation for physical systems of our interest.

In this paper, we propose a $  \hbar $-counting scheme to circumvent the aforementioned difficulty
and construct the quantum kinetic theory with collisions and background electromagnetic fields for spin transport of spin-1/2 fermions with arbitrary mass. Our power counting entails that the axial-charge distribution function $f_A$ is at $\mathcal{O}(\hbar^1)$ as opposed to the vector-charge distribution function $f_V$ at $\mathcal{O}(\hbar^0)$ in the $\hbar$ expansion applied in the Wigner-function formalism. 
However, this power counting may be applicable in most of physical systems such as HIC. 
The consequent quantum kinetic theory can be regarded as an ``effective" axial kinetic theory with collisions, 
for which the free-streaming part has been established in Ref.~\cite{Hattori:2019ahi}. 
%In our approach, we solve for the Wigner functions and derive the kinetic equations from the Kadanoff-Baym equations with collisions characterized by the self-energy based on our power-counting scheme. 
Once deriving the full master equations with the self-energy terms in the Kadanoff-Baym equations, 
we apply our power-counting scheme and solve the reduced master equations. 
At the end of the day, we obtain the quantum kinetic equations with collisions for 
the spin polarization and axial-charge evolution. The collision term therein can be decomposed into the classical part responsible for the spin diffusion 
and the quantum part contributing to the spin polarization, 
which importantly stems from the spin-orbit interaction through the entangled dynamics of vector-charge distributions.

Applications of our general framework are not limited to QCD 
although we emphasize the motivation for the quark transport in QGP. 
As an example, 
we further evaluate the classical part of collisions for the spin-diffusion dynamics in the weakly-coupled QGP 
with specific forms of the self-energies from the hard thermal loop approxmation. 
Computation of the quantum corrections demand the so-far unknown self-energies with polarized gluons, 
which needs to be further investigated in the future.

The paper is organized as follows: In Sec.~\ref{sec_WF_approach}, we briefly review the Wigner-function approach 
and derive the master equations led by Kadanoff-Baym equations with collisions. 
Then, we introduce our power-counting scheme in $  \hbar$. 
Based on this scheme, we find the perturbative solutions for Wigner functions. 
In Sec.~\ref{sec_AAKT}, we then derive the SKE and AKE in an effective axial kinetic theory with collisions and background electromagnetic fields as a generic formalism for studying quantum transport of fermions. Moreover, we highlight the simplified version suitable for the application to heavy-ion physics and make more detailed discussions upon the collision terms. As a concrete example for applications, in Sec.~\ref{sec_diffusion_QGP}, we further investigate the spin-diffusion term in weakly-coupled QGP by utilizing our formalism. 
Finally, we make concluding remarks and outlook in Sec.~\ref{sec_conclusion}. 
For completeness, we present critical steps in the derivations/computations in Appendices.

\section{Wigner-function approach with collisions}\label{sec_WF_approach}

In this section, we first derive the ``master equations'' by applying 
the spinor decomposition and the $ \hbar  $ expansion to the Kadanoff-Baym equation. 
Those master equations are subsequently used to derive the quantum kinetic theories. 
We systematically include the collisional effects that are necessary for describing the relaxation dynamics. 
%\com{[Are there any clear advantages over Ho-Ung's framework in general systematics 
%when going ahead to the first order in hbar?]}

The derivation of quantum kinetic theories in this framework 
has been investigated for massless fermions \cite{Hidaka:2016yjf} 
and for massive fermions without collisions, that is, the ``free-streaming'' case \cite{Hattori:2019ahi}. 
Below, we follow the basic flow of the derivation established in those preceding studies 
and summarize the crucial intermediate steps below. 
One will find that the master equations (\ref{full_Master_eq}) with the collisional effects 
have quite involved structures originating from the spinor structures of fermions. 
Those involved structures yield versatile transport phenomena 
that need to be investigated with appropriate strategies for further developments. 
Therefore, we propose an $ \hbar  $ counting scheme in Sec.~\ref{sec:counting} 
that enables one to cure the pathological complication of the collisional effects 
and to extract the entangled dynamics between the vector charge and the spin polarization.

\subsection{Full master equations}

We shall start with the Wigner transformation applied to quantum expectation values of correlation functions of fermionic fields,
\begin{eqnarray}
S_{\alpha\beta}^{<(>)}(q,X)=\int d^4Ye^{\frac{iq\cdot Y}{\hbar}}\tilde{S}_{\alpha\beta}^{<(>)}(x,y),
\end{eqnarray}
where $X=(x+y)/2$ and $Y=x-y$ and we work in the Minkowski spacetime with the mostly negative spacetime metric. 
Here, $\tilde{S}_{\alpha\beta}^<(x,y)=\langle\bar{\psi}_{\beta}(y)U(y,x)\psi_{\alpha}(x)\rangle$ and $\tilde{S}_{\alpha\beta}^>(x,y)=\langle\psi_{\alpha}(x)U^{\dagger}(x,y)\bar{\psi}_{\beta}(y)\rangle$ 
are lessor and greater propagators, respectively. To maintain the gauge invariance, we also insert the gauge link, e.g., $U(y,x)=\exp\big(-i\int^y_xdz^{\rho}A_{\rho}(z)\big)$ for QED with $A_{\rho}$ denoting the $U(1)$ electromagnetic gauge field. 
Note that $q^{\mu}$ thus represents the kinetic momentum. 
Hereafter, we focus on $S^<(q,X)$ and suppress the indices of spinors.
After the Wigner transformation, the lessor propagator obeys the Kadanoff-Baym equations derived from the Schwinger-Dyson equation,
\begin{eqnarray}\nonumber
		\label{eq:KBequation}
		&&(\slashed{\Pi}-m)S^{<} 
		+\gamma^{\mu}i\frac{\hbar}{2}\nabla_{\mu}S^{<}
		=\frac{i\hbar}{2}\Big(\Sigma^<\star S^>-\Sigma^>\star S^<\Big),
		\\
		&&S^{<}(\slashed{\Pi}-m)-i\frac{\hbar}{2}\nabla_{\mu}S^{<}\gamma^{\mu}  =-\frac{i\hbar}{2}\Big(S^>\star\Sigma^<-S^<\star\Sigma^>\Big),
\end{eqnarray}
where $\Sigma^{<(>)}$ represents the lessor (greater) self-energy.
Since we only focus on the scattering process, here we drop the real parts of the retarded and advanced self-energies and of the retarded propagators. See Ref.~\cite{Hidaka:2016yjf} for the same setup to derive the equations for Weyl fermions. The symbol of $\star$ represents the Moyal product incorporating higher-order corrections in $\hbar$. The star product of two functions $A(q,X)$ and $B(q, X)$ are expanded as
\begin{eqnarray}
A \star B=AB+\frac{i\hbar}{2}\{A, B\}_{\text{P.B.}}+\mathcal{O}(\hbar^2),
\end{eqnarray}
where we define the Poisson bracket as $\{A,B\}_{\text{P.B.}}\equiv(\partial^{\mu}_{q}A)(\partial_{\mu}B)-(\partial_{\mu}A)(\partial^{\mu}_qB)$.  The sum and difference of Eq.~\eqref{eq:KBequation} read 
\begin{eqnarray}\nonumber\label{KB_eq}
	&&\{(\slashed{\Pi}-m), S^< \} +\frac{i\hbar}{2}\Big([\gamma^{\mu}, \nabla_{\mu}S^{<}]-[\Sigma^<, S^>]_{{\star}}+[\Sigma^>, S^<]_{{\star}}\Big)=0,
	\\
	&&[(\slashed{\Pi}-m), S^{<}]+\frac{i\hbar}{2}\Big(\{\gamma^{\mu}, \nabla_{\mu}S^{<}\}-\{\Sigma^<,  S^{>}\}_{{\star}}+\{\Sigma^>,  S^{<}\}_{{\star}}\Big)
	=0
	.
\end{eqnarray}
{Here, we introduced $\{F,G\}\equiv FG+GF$, $[F,G]\equiv FG-GF$,  $\{F,G\}_{\star}\equiv F\star G+G\star F$ and $[F,G]_{\star}\equiv F\star G-G\star F$, where $F$ and $G$ are arbitrary matrix-valued functions.}

The notations and conventions in the above equations are as follows. 
First, the derivative operators are given as \cite{Vasak:1987um}
\begin{eqnarray}
	\nabla_{\mu}=\partial_{\mu}+j_0(\Box)F_{\nu\mu}\partial^{\nu}_{q},\quad \Pi_{\mu}=q_{\mu}+\frac{\hbar}{2}j_1(\Box)F_{\nu\mu}\partial^{\nu}_{q},\quad\Box=\frac{\hbar}{2}\partial_{\rho}\partial^{\rho}_q
	.
\end{eqnarray} 
We will hereafter use $\partial_{\mu}\equiv\partial/\partial X^{\mu}$ for convenience.
Here $j_0(\Box),j_1(\Box)$ are 
spherical Bessel functions and $\partial_{\rho}$ in $\Box$ only act on the field strength $F_{\nu\mu}$ when having spacetime-dependent background fields. 
Making the $\hbar$ expansion, which corresponds to the gradient expansion for $\partial_{\mu}\ll q_{\mu}$, one finds
\begin{eqnarray}\nonumber
	\nabla_{\mu}&=&\partial_{\mu}+F_{\nu\mu}\partial^{\nu}_q-\frac{\hbar^2}{24}(\partial_{\rho}\partial_{\lambda}F_{\nu\mu})\partial^{\rho}_q\partial^{\lambda}_q\partial^{\nu}_{q}+\mathcal{O}(\hbar^4), 
	\\
	\Pi_{\mu}&=&q_{\mu}+\frac{\hbar^2}{12}(\partial_{\rho}F_{\nu\mu})\partial^{\rho}_q\partial^{\nu}_{q}+\mathcal{O}(\hbar^4).
\end{eqnarray}

By using the complete basis for the Clifford algebra \cite{Elze:1986qd,Vasak:1987um}, we may decompose the Wigner functions into
\begin{align}
&
S^<=\mathcal{S}+ i\mathcal{P}\gamma^5+ \mathcal{V}_{\mu}\gamma^\mu+\mathcal{A}_{\mu}\gamma^5\gamma^{\mu}+ \frac{\mathcal{S}_{\mu\nu}}{2}\spinGamma^{\mu\nu},
\nn
\\
&
S^>=\bar{\mathcal{S}}+ i\bar{\mathcal{P}}\gamma^5+ \bar{\mathcal{V}}_{\mu}\gamma^\mu+\bar{\mathcal{A}}_{\mu}\gamma^5\gamma^{\mu}+ \frac{\bar{\mathcal{S}}_{\mu\nu}}{2}\spinGamma^{\mu\nu}
\label{spinor-decomp}
,
\end{align}
where $\spinGamma^{\mu\nu}=i[\gamma^{\mu},\gamma^{\nu}]/2$ and $\gamma^5=i\gamma^0\gamma^1\gamma^2\gamma^3$. 
We shall then focus on $\mathcal{V}^{\mu}$ and $\mathcal{A}^{\mu}$, which give rise to the vector-charge and axial-charge currents through $J_{V}^{\mu}=4\int_q\mathcal{V}^{\mu}$ and $J_{5}^{\mu}=4\int_q\mathcal{A}^{\mu}$, where $\int_{q}\equiv\int d^4q/(2\pi)^4$. In fact, from field theory, the axial-charge current can be regarded as a spin current for fermions. One may also establish a direct connection between $\mathcal{A}^{\mu}$ and the momentum spectrum of spin polarization from the modified Cooper-Frye formula \cite{Becattini2013a}. See e.g., Refs.~\cite{Fang:2016vpj,Yang:2018lew} and Appendix \ref{app_AM_decomp} for references. It is worthwhile to note that the axial-charge currents engendered by magnetic fields and vorticity in equilibrium, known as the chiral separation effect and axial vortical effect, could be thus pertinent to the spin polarization particularly in the case with mass corrections \cite{Gorbar2013,Buzzegoli:2017cqy,Lin:2018aon}. See also Refs.~\cite{Huang:2013iia,Jiang:2014ura,Pu:2014cwa,Pu:2014fva} for axial-charge currents triggered by electric fields. 
Similarly, it is useful to carry out the same spinor-basis decomposition for the self-energies, 
\begin{eqnarray}\nonumber
&&\Sigma^<=\Sigma_S+ i\Sigma_P\gamma^5+ \Sigma_{V\mu}\gamma^\mu+\Sigma_{A\mu}\gamma^5\gamma^{\mu}+ \frac{\Sigma_{T\mu\nu}}{2}\spinGamma^{\mu\nu},
\\
&&\Sigma^>=\bar{\Sigma}_S+ i\bar{\Sigma}_P\gamma^5+ \bar{\Sigma}_{V\mu}\gamma^\mu+\bar{\Sigma}_{A\mu}\gamma^5\gamma^{\mu}+ \frac{\bar{\Sigma}_{T\mu\nu}}{2}\spinGamma^{\mu\nu}
.
\label{decomp_Sigma_general}
\end{eqnarray}
From the Kadanoff-Baym equations and decomposition of the Wigner functions and of the self-energies, 
one can derive the master equations leading to the derivation of axial kinetic theory. 

Now, inserting the spinor decompositions (\ref{spinor-decomp}) and (\ref{decomp_Sigma_general}) 
into the Kadanoff-Baym equation (\ref{KB_eq}), 
we find ten equations according to the orthogonal property of the spinor basis. 
This calculation is tedious mostly due to the manipulation of the spinor structures. 
Nevertheless, it is a straightforward substitution and further decomposition of the products of the gamma matrices. 
For brevity, we provide the intermediate steps in Appendix~\ref{sec:spinor-decomp} 
and here show the results that serve as a set of the master equations 
for the derivation of quantum kinetic theories [same as Eqs.~(\ref{eq:Eq-sum0}) and (\ref{eq:Eq-diff0})]: 
\begin{subequations}
	\begin{eqnarray}
	\label{eq:Eq-sum0-main}
	&&\label{M2-main} 
	m \cS = \Pi^{\mu} \mathcal{V}_{\mu} 
	{-\frac{\hbar^2}{4}[ \widehat{ \Sigma_{S} \cS} - \widehat{\Sigma_{P}  \cP}  +  \widehat{\Sigma_{V\mu}  \cV^\mu} -   \widehat{\Sigma_{A\mu}  \cA^\mu} 
		+ \frac{1}{2} \widehat{\Sigma_{T\mu\nu} \cS^{\mu\nu}}   ]_\text{P.B.}+\mathcal{O}(\hbar^3)}
	,
	\\
	&&\label{M5-main} 
	m\cP =  - \frac{ \hbar}{2} ( \tilde \cD_{\mu}\mathcal{A}^{\mu} - \widehat {  \Sigma_{A\mu} \cV^\mu}) 
	{-\frac{\hbar^2}{4}[  \widehat{\Sigma_{S}   \cP} +   \widehat{\Sigma_P  \cS} 
		+  \frac{1}{4} \epsilon_{\mu\nu\alpha\beta} \widehat{ \Sigma_{T}^{\mu\nu} \cS^{\alpha\beta}} ]_\text{P.B.}+\mathcal{O}(\hbar^3)}
	,
	\\\nonumber
	&&\label{M7-main}
	2 \Pi_{\alpha}\mathcal{S}  - \hbar \tilde \cD^{\nu}\mathcal{S}_{\nu\alpha} - 2m \cV_\alpha
	- \hbar (  \widehat { \Sigma_P  \cA_\alpha } -\widehat { \Sigma_{A\alpha}  \cP } -  \widehat {\Sigma_{T\mu\alpha} \cV^\mu } ) 
	\\
	&&={\frac{\hbar^2}{2}[\widehat{ \Sigma_{S} \cV_\alpha}  + \widehat{ \Sigma_{V \alpha} \cS} 
		+  \frac{1}{2}  \epsilon_{\mu\nu\lambda\alpha}
		( \widehat{ \Sigma_{A}^{\mu}  \cS^{\nu\lambda}}  + \widehat{ \Sigma_{T}^{\mu\nu}  \cA^{\lambda}} )
		]_\text{P.B.}+\mathcal{O}(\hbar^3)}
	,
	\\
	&&\label{M9-main}
	\hbar \tilde \cD_{\alpha}\mathcal{P}  - \epsilon_{\alpha\nu\rho\sigma}\Pi^{\sigma}\mathcal{S}^{\nu\rho} 
	-  2m \cA_\alpha
	-\hbar (   \widehat { \Sigma_P\cV_\alpha }   +\widehat {\Sigma_{A}^\mu \cS_{\mu\alpha} }
	- \widehat { \Sigma_{T\mu\alpha} \cA^{\mu} } )\notag
	\\
	&&={\frac{\hbar^2}{2}[ \widehat{ \Sigma_{S}  \cA_\alpha}   + \widehat{ \Sigma_{A\alpha}   \cS}
		+ \frac{1}{2}  \epsilon_{\mu\nu\lambda\alpha}
		( \widehat{ \Sigma_{V}^{\mu}  \cS^{\nu\lambda}}  + \widehat{ \Sigma_{T}^{\mu\nu}  \cV^{\lambda}} )
		]_\text{P.B.}+\mathcal{O}(\hbar^3)}
	,
	\\
	&&\label{M4-main} \label{replace_3-main}\nonumber
	m \cS_{\alpha\beta} 
	+ \epsilon_{\alpha\beta\rho\sigma} \Pi^{\rho}\mathcal{A}^{\sigma} 
	-\frac{ \hbar }{2} ( \tilde \cD_{[\alpha}\mathcal{V}_{\beta]}
	- \widehat { \Sigma_{A[\alpha}   \cA_{\beta]} } +  \widehat {\Sigma_{T\mu[\alpha} \cS^\mu_{\ \,\beta]} } )
	\\
	&&=
	{-\frac{\hbar^2}{4}
		[ \widehat{ \Sigma_{S}   \cS_{\alpha\beta}} +  \widehat{ \Sigma_{T\alpha\beta}   \cS}
		-  \epsilon_{\mu\nu\alpha\beta} \widehat{ \Sigma_{V}^{\mu} \cA^\nu}
		+   \epsilon_{\mu\nu\alpha\beta} \widehat{ \Sigma_{A}^{\mu}   \cV^\nu} 
		-  \frac{1}{2}  \epsilon_{\mu\nu\alpha\beta}  (\widehat{ \Sigma_{P}  \cS^{\mu\nu}}  
		+  \widehat{ \Sigma_{T}^{\mu\nu}  \cP} )
		]_\text{P.B.}} 
			+\mathcal{O}(\hbar^3)
	,\notag\\
	\end{eqnarray}
\end{subequations}
and 
\begin{subequations}
	\label{eq:Eq-diff0-main}
	\begin{eqnarray}
	&&\label{M1-main} 
	\tilde \cD^{\mu} \cV_\mu =
	-  \widehat{ \Sigma_{S}  \cS} + \widehat{ \Sigma_{P} \cP } 
	+  \widehat{ \Sigma_{A\mu} \cA^\mu }
	- \frac{1}{2} \widehat{ \Sigma_{T\mu\nu} \cS^{\mu\nu}  } +\mathcal{O}(\hbar^3)
	,
	\\
	&&\label{M6-main}
	2\Pi^{\nu}\mathcal{A}_{\nu} =
	- \hbar(  \widehat{ \Sigma_{S}  \cP} + \widehat{ \Sigma_P \cS }
	+  \frac{1}{ 4 } \epsilon_{\mu\nu\alpha\beta} \widehat{\Sigma_{T}^{\mu\nu} \cS^{\alpha\beta}} ) 
	{-\frac{\hbar^2}{2}[ -  \widehat{ \Sigma_{V\alpha} \cA^\alpha} +  \widehat{ \Sigma_{A\alpha}  \cV^\alpha}]_\text{P.B.}+\mathcal{O}(\hbar^3)}
	,
	\\
	&&\label{M8-main}\nonumber
	2 \Pi^{\nu} \mathcal{S}_{\nu\alpha}+  \hbar \tilde \cD_{\alpha}\mathcal{S}
	+\hbar (  \widehat{ \bar \Sigma_{S} \cV_\alpha}
	+  \frac{1}{2}  \epsilon_{\mu\nu\lambda\alpha}
	( \widehat{ \Sigma_{A}^{\mu} \cS^{\nu\lambda} } + \widehat{ \Sigma_{T}^{\mu\nu} \cA^{\lambda} } )
	)
	\\
	&&=
	{\frac{\hbar^2}{2}[   \widehat{ \Sigma_P  \cA_\alpha} + \widehat{ \Sigma_{V}^{\mu}  \cS_{\mu\alpha}} - \widehat{ \Sigma_{A{\alpha}}  \cP} +  \widehat{ \Sigma_{T\alpha\beta}  \cV^\beta}  ]_\text{P.B.}+\mathcal{O}(\hbar^3)}
	,
	\\
	&& \label{M10-main}\nonumber
	2 \Pi_{\alpha}\mathcal{P}  
	+ \frac{\hbar}{2}\epsilon_{\alpha\nu\rho\sigma} 
	( \tilde \cD^{\nu}\mathcal{S}^{\rho\sigma} +   \widehat{ \Sigma_{T}^{\nu\rho} \cV^{\sigma}} )
	-\hbar ( \widehat{ \Sigma_{S} \cA_\alpha }  + \widehat{ \Sigma_{A\alpha}  \cS} )
	\\
	&&=1-{\frac{\hbar^2}{2}[  \widehat{ \Sigma_P \cV_\alpha} -   \widehat{ \Sigma_{V{\alpha}}   \cP} 
		+  \widehat{\Sigma_{A}^\mu   \cS_{\mu\alpha}} 
		+  \widehat{\Sigma_{T\alpha\beta}  \cA^{\beta}}  ]_\text{P.B.}+\mathcal{O}(\hbar^3)}
	,
	\\
	&&\label{M3-main}\nonumber
	\Pi_{[\alpha}\mathcal{V}_{\beta]} 
	+ \frac{\hbar}{2}\epsilon_{\alpha\beta\mu\nu} 
	( \tilde \cD^{\mu}\mathcal{A}^{\nu}    -   \widehat{ \Sigma_{A}^{\mu}  \cV^\nu} )
	- \frac{\hbar}{2} ( \widehat{ \Sigma_{S} \cS_{\alpha\beta} } +  \widehat{ \Sigma_{T\alpha\beta}  \cS} )
	+ \frac{\hbar}{4}  \epsilon_{\mu\nu\alpha\beta} (  \widehat{ \Sigma_{P}  \cS^{\mu\nu}  }  + \widehat{ \Sigma_{T}^{\mu\nu} \cP} )
	\\
	&&={\frac{\hbar^2}{4}[  \widehat{ \Sigma_{V{[}\alpha}    \cV_{\beta{]}}} {-}  \widehat{ \Sigma_{A{[}\alpha}    \cA_{\beta{]}}} 
		{+}  \widehat{ \Sigma_{T\mu{[}\alpha}  \cS^\mu_{\ \,\beta{]}}}]_\text{P.B.}+\mathcal{O}(\hbar^3)}
	,
	\end{eqnarray}
\end{subequations}
where $\epsilon_{\mu\nu\rho\sigma}$ is the totally antisymmetric tensor with $\epsilon_{0123}=-1$.
In the above, we introduced a few shorthand notations. 
$[AB]_\text{P.B.}=\{A(q,X),B(q,X)\}_\text{P.B.}$ is a shorthand notation for the Poisson bracket. 
We use $   \widehat { X Y  }=  \bar X Y - X \bar Y   $, where $  X$ and $Y  $ are the coefficients of the Clifford decomposition of the propagators and self-energies. %, e.g., $ \widehat{ \{\Sigma_{V\mu}, \cV_\nu\}_{\star}} = \{\bar \Sigma_{V\mu}, \cV_\nu\}_{\star} -  \{\Sigma_{V\mu}, \bar \cV_\nu\}_{\star}$. 
We also use $\tilde \cD_\mu \M = \nabla_\mu \M + \widehat { \Sigma_{V\mu} \M\ }+\mathcal{O}(\hbar^{2})$.\footnote{
The original definition before the $hbar $ expansion is given below Eq.~(\ref{eq:A16}). 
}

As has been known in light of the derivation of the free-streaming case without the collision term~\cite{Hattori:2019ahi}, 
one can first reduce the number of variables to solve the above master equations. 
Namely, one can replace $\mathcal{S}$, $\mathcal{P}$ and $\mathcal{S}_{\mu\nu}$ 
by $\mathcal{V}^{\mu}$ and $\mathcal{A}^{\mu}$ by the use of Eqs.~(\ref{M2-main}), (\ref{M5-main}) and (\ref{M4-main}). 
This procedure is explained in Appendix~\ref{sec:elimination}. 
Since $ \cS^{\mu\nu} $ is contracted with $\Sigma_{T\mu\nu}$ and $\bar{\Sigma}_{T\mu\nu}$ on the right-hand side, 
one may not express $\mathcal{S}_{\mu\nu}$ as an explicit function of $\mathcal{V}_{\mu}$ and $\mathcal{A}_{\mu}$ only. 
Nevertheless, assuming that the interaction is sufficiently weak, we may drop the nonlinear terms in the self-energy. 
Maintaining the linear terms in the self-energies 
and the explicit $  \hbar $ dependence up to $ {\mathcal O}(\hbar^1) $, 
$\mathcal{S}$, $\mathcal{P}$, and $\mathcal{S}_{\mu\nu}$ are expressed as [cf. Eq.~(\ref{eq:SPS})]
\begin{subequations}
\begin{eqnarray}
\label{replace_Smunu}
\cS_{\alpha\beta} 
&=&
- \frac{1}{m} \epsilon_{\alpha\beta\rho\sigma} q^{\rho}\mathcal{A}^{\sigma}
+ \frac{ \hbar }{2m} \{  \cD_{[\alpha}\mathcal{V}_{\beta]}
- \widehat { \Sigma_{A[\alpha}   \cA_{\beta]} } 
+  {\frac{q_{\mu}}{m}}  \epsilon_{\ \ \ \, [\alpha}^{\mu \rho\sigma} \widehat{ \Sigma_{T\beta] \rho}  \mathcal{A}_{\sigma} }  \,\},
\\\label{replace_S}
\cS &=& \frac{q^{\mu}}{m} \mathcal{V}_{\mu},
\\\label{replace_P} 
\cP &=&  - \frac{ \hbar}{2m} ( \cD_{\mu}\mathcal{A}^{\mu} - \widehat {  \Sigma_{A\mu} \cV^\mu})
,
\end{eqnarray}
\end{subequations}
where we define $ \cD_\mu \M = \nabla_\mu \M + \widehat { \Sigma_{V\mu} \M\ }$ and the antisymmetrization $ T_{[\mu\nu ]} = T_{\mu\nu} - T_{\nu\mu} $.
After elimination of those variables, 
the other master equations up to the linear order in the self-energies and $\mathcal{O}(\hbar)$ read 
[same as Eqs.~(\ref{eq:Eq-sum}) and (\ref{eq:Eq-diff})]

\begin{subequations}\label{full_Master_eq}
	\begin{eqnarray}
	&&q^{\mu}  ( q \cdot \mathcal{V})    - m ^2 \cV^\mu
	= \frac{ \hbar }{2}  [ m (  \widehat { \Sigma_P  \cA^\mu }  + \widehat {\Sigma_{T}^{\mu\alpha} \cV_\alpha } ) 
	- 2 ( \tilde F^{\mu\beta} {\mathcal{A}}_\beta 
	+ \frac{1}{2}  \epsilon^{\mu \alpha\beta\gamma}  q_{\alpha} \Delta_\beta \mathcal{A}_{\gam} ) 
	-   \epsilon^{\mu \alpha \beta\gam} q_\alpha \widehat{ \Sigma_{V\beta} \cA_\gam }
	]
	,
	\label{M7_N}
	\\\nonumber
	&&q^2 \mathcal{A}^\mu -  q^\mu q \cdot \mathcal{A}  -  m^2 \cA^\mu
	\\\label{M9_N}
	&& 
	=  
	\frac{1}{2}\hbar \Bigl[ m (   \widehat { \Sigma_P\cV^\mu }   + \widehat { \Sigma_{T}^{\mu\alpha} \cA_{\alpha} } ) 
	+ \epsilon^{\mu\alpha\beta\gam} q_{\alpha} \cD_{\beta}\mathcal{V}_\gam
	- 2  \epsilon^{\mu\alpha\beta\gam} q_{\alpha} \widehat {\Sigma_{A\beta}  \mathcal{A}_\gam }
	-{\frac{1}{m}} \epsilon^{\mu}_{\ \, \alpha\beta\gamma}    \epsilon^{ \beta\lambda\rho\sigma } q^{\alpha}  q_{\lambda}
	\widehat{   \Sigma_{T \rho}^{\gam}   \mathcal{A}_{\sigma} }
	\Bigr]
	,
	\\
	&&\label{M1_N} 
	\cD^{\mu} \cV_\mu =
	\widehat{ \Sigma_{A\mu} \cA^\mu }
	-  \frac{1}{2m} [ 2 q_\mu \widehat{ \Sigma_{S}   \mathcal{V}  ^\mu } 
	-  \epsilon_{\mu\nu\alpha\beta} q^{\alpha} \widehat{ \Sigma_{T}^{\mu\nu}  \mathcal{A}^{\beta}   } 
	+ \hbar \widehat{ \Sigma_{P}  (\nabla \cdot  \mathcal{A} )} 
	+ \frac{\hbar}{2} \widehat{ \Sigma_{T}^{\mu\nu}  \nabla_{[\mu}\mathcal{V}_{\nu]}  } 
	],
	\\
	&&\label{M6_N}
	q \cdot \mathcal{A}  =
	- \frac{\hbar}{2m}  q_\mu ( \widehat{ \Sigma_P  \mathcal{V}^\mu }
	+  \widehat{\Sigma_{T}^{\mu\nu}  \mathcal{A}_{\nu}  } ),
	\\
	&&\label{M8_N}
	\cD_{\mu}  (q \cdot \cV) - q^{\nu}  D_{[\mu}\mathcal{V}_{\nu]} 
	\nn
	\\\nonumber
	&&%\hspace{5cm}
	= -  m  \widehat{ \Sigma_{S} \cV_\mu}
	+ q_{[\mu}  \widehat { \Sigma_{A \alpha]}   \cA^{\alpha} } 
	+ {\frac{1}{m}} q^{\nu}   q_{\alpha}  \epsilon_{\ \ \ \, [\mu}^{\alpha \beta\gam} \widehat{ \Sigma_{T\nu] \beta}  \mathcal{A}_{\gam} } 
	+ \frac{1 }{2}  \epsilon_{\mu\nu\rho\sigma} 
	\{
	m \widehat{ \Sigma_{T}^{\nu\rho} \cA^{\sigma} } 
	- \hbar (\partial^\nu F^{ \rho\beta}) \partial_{q\beta} \mathcal{A}^{\sigma} 
	+  \hbar\widehat{ \Sigma_{A}^{\nu} ( \nabla^{\rho}\mathcal{V}^{\sigma}) }
	\}
	\\\nonumber
	&&\quad
	{-\frac{ \hbar  q^{\nu}}{2m} (  
		\widehat {\Sigma_{T\gamma[\nu}\Delta^{\gamma}\mathcal{V}_{\mu]}}
		-\widehat {\Sigma_{T\gamma[\nu}\Delta_{\mu]}\mathcal{V}^{\gamma}
		} )}
	+{\frac{\hbar q^{\nu}}{2}
		[   \frac{1}{m}\widehat{ \Sigma_{T\nu\mu}(q\cdot\mathcal{V})}}
	{	-  \epsilon_{\nu\mu\rho\sigma} \widehat{ \Sigma_{V}^{\rho} \cA^\sigma}
		+   \epsilon_{\nu\mu\rho\sigma} \widehat{ \Sigma_{A}^{\rho}   \cV^\sigma} 
		-  \frac{1}{m}\widehat{ \Sigma_{P} (q_{[\nu}\mathcal{A}_{\mu]})}  
		]_\text{P.B.}}
	\\
	&&\quad+{\frac{\hbar m}{2}[   \widehat{ \Sigma_P  \cA_\mu} -\frac{\epsilon_{\nu\mu\rho\sigma}}{m} \widehat{ \Sigma_{V}^{\nu}(q^{\rho}\mathcal{A}^{\sigma})} +  \widehat{ \Sigma_{T\mu\beta}  \cV^\beta}  ]_\text{P.B.}},
	\\\nonumber
	&& 
	F_{\mu\nu} \cA^\nu - q\cdot \cD \cA_\mu
	+ \frac{ \hbar }{4}  \epsilon_{\mu\nu\rho\sigma}  [ \cD^{\nu} ,   \cD^{\rho} ] \mathcal{V}^{\sigma}
	-{\frac{\hbar}{2m}q_{\mu}[ \widehat{\Sigma_P (q\cdot\mathcal{V})} 
		+  \frac{1}{2} \widehat{ \Sigma_{T}^{\rho\sigma} (q_{[\rho}\mathcal{A}_{\sigma]})} ]_\text{P.B.}} 
	\\\nonumber
	&&
	= m [ \widehat{ \Sigma_{S} \cA_{\mu} } 
	- \frac{1}{2}  \epsilon_{{\mu}\nu\rho\sigma}   \widehat{ \Sigma_{T}^{\nu\rho} \cV^{\sigma}} ]
	+  q_\alpha\widehat{ \Sigma_{A{\mu}} \mathcal{V}^\alpha } - q_\mu\widehat {  \Sigma_{A\alpha} \cV^\alpha} 
	+ \frac{\hbar}{2}\epsilon_{\mu\nu\rho\sigma}  \Delta^{\nu}
	[ 
	\widehat { \Sigma_{A}^{\rho}   \cA^{\sigma} } 
	-  {\frac{1}{m}}q^{\alpha}  \epsilon^{\ \ \ \ \rho}_{\alpha\beta\gamma} \widehat{ \Sigma_{T}^{{\sigma\beta}}  \mathcal{A}^{\gamma} }  
	]
	\\\label{M10_N}
	&&\quad{-}{\frac{\hbar m}{2}[  \widehat{ \Sigma_P \cV_{\mu}} 
		{+}\frac{1}{m}\epsilon_{\mu{\nu}\rho\sigma}  \widehat{\Sigma_{A}^{\nu}  (q^{{\rho}}\mathcal{A}^{{\sigma}})} 
		+  \widehat{\Sigma_{T{\mu\nu}}  \cA^{\nu}}  ]_\text{P.B.}},
	\\
	&&\label{M3_N}
	q^{[\mu}\mathcal{V}^{\nu]} 
	+ \frac{\hbar}{2}\epsilon^{\mu\nu\alpha\beta} 
	( \cD_{\alpha}\mathcal{A}_{\beta}    -   \widehat{ \Sigma_{A\alpha}  \cV_\beta} )
	= - \frac{\hbar}{2m} [  \epsilon^{\mu\nu\alpha\beta} q_{\alpha} \widehat{ \Sigma_{S} \mathcal{A}_{\beta} } 
	- q_\alpha  \widehat{ \Sigma_{T}^{\mu\nu}    \mathcal{V}^\alpha } 
	+ q^{[\mu}  \widehat{ \Sigma_{P}  \mathcal{A}^{\nu]}  
		% - \frac{ \hbar }{4}  \widehat{ \Sigma_{T}^{\alpha\beta} \nabla \cdot  \mathcal{A} }
	}
	]
	.
	\end{eqnarray}
\end{subequations}

As opposed to the free-streaming case, 
the inclusion of the collision terms cause the drastic complication in the master equations 
as seen in the above expressions. 
This hinders finding the perturbative solution for the Wigner functions 
and deriving the consequent quantum kinetic equations. 
In the subsequent sections, we tackle this problem %in the implementation of the collision terms 
that needs to be addressed at both formal and practical levels before describing any kinds of the relaxation dynamics. 
Our main proposal is an appropriate $ \hbar $ counting scheme 
that greatly simplifies the collision terms and 
may work in some applications such as the spin transport in the quark-gluon plasma.

\subsection{Setting the power-counting scheme}

\label{sec:counting}

Thus far, we have not specified the orders of $\mathcal{V}^{\mu}$ and $\mathcal{A}^{\mu}$ in the $ \hbar $ expansion. 
In general, they could be both comparable and different in magnitudes. 
In most of practical situations such as in heavy ion collisions, 
the axial-charge current is usually smaller than the vector-charge current 
since the spin polarization is basically generated by quantum effects. 
This observation motivates us to introduce the power counting: 
%such that $\mathcal{V}^{\mu}\sim\mathcal{O}(\hbar^0)$ and $\mathcal{A}^{\mu}\sim\mathcal{O}(\hbar)$. 
\begin{eqnarray}
\mathcal{V}^{\mu}\sim\mathcal{O}(\hbar^0)
 \quad {\rm and } \quad
\mathcal{A}^{\mu}\sim\mathcal{O}(\hbar)
.
\end{eqnarray}

In light of Eq.~(\ref{replace_S}) and Eq.~(\ref{replace_Smunu}), 
these assignments lead to %$\mathcal{S}\sim\mathcal{O}(\hbar^0)$ and $\mathcal{S}^{\mu\nu}\sim\mathcal{O}(\hbar)$ 
the order counting: 
\begin{eqnarray}
\mathcal{S}\sim\mathcal{O}(\hbar^0)
 \quad {\rm and } \quad
 \mathcal{S}^{\mu\nu}\sim\mathcal{O}(\hbar).
\end{eqnarray}
Also, in the free case, from Eq.~(\ref{replace_P}), we find %$\mathcal{P}\sim\mathcal{O}(\hbar^2)$. 
\begin{eqnarray}
\mathcal{P}\sim\mathcal{O}(\hbar^2)
.
\end{eqnarray}
Consequently, similar power counting will be applied to $\Sigma^{<(>)}$. 
From Eq.~(\ref{M1_N}), it is clear that the ordinary Boltzmann equation at $\mathcal{O}(\hbar^0)$ comprises only $\Sigma_S$ and $\Sigma_{V\mu}$ in collisions when $\mathcal{V}^{\mu}\sim\mathcal{O}(\hbar^0)$ and $\mathcal{A}^{\mu}\sim\mathcal{O}(\hbar)$. 
This implies %$\Sigma_S\sim\mathcal{O}(\hbar^0)$ and  $\Sigma^{\mu}_V\sim\mathcal{O}(\hbar^0)$ 
\begin{eqnarray}
\Sigma_S\sim\mathcal{O}(\hbar^0)
 \quad {\rm and } \quad
 \Sigma^{\mu}_V\sim\mathcal{O}(\hbar^0)
 ,
\end{eqnarray}
and the same counting applies to $\bar{\Sigma}_S$ and  $\bar{\Sigma}^{\mu}_V$. 
Those self-energies are responsible for the ``classical'' collision term. 
On the contrary, other components in $\Sigma^{<(>)}$ come from quantum corrections. 
In light of Eq.~(\ref{M10_N}), one finds that at most 
%$\Sigma^{\mu}_A\sim\mathcal{O}(\hbar)$ and  $\Sigma^{\mu\nu}_T\sim\mathcal{O}(\hbar)$
\begin{eqnarray}
\Sigma^{\mu}_A\sim\mathcal{O}(\hbar^1) 
 \quad {\rm and } \quad
 \Sigma^{\mu\nu}_T\sim\mathcal{O}(\hbar^1)
 .
\end{eqnarray}
This is required to balance the orders of free-streaming and collision parts, 
and is also consistent with Eq.~(\ref{replace_P}).

Physically, the ``classical'' Boltzmann equation only incorporates the vector-charge conservation. In order to have nonzero $\Sigma_{A}^{\mu}$ or $\Sigma^{\mu\nu}_T$, either the scattered fermion or gluon should carry nonzero chirality imbalance or spin (angular-momentum), which has to come from quantum corrections at least at $\mathcal{O}(\hbar)$ since our power counting has implied the suppression of spin currents compared to the vector-charge currents. Albeit there exists no explicit restriction for $\Sigma_P$ from generic master equations, it is expected that the presence of $\Sigma_P$ has to be induced by nonzero pseudo-scalar condensate, which should be at $\mathcal{O}(\hbar^2)$ from the consistency with the anomaly equation (mass correction upon the chiral anomaly). Nonetheless, due to the lack of a rigorous proof for the order of $\Sigma_P$ from Kadanoff-Baym equations, we will naively take $\Sigma_{P}\sim\mathcal{O}(\hbar^0)$ for completeness. 
We hence establish our power counting in that we have 
%$\Sigma_S\sim\mathcal{O}(\hbar^0)$ and  $\Sigma^{\mu}_V\sim\mathcal{O}(\hbar^0)$, 
\begin{eqnarray}
\Sigma_S\sim\mathcal{O}(\hbar^0) 
 \quad {\rm and } \quad
\Sigma^{\mu}_V\sim\mathcal{O}(\hbar^0)
,
\end{eqnarray}
while 
%$\Sigma^{\mu}_A\sim\mathcal{O}(\hbar)$,  $\Sigma^{\mu\nu}_T\sim\mathcal{O}(\hbar)$, 
%and $\Sigma_{P}\sim\mathcal{O}(\hbar^0)$ in our power counting. 
\begin{eqnarray}
\Sigma^{\mu}_A\sim\mathcal{O}(\hbar^1),  
\quad 
\Sigma^{\mu\nu}_T\sim\mathcal{O}(\hbar^1) 
 \quad {\rm and } \quad
 \Sigma_{P}\sim\mathcal{O}(\hbar^0)
 .
\end{eqnarray}
All components certainly can also have higher-order corrections in $\hbar$ 
depending on the details of collisions for different systems.

\subsection{Effective master equations and constrains}

\label{sec:effective-eqs}

By implementing the power counting and the results shown in Eq.~(\ref{full_Master_eq}), the master equations obtained from the Kadanoff-Baym equations at ``the leading order $\mathcal{O}(\hbar^0)$'' now read    
\begin{subequations}\label{eq:master-eqs}
	\begin{eqnarray}
	\label{MA_1}
	&&
	\mathcal{D}^{\mu} \cV_\mu = -  \frac{1}{m}  q_\mu \widehat{ \Sigma_{S}   \mathcal{V} ^\mu } ,
	\\\label{MA_2}
	&& 
	q^{[\mu}\mathcal{V}^{\nu]} = 0 ,
	\\\label{MA_3}
	&&(q^2-m^2)\mathcal{V}_{\mu}=0,
	\\\label{MA_4}
	&&
	q \cdot \mathcal{A} =   - \frac{\hbar}{2m}  q_\mu  \widehat{ \Sigma_P  \mathcal{V}^\mu } 
	,
	\\\label{MA_5}
	&&(q^2-m^2)\mathcal{A}^{\mu}
	=
	\frac{ \hbar}{2} \epsilon^{\mu\alpha\beta\gam} q_{\alpha} D_{\beta}\mathcal{V}_\gam,
	\\\nonumber
	&&
	q\cdot \cD \cA_\mu - F_{\mu\nu} \cA^\nu+{\frac{\hbar}{2m}q_{\mu}[ \widehat{\Sigma_P (q\cdot\mathcal{V})}]_\text{P.B.}-\frac{\hbar m}{2}[\widehat{ \Sigma_P \cV_\mu}]_\text{P.B.}}
	\\
	&&=  \frac{ \hbar }{4}  \epsilon_{\mu\nu\rho\sigma}  [ \cD^{\nu} ,   \cD^{\rho} ] \mathcal{V}^{\sigma}
	- m \big( \widehat{ \Sigma_{S} \cA_{\mu} } 
	- \frac{1}{2}  \epsilon_{\mu\nu\rho\sigma}   \widehat{ \Sigma_{T}^{\nu\rho} \cV^{\sigma}} \big)
	-  q_\alpha \widehat{ \Sigma_{A\mu} \mathcal{V}^\alpha } +q_{\mu} \widehat {  \Sigma_{A\alpha} \cV^\alpha} 
	,
	\label{MA_6}
	\\\label{MA_red}
	&&
	q \cdot \cD \cV_\mu - F_{\mu\nu} \cV^\nu= -  m  \widehat{ \bar \Sigma_{S} \cV_\mu}  
	.
	\end{eqnarray}
\end{subequations}
%where we introduced shorthand notations $\mathcal{D}^{\mu}\mathcal{M}\equiv \Delta^{\mu}\mathcal{M}+ \widehat{\Sigma_{V}^{\mu}\mathcal{M}} $ and $   \widehat { X Y  } \equiv  \bar X Y - X \bar Y   $ which goes like $ \widehat{ \Sigma_{V\mu} \cV_\nu} = \bar \Sigma_{V\mu} \cV_\nu -  \Sigma_{V\mu} \bar \cV_\nu$ (see Appendix~\ref{sec:master-eqs}). In addition, $[ \widehat{\Sigma_P (q\cdot\mathcal{V})}]_\text{P.B.}=\{\bar{\Sigma}_P,q\cdot\mathcal{V} \}_\text{P.B.}-\{\Sigma_P,q\cdot\bar{\mathcal{V}} \}_\text{P.B.}$,  $[ \widehat{\Sigma_P \mathcal{V}_{\mu}}]_\text{P.B.}=\{\bar{\Sigma}_P,\mathcal{V}_{\mu} \}_\text{P.B.}-\{\Sigma_P,\bar{\mathcal{V}}_{\mu} \}_\text{P.B.}$, and $\epsilon_{\mu\nu\rho\sigma}$ is the totally antisymmetric tensor with $\epsilon_{0123}=-1$.
Note that we have to keep the $\mathcal{O}(\hbar)$ terms linear to $\mathcal{V}^{\mu}$ in Eqs.~(\ref{MA_5}) and (\ref{MA_6}) since $\mathcal{A}^{\mu}\sim\mathcal{O}(\hbar)$. In addition, we consider weakly coupled systems and thus drop the $\mathcal{O}(\Sigma^2)$ terms. 
One can further implement 
\begin{eqnarray}
\frac{\hbar}{2m}q_{\mu}[ \widehat{\Sigma_P (q\cdot\mathcal{V})}]_\text{P.B.}-\frac{\hbar m}{2}[\widehat{ \Sigma_P \cV_\mu}]_\text{P.B.}=\frac{\hbar}{2m}\widehat{(\partial_{\mu}\Sigma_P)q\cdot\mathcal{V}}
\end{eqnarray}
to simplify Eq.~(\ref{MA_6}).

From Eqs.~(\ref{MA_2}) and (\ref{MA_3}), one immediately obtains the leading-order solution for the vector part, 
\begin{eqnarray}
\mathcal{V}^{\mu}=2\pi \delta(q^2-m^2)q^{\mu}f_V
\label{eq:V-LO}
,
\end{eqnarray}
where $f_V(q,X)$ denotes the vector-charge distribution function. For completeness, we should, in principle, multiply $\mathcal{V}^{\mu}$ by the sign function for energy 
to include antiparticles. For notational simplicity, we will mostly focus on just the positive-energy solution thorough out the paper.
 For $\bar{V}^{\mu}$, we simply replace $f_V$ by $\bar{f}_V=1-f_V$ as the vector-charge distribution function for an outgoing fermion. We may also easily show that Eq.~(\ref{MA_red}) is a redundant equation, which can be derived from Eqs.~(\ref{MA_1})-(\ref{MA_3}). Plugging the leading-order solution for $\mathcal{V}^{\mu}$ into Eq.~(\ref{MA_1}), one acquires the SKE as the usual Boltzmann equation 
\begin{eqnarray}\label{SKE}
\delta(q^2-m^2)\Big(q\cdot\Delta f_V
+ 
q_\mu \widehat{ \Sigma_V^\mu  f_{V} } + m  \widehat{  \Sigma_S   f_{V} } 
\Big)
+ \mathcal{O}(\hbar) = 0
.
\end{eqnarray}
For the axial part $\mathcal{A}^{\mu}$, we have to solve Eqs.~(\ref{MA_4}) and (\ref{MA_5}) up to $\mathcal{O}(\hbar)$. 
With the leading-order solution (\ref{eq:V-LO}) constrained above, 
one however finds that the collisional term vanishes on the left-hand side of Eq.~(\ref{MA_5}). 
This means that the collisional effects do not modify the dispersion relation for $\mathcal{A}^{\mu}$ at the leading order. 
Nevertheless, the same as the collisionless case \cite{Hattori:2019ahi}, 
we are not able to uniquely determine the magnetization-current term 
at the next-to-leading order from the master equations. In the absence of a background field, one can determine the corresponding term 
by explicitly solving the free Dirac equation \cite{Hattori:2019ahi}. 
In addition, we refer to the correspondence between the cases 
with and without a background field in the massless limit, 
where the latter is given by the former with the simple replacement of the partial derivative 
by the derivative operator with a background field. 
Based on those observations, we could generalize the free-theory form to the case with background fields and collisions. 

Namely, we ``postulate''{\footnote{Thanks to the power-counting scheme, the generalization of the magnetization-current term with collisions here is rather natural since $\Sigma_{V}^{\mu}$ and $\bar{\Sigma}_{V}^{\mu}$ are only vectors at $\mathcal{O}(\hbar^0)$ in self-energies, which can then be coupled to $S^{\mu\nu}_{m(n)}$.}}
\begin{eqnarray}\label{axial_sol}
\mathcal{A}^{\mu}=2\pi\Big[\delta(q^2-m^2)\Big(a^{\mu}f_A+\hbar S^{\mu\nu}_{m(n)}\mathcal{D}_{\nu}f_V+{\frac{\hbar}{2m}q^{\mu}C_P[f_V]}\Big)
+\hbar\tilde{F}^{\mu\nu}q_{\nu}\delta'(q^2-m^2)f_V\Big],
\end{eqnarray} 
where $C_{P}[f_V]= - \widehat{ \Sigma_{P} {f}_V } $ 
and 
\begin{eqnarray}
S^{\mu\nu}_{m(n)}=\frac{\epsilon^{\mu\nu\alpha\beta}q_{\alpha}n_{\beta}}{2a\cdot n}=\frac{\epsilon^{\mu\nu\alpha\beta}q_{\alpha}n_{\beta}}{2(q\cdot n+m)},
\end{eqnarray}
which is generalization of the $\mathcal{A}^{\mu}$ found in Ref.~\cite{Hattori:2019ahi} 
with the replacement of $\Delta_{\nu}f_V$ by $\mathcal{D}_{\nu}f_V$. 
Here, $a^{\mu}$ represents a spin four-vector satisfying $q\cdot a=q^2-m^2$ and $f_A$ denotes the axial-charge distribution function. 

At $\mathcal{O}(\hbar)$, the dispersion relation is modified by, e.g., the magnetic-moment coupling from the last term in Eq.~(\ref{axial_sol}) with $\delta'(q^2-m^2)=d\delta(q^2-m^2)/dq^2$. On the other hand, the $S^{\mu\nu}_{m(n)}$ term corresponds to the so-called magnetization-current term led by spin-orbit interaction, which depends on a frame vector $n^{\mu}$ specifying the spin basis. The presence of such a term implies the frame dependence of $f_V$ because of the frame invariance of full $\mathcal{A}^{\mu}$. In the massless limit, $a^{\mu}=q^{\mu}$ 
according to the spin enslavement specified by the helicity. 
The expression in Eq.~(\ref{axial_sol}) then agrees with the solution directly solved from Kadanoff-Baym equations of Weyl fermions \cite{Hidaka:2016yjf}. Furthermore, for the solution of $\bar{\mathcal{A}}^{\mu}$, we simply replace $f_A$ and $f_V$ in Eq.~(\ref{axial_sol}) by $\bar{f}_A$ and $\bar{f}_V$, respectively. Nonetheless, unlike $\bar{f}_V=1-f_V$, we have $\bar{f}_A=-f_A$ due to its origin from the expectation value of the fermionic density operator in spinor space (see Ref.~\cite{Hattori:2019ahi} for a detailed definition of $f_A$ in field theory).     
As a consequence of our power counting, Eq.~(\ref{axial_sol}) corresponds to the leading-order solution for $\mathcal{A}^{\mu}$ 
starting at $\mathcal{O}(\hbar)$. Albeit the power counting we apply, we will still dub the terms with $\hbar$ prefactors as the ``quantum corrections'' at $\mathcal{O}(\hbar)$ throughout this paper for convenience. That is, in the rest part of the paper, $\hbar$ is simply a parameter indicating the quantum origin of certain terms, while these terms are in the same order of magnitudes as the classical terms without $\hbar$ prefactors. More precisely, one should distinguish the difference between $\hbar$ terms and $\mathcal{O}(\hbar)$ terms.

\section{Effective axial kinetic theory with collisions}\label{sec_AAKT}
In this section, we elaborate the collision terms for the effective axial kinetic theory 
arising from our $  \hbar$ counting scheme introduced in Sec.~\ref{sec:counting}. 
Assembling the ingredients obtained in Sec.~\ref{sec:effective-eqs}, 
the general expression will be given in Eq.~(\ref{AKE_n}) 
with the collision terms $\hat{\mathcal{C}}^{(n)\mu}_{1}$ and  $\hbar\hat{\mathcal{C}}^{(n)\mu}_{2}$. 
Here, we maintain the general frame vector $ n^\mu $.

We provide a simpler form in Eq.~(\ref{AKE_v1}) by choosing a specific frame vector $ n^\mu=(1,0) $ 
and dropping the background electromagnetic field, which will be useful for applications. 
A further alternative expression (\ref{AKE_v2}) enables us to take the massless limit 
and to confirm the agreement to the (massless) chiral kinetic theory obtained in Ref.~\cite{Hidaka:2016yjf}. 
In Sec.~\ref{subsec_AKE_vr}, we take another frame vector $n^{\mu}=n^{\mu}_{r}(q)=q^{\mu}/m$ 
specifying the rest frame of a massive fermion. 
$ \hat{\mathcal{C}}^{(n_r)\mu}_{1,2} $ in Eq.~(\ref{AKE_nr}) 
adds collisional effects to the free-streaming part investigated in Refs.~\cite{Weickgenannt:2019dks,Gao:2019znl}, 
and leads to a useful expression of the effective axial kinetic theory (\ref{AKE_vr}).

Besides the explicit $  \hbar $ dependences originating from those in the Kadanoff-Baym equation (\ref{eq:KBequation}), 
there are implicit $ \hbar $ dependences contained in the self-energies $ \Sigma $. 
In Sec.~\ref{subsec_decompCQ}, we therefore sort the $ \hbar $ dependences 
in the actual order of our counting scheme, which results in the form of 
$\hat{\mathcal{C}}^{\mu}_{\text{cl}} $ and $\hbar\hat{\mathcal{C}}^{(n)\mu}_{\text{Q}}$ in Eq.~(\ref{C_decomp}). 
Finally, in Sec.~\ref{sec:discussions}, we discus the physical meaning of the quantum corrections to the collision terms 
and the conditions to make efficient applications of our effective axial kinetic theories. As noted in the end of previous section, one should bear in mind that $\hat{\mathcal{C}}^{\mu}_{\text{cl}} $ and $\hbar\hat{\mathcal{C}}^{(n)\mu}_{\text{Q}}$ have the same order of magnitude and we hereafter use $\hbar$ as just a parameter indicating the quantum origin of attached terms unless specified.

\subsection{Axial kinetic equation with general frame vector and its massless limit}\label{subsec_AKE_v12}

We now utilize Eqs.~(\ref{MA_6}) and (\ref{axial_sol}) to derive the effective AKE. 
Given that the collisionless (free-streaming) part up to $\mathcal{O}(\hbar)$ has been obtained in Ref.~\cite{Hattori:2019ahi}, we only need to further work out the collisional part. 
This can be carried out by inserting the solution (\ref{axial_sol}) of the constraint equations into Eq.~(\ref{MA_6}). 
By the use of an identity 
\begin{eqnarray}
 \epsilon_{\mu\nu\rho\sigma}  [ \cD^{\nu} ,   \cD^{\rho} ] \mathcal{V}^{\sigma}
 =  \epsilon_{\mu\nu\rho\sigma} 
 \left( \, [ \Delta^{\nu} ,   \Delta^{\rho} ] \mathcal{V}^{\sigma}
+ 2(\Delta^\nu \widehat{  \Sigma_V^\rho) \cV^{\sigma} }
 \right)
 ,
\end{eqnarray}
the AKE with the collisional effects and the general frame vector $n^{\mu}=n^{\mu}(X)$ takes the form 
\begin{eqnarray}\label{AKE_n}
\Box^{(n)}\mathcal{A}^{\mu}=\hat{\mathcal{C}}^{(n)\mu}_1+\hbar\hat{\mathcal{C}}^{(n)\mu}_2.
\end{eqnarray}
Here the free-streaming part is given by \cite{Hattori:2019ahi}
\begin{eqnarray}\nonumber\label{CKT_axial_final}
\Box^{(n)}\mathcal{A}^{\mu}&=&\delta(q^2-m^2)
\Big(q\cdot\Delta (a^{\mu}f_A)+F^{\nu\mu}a_{\nu}f_A\Big)
+\hbar q^{\mu}\Bigg\{\delta(q^2-m^2)\Bigg[(\partial_{\alpha}S_{m(n)}^{\alpha\nu})\Delta_{\nu}+\frac{S^{\alpha\nu}_{m(n)}F_{\alpha\beta}n^{\beta}\Delta_{\nu}}{q\cdot n+m}
\\\nonumber
&&
+S^{\rho\nu}_{m(n)}(\partial_{\rho}F_{\beta\nu})\partial_q^{\beta}\Bigg]
-\delta'(q^2-m^2)\frac{q^{\alpha}\tilde{F}_{\alpha\beta}n^{\beta}}{q\cdot n+m}q\cdot\Delta\Bigg\}f_V
\\\nonumber
&&+\hbar m\Bigg\{\frac{\delta(q^2-m^2)\epsilon^{\mu\nu\alpha\beta}}{2(q\cdot n+m)}\Bigg[m(\partial_{\alpha}n_{\beta})\Delta_{\nu}
+(mn_{\beta}+q_{\beta})\Bigg(\frac{\big(F_{\alpha\rho}n^{\rho}-\partial_{\alpha}(q\cdot n)\big)}{q\cdot n+m}\Delta_{\nu}
\\
&&-(\partial_{\nu}F_{\rho\alpha})\partial_q^{\rho}\Bigg)\Bigg]
+\delta'(q^2-m^2)\frac{(mn_{\beta}+q_{\beta})\tilde{F}^{\mu\beta}}{q\cdot n+m}q\cdot\Delta\Bigg\} f_V.
\end{eqnarray}
Taking $\hbar=0$, the equation above reproduces the so-called Bargmann-Michel-Telegdi (BMT) equation  as a classical kinetic equation for spin transport \cite{BLT_spin}.
On the other hand, the collision terms are
\begin{eqnarray}\nonumber
\hat{\mathcal{C}}^{(n)\mu}_1&=&\delta(q^2-m^2)\Big[
- a^{\mu} q_\nu \widehat{ \Sigma_{V}^\nu {f}_A} 
- m^2 \widehat{ \Sigma_A^{\mu} f_V}  + q^{\mu} q_\nu \widehat{ {\Sigma}_A ^\nu f_V} 
- m \Big(a^{\mu} \widehat{ {\Sigma}_S f_A} 
-\frac{1}{2} \epsilon^{\mu\nu\rho\sigma}  q_{\nu} \widehat{ {\Sigma}_{T{\rho\sigma}} f_V}
\Big)
\Big],
\\
\label{C_n1}
\end{eqnarray}
and
\begin{eqnarray}\nonumber\label{C_nQ}
\hat{\mathcal{C}}^{(n)\mu}_{2}&=&\frac{\delta(q^2-m^2)}{2}\Bigg [\epsilon^{\mu\nu\rho\sigma}
q_{\nu} 
(\Delta_\rho \widehat{  \Sigma_{V\sigma}) f_V }
-m  (\partial^{\mu} \widehat{ \Sigma_P)f_{V} } 
- 2 \widehat{ \Sigma_{V\nu}f_V}
\Big(q\cdot\Delta S^{\mu\nu}_{m(n)}-F^{\mu}_{\,\,\,\lambda}S^{\lambda\nu}_{m(n)}\Big)
\\ \notag
&&
- 2S^{\mu\nu}_{m(n)}\Big(
(q\cdot \widehat{ {\Sigma}_{V}  (\Delta_{\nu}f_V ) } + m \widehat{ {\Sigma}_S  (\Delta_{\nu}f_V ) }
+ (q\cdot\Delta \widehat{ \Sigma_{V\nu}) f_V} \Big)
+ \frac{1}{m}q^{\mu} ( q\cdot\Delta  \widehat{ \Sigma_P) {f}_V}
\Bigg]\\ \notag
&&-S^{\mu\nu}_{m(n)}\Delta_{\nu}\Big(\delta(q^2-m^2)q\cdot\Delta f_V\Big)
	-\delta(q^2-m^2)\big(\Delta_{\nu}S^{\mu\nu}_{m(n)}\big)q\cdot\Delta f_V
\\
&&
- \tilde{F}^{\mu\nu}q_{\nu}\delta'(q^2-m^2) (q\cdot \widehat{ \Sigma_V f_V}+m \widehat{ \Sigma_S f_V} )
.
\end{eqnarray}
Note that $\hat{\mathcal{C}}^{(n)\mu}_1$ implicitly contains the $\hbar$ terms from, e.g., $\Sigma^{\mu}_A$ and $\Sigma_{T\rho\sigma}$, which hence implicitly depends on the frame choice. 

Implementing the leading-order SKE in Eq.~(\ref{SKE}), the following term in $\hat{\mathcal{C}}^{(n)\mu}_{2}$ can be also written as
\begin{eqnarray}\nonumber
&&-S^{\mu\nu}_{m(n)}\Delta_{\nu}\Big(\delta(q^2-m^2)q\cdot\Delta f_V\Big)
\\
&&=-S^{\mu\nu}_{m(n)}\Bigg[\delta(q^2-m^2)\Delta_{\nu}+2q^{\lambda}F_{\lambda\nu}\delta'(q^2-m^2)\Bigg]\big(C_{V}[f_V]+mC_S[f_V]\big),
\end{eqnarray}
where $C_{V}[f_V]=q\cdot \Sigma_{V}\bar{f}_{V}-q\cdot \bar{\Sigma}_{V}f_{V}$ and $C_S[f_V]=\Sigma_S\bar{f}_V-\bar{\Sigma}_Sf_V$. Moreover, one may rewrite
\begin{eqnarray}\nonumber
&&-\big(q\cdot \widehat{ {\Sigma}_{V}  (\Delta_{\nu}f_V ) } + m \widehat{ {\Sigma}_S  (\Delta_{\nu}f_V ) }\big)
\\
&&=\Delta_{\nu} C_{V}[f_V]+q^{\rho}(\Delta_{\nu}\widehat{\Sigma_{V\rho})f_V}
+F^{\rho}_{\,\,\,\nu}\widehat{\Sigma_{V\rho}f_V}
+m(\Delta_{\nu}\widehat{f_V)\Sigma_S}.
\end{eqnarray}
Then we may re-express $\hat{\mathcal{C}}^{(n)\mu}_{2}$ as 
\begin{eqnarray}\nonumber\label{C_nQ_2}
\hat{\mathcal{C}}^{(n)\mu}_{2}&=&\frac{\delta(q^2-m^2)}{2}\Bigg\{\epsilon^{\mu\nu\rho\sigma}
q_{\nu}(\Delta_{\rho}\widehat{\Sigma_{V\sigma})f_V}
+2S^{\mu\nu}_{m(n)}\Big[m(\Delta_{\nu}\widehat{\Sigma_S)f_V}
+F^{\rho}_{\,\,\,\nu}\widehat{\Sigma_{V\rho}f_V}
+q^{\rho}(\Delta_{\nu}\widehat{\Sigma_{V\rho})f_V}
\\\nonumber
&&-(q\cdot\Delta\widehat{\Sigma_{V\nu})f_V}\Big]
+2\widehat{f_V\Sigma_{V\nu}}
\big(q\cdot\Delta S^{\mu\nu}_{m(n)}-F^{\mu}_{\,\,\,\lambda}S^{\lambda\nu}_{m(n)}\big)\Bigg\}
\\\nonumber
&&-\Big[\big(2S^{\mu\nu}_{m(n)}q^{\lambda}F_{\lambda\nu}-\tilde{F}^{\mu\nu}q_{\nu}\big)\delta'(q^2-m^2)+\delta(q^2-m^2)\big(\Delta_{\nu}S^{\mu\nu}_{m(n)}\big)\Big]\big(C_{V}[f_V]+mC_S[f_V]\big)
\\
&&+\frac{\delta(q^2-m^2)}{2m}\big(q^{\mu}(q\cdot\Delta\widehat{\Sigma_{P})f_V}
-m^2(\partial^{\mu}\widehat{\Sigma_P)f_{V}}\big)
.
\end{eqnarray}
In fact, we can further decompose $\hat{\mathcal{C}}^{(n)\mu}_2$ into the piece proportional to $q^{\mu}$, which survives in the massless limit and reproduces the collision term in CKT, and another piece proportional to $m$, which stems from the purely finite-mass correction. This decomposition can be used as a consistency check in the massless limit. Note that the procedure is simply to rewrite $\hat{C}_2^{(n)\mu}$ in the form with a more apparent connection to the CKT despite some of the computational complexity. A similar procedure and comparison are also performed for the free-streaming AKE in Ref.~\cite{Hattori:2019ahi}.  
%although it is expected since both the Wigner functions and Kadanoff-Baym equations derived previously have already had a smooth connection to the massless ones. 
We hence present such a complicated yet straightforward check in Appendix~\ref{app_decomp_AKE} (see Eq.~\ref{C_2n_decomp}), while we will later show such an alternative expression for $\hat{C}_2^{(n)\mu}$ in a simpler case suitable for the application in HIC.  

Although $\hat{\mathcal{C}}^{(n)\mu}_1+\hbar\hat{\mathcal{C}}^{(n)\mu}_2$ given by Eq.~(\ref{C_n1}) and Eq.~(\ref{C_nQ_2}) serves as the generic collision term in the presence of spacetime-dependent background electromagnetic fields and an arbitrary spacetime-dependent frame vector $n^{\mu}(X)$, we may drop unnecessary terms for practical applications in HIC. First, it is more convenient to work with a constant frame vector such that $\partial_{\mu}n^{\nu}=0$. More precisely, we could simply take $n^{\mu}=(1,{\bf 0})$, which also corresponds to the frame choice for the CKT presented in early works obtained from the Berry phase  \cite{Stephanov:2012ki,Son:2012wh}. Note that the choice of a frame vector is analogous to the choice of a gauge, which does not affect the physics in the end (see e.g. Refs.~\cite{Hidaka:2016yjf,Hattori:2019ahi} for comprehensive discussions). Second, the background electromagnetic fields and particularly the magnetic field may only exist in HIC for a rather short period in the pre-equilibrium phase although the finite electric conductivity of QGP may slightly mitigate the decrease of magnetic fields in time \cite{Deng:2012pc,McLerran:2013hla}. It is thus more practical to drop the contributions from electromagnetic fields in the QGP phase. In such a case, the explicit form of $\hat{\mathcal{C}}^{(n)\mu}_1$ remains unchanged. On the other hand, not only $\hbar\hat{\mathcal{C}}^{(n)\mu}_2$ but also the free-streaming part in the AKE become much simpler. The AKE now is given by
\begin{eqnarray}\nonumber\label{AKE_v1}
&&\delta(q^2-m^2)\Bigg\{q\cdot \partial (a^{\mu}f_A)
+a^{\mu} q_\nu \widehat{ \Sigma_{V}^\nu {f}_A} 
+m^2 \widehat{ \Sigma_A^{\mu} f_V}  - q^{\mu} q_\nu \widehat{ {\Sigma}_A ^\nu f_V} 
+m \Big(a^{\mu} \widehat{ {\Sigma}_S f_A} 
-\frac{1}{2} \epsilon^{\mu\nu\rho\sigma}  q_{\nu} \widehat{ {\Sigma}_{T{\rho\sigma}} f_V}
\Big)
\\
&&-\frac{\hbar}{2}\Big[\epsilon^{\mu\nu\rho\sigma}
q_{\nu}(\partial_{\rho}\widehat{\Sigma_{V\sigma})f_V}
+2S^{\mu\nu}_{m(n)}\Big(m(\partial_{\nu}\widehat{\Sigma_S)f_V}
+q^{\rho}(\partial_{\nu}\widehat{\Sigma_{V\rho})f_V}
-(q\cdot\partial\widehat{\Sigma_{V\nu})f_V}\Big)\Big]\Bigg\}=0,
\end{eqnarray}    
where we also took $\Sigma_{P}=\bar{\Sigma}_P=0$ as expected higher-order contributions. Note that the $\hbar$ terms shown above only come from $\hbar\hat{\mathcal{C}}^{(n)\mu}_2$.

By employing the Schouton identity
\begin{eqnarray}\label{Schouten}\nonumber
\eta^{\lambda}_{\mu}\epsilon_{\rho\nu\alpha\beta}-\eta^{\lambda}_{\rho}\epsilon_{\mu\nu\alpha\beta}-\eta^{\lambda}_{\nu}\epsilon_{\rho\mu\alpha\beta}-\eta^{\lambda}_{\alpha}\epsilon_{\rho\nu\mu\beta}-\eta^{\lambda}_{\beta}\epsilon_{\rho\nu\alpha\mu}=0,
\\
\end{eqnarray}
one finds
\begin{eqnarray}
S^{\mu\nu}_{m(n)}q^{\alpha}=q^{\mu}S^{\alpha\nu}_{m(n)}+q^{\nu}S^{\mu\alpha}_{m(n)}+\frac{\epsilon^{\mu\nu\rho\alpha}q_{\rho}}{2}-\frac{\epsilon^{\mu\nu\rho\alpha}}{2(q\cdot n+m)}\big(mq_{\rho}+q^2n_{\rho}\big).
\end{eqnarray}
Given the relation above, the AKE in Eq.~(\ref{AKE_v1}) can be alternatively written as
\begin{eqnarray}\nonumber\label{AKE_v2}
&&\delta(q^2-m^2)\Bigg\{q\cdot \partial (a^{\mu}f_A)
+a^{\mu} q_\nu \widehat{ \Sigma_{V}^\nu {f}_A} 
+m^2 \widehat{ \Sigma_A^{\mu} f_V}  - q^{\mu} q_\nu \widehat{ {\Sigma}_A ^\nu f_V} 
+m \Big(a^{\mu} \widehat{ {\Sigma}_S f_A} 
-\frac{1}{2} \epsilon^{\mu\nu\rho\sigma}  q_{\nu} \widehat{ {\Sigma}_{T{\rho\sigma}} f_V}
\Big)
\\
&&+\hbar\Bigg[q^{\mu}S^{\rho\nu}_{m(n)} \widehat{ (\partial_{\rho} \Sigma_{V\nu})f_V}
-m\Bigg(S^{\mu\nu}_{m(n)}  (\partial_{\nu} \widehat{ \Sigma_S) f_V} 
+ \frac{\epsilon^{\mu\nu\rho\sigma}(q_{\rho}+mn_{\rho})}{2(q\cdot n+m)}
( \partial_{\sigma} \widehat{  \Sigma_{V\nu}) f_V }\Bigg)\Bigg]
\Bigg\}=0,
\end{eqnarray}   
where we rearrange the explicit $\hbar$ corrections in the collision term based on the decomposition for the terms proportional to $q^{\mu}$ and to $m$, respectively. One can now more easily check that Eq.~(\ref{AKE_v2}) reduces to the CKT with a constant frame vector in the absence of electromagnetic fields \cite{Hidaka:2016yjf} in the massless limit by taking $m=0$ and $a^{\mu}=q^{\mu}$. We shall discuss later about the physical interpretations of the explicit $\hbar$ corrections in Eq.~(\ref{AKE_v1}) and in Eq.~(\ref{AKE_v2}) as two mathematically equivalent expressions.

\subsection{Rest-frame expression}\label{subsec_AKE_vr}

Notably, when focusing on massive fermions with mass much greater than the gradient scale, we can set the frame vector at their rest frame $n^{\mu}=n^{\mu}_{r}(q)=q^{\mu}/m$ to simplify both the Wigner functions and AKE. Such a frame choice is also applied in Refs.~\cite{Weickgenannt:2019dks,Gao:2019znl}. Nonetheless, this frame choice is rather different from the previous one when $n^{\mu}(X)$ only depends on spacetime coordinates. In such a case, the magnetization-current term in $\mathcal{A}^{\mu}$ vanishes and the $\mathcal{A}^{\mu}$ reduces to
\begin{eqnarray}
\mathcal{A}^{\mu}=2\pi\Big[\delta(q^2-m^2)a^{\mu}f_A
+\hbar\tilde{F}^{\mu\nu}q_{\nu}\delta'(q^2-m^2)f_V
+\frac{\hbar\delta(q^2-m^2)}{2m}q^{\mu}C_P[f_V]
\Big].
\end{eqnarray}
Accordingly, the AKE from Eq.~(\ref{MA_6}) becomes
	\begin{eqnarray}\label{AKE_nr}
	\Box^{(n_r)}\mathcal{A}^{\mu}=\hat{\mathcal{C}}^{(n_r)\mu}_{1}+\hbar\hat{\mathcal{C}}^{(n_r)\mu}_{2},
	\end{eqnarray}
where $\hat{\mathcal{C}}^{(n_r)\mu}_{\text{1}}$ is the same as Eq.~(\ref{C_n1}) by taking $n^{\mu}=n^{\mu}_r$, while
\begin{eqnarray}\notag
\Box^{(n_r)}\mathcal{A}^{\mu}&=&\delta(q^2-m^2)
\Big(q\cdot\Delta (a^{\mu}f_A)+F^{\nu\mu}a_{\nu}f_A-\frac{1}{2}{\hbar\epsilon^{\mu\nu\rho\sigma}q_{\rho}(\partial_{\sigma}F_{\beta\nu})
\partial_{q}^{\beta}f_{V}}\Big)
\\
&&+\hbar\tilde{F}^{\mu\nu}q_{\nu}\delta'(q^2-m^2)q\cdot\Delta f_V
\end{eqnarray}
and 
\begin{eqnarray}\label{C_rQ}\nonumber
\hat{\mathcal{C}}^{(n_r)\mu}_{2}&=&\frac{\delta(q^2-m^2)}{2}
\Bigl(\epsilon^{\mu\nu\rho\sigma} q_{\nu}
 (\Delta_{\rho} \widehat{ \Sigma _{V\sigma})f_V}
+\frac{1}{m}q^{\mu}
( q\cdot\Delta \widehat{ \Sigma_{P} ) f_V }
- m(\partial^{\mu}\widehat{ \Sigma_P)f_{V} } \Bigr)
\\\
&&
- \tilde{F}^{\mu\nu}q_{\nu}\delta'(q^2-m^2)
 (q\cdot \widehat{ \Sigma_V f_V}+m \widehat{ \Sigma_S f_V} ).
\end{eqnarray}
Finally, when considering the application to a heavy quark traveling in QGP, the AKE in the rest frame could be simplified as
\begin{eqnarray}\nonumber\label{AKE_vr}
&&\delta(q^2-m^2)\Bigg\{q\cdot \partial (a^{\mu}f_A)
+a^{\mu} q_\nu \widehat{ \Sigma_{V}^\nu {f}_A} 
+m^2 \widehat{ \Sigma_A^{\mu} f_V}  - q^{\mu} q_\nu \widehat{ {\Sigma}_A ^\nu f_V} 
+m \Big(a^{\mu} \widehat{ {\Sigma}_S f_A} 
-\frac{1}{2} \epsilon^{\mu\nu\rho\sigma}  q_{\nu} \widehat{ {\Sigma}_{T{\rho\sigma}} f_V}
\Big)
\\
&&-\frac{\hbar}{2}\epsilon^{\mu\nu\rho\sigma}
q_{\nu}(\partial_{\rho}\widehat{\Sigma_{V\sigma})f_V}
\Bigg\}=0,
\end{eqnarray}   
by taking $\Sigma_P=\bar{\Sigma}_{P}=0$ and $F_{\mu\nu}=0$.

In general, when involving also the $\hbar$ corrections in $\mathcal{V}^{\mu}$, such a frame choice is  
only valid when $m$ is much larger than the gradient and electromagnetic scales in the system. The magnetization-current term in $\mathcal{V}^{\mu}$ explicitly reveals the breakdown for the choice of a rest frame away from the aforementioned regime. Although we dot not explicitly include the $\hbar$ term in $\mathcal{V}^{\mu}$ based on our power counting, it is still essential to be aware of the valid regime for the frame choice $n^{\mu}=n^{\mu}_r$. In heavy-ion phenomenology, one may assume the validity is held, which could be "somewhat applicable" for the spin transport of strange quarks in QGP. For up and down quarks or other applications, it is inevitable to maintain the general frame vector $n^{\mu}=n^{\mu}(X)$. Note that the quantum correction on collisions in Eq.~(\ref{C_rQ}) is also presented in Eq.~(\ref{C_nQ}). We may regard other terms in Eq.~(\ref{C_nQ}) as the $\mathcal{O}(|{\bf q}|/m)$ corrections on top of Eq.~(\ref{C_rQ}), where $\bm q$ here denotes the spatial momentum.

\subsection{$  \hbar $ sorting with the present order counting}\label{subsec_decompCQ}
In order to explicitly disentangle the (classical) spin-diffusion and (quantum) spin-polarization parts in collisions, %except for $\Sigma_{V\rho}$ and $\Sigma_S$, 
we have to retrieve the $\hbar$ terms in $\Sigma_{V\rho}$, $\Sigma_S$, $\Sigma_{A\rho}$ and $\Sigma_{T\rho\sigma}$. That is, we have to further make the decomposition $\bar{\Sigma}_{V\rho}=\bar{\Sigma}^{\text{cl}}_{V\rho}+\hbar \bar{\Sigma}^{Q(n)}_{V\rho}$, $\bar{\Sigma}_{S}=\bar{\Sigma}^{\text{cl}}_{S}+\hbar \bar{\Sigma}^{Q(n)}_{S}$, $\bar{\Sigma}_{A\rho}=\bar{\Sigma}^{\text{cl}}_{A\rho}+\hbar \bar{\Sigma}^{Q(n)}_{A\rho}$ and $\bar{\Sigma}_{T\rho\sigma}=\bar{\Sigma}^{\text{cl}}_{T\rho\sigma}+\hbar\bar{\Sigma}^{\text{Q}}_{T\rho\sigma}$, where their explicit forms depend on the details of collisions in systems. Nonetheless, $\hbar \bar{\Sigma}^{Q(n)}_{V\rho}$ and $\hbar \bar{\Sigma}^{Q(n)}_{S}$ are coupled to $f_A$ in $\hat{\mathcal{C}}^{(n)\mu}_{1}$, which actually contribute to $\mathcal{O}(\hbar^2)$ (in the order of magnitude) corrections from our power counting. Consequently, we only have to retain the quantum corrections from $\hbar \bar{\Sigma}^{Q(n)}_{A\rho}$ and $\hbar\bar{\Sigma}^{\text{Q}}_{T\rho\sigma}$. One can then rewrite the collision term in the AKE as 
	\begin{eqnarray}\label{C_decomp}
	\hat{\mathcal{C}}^{(n)\mu}_{1}+\hbar\hat{\mathcal{C}}^{(n)\mu}_{2}=\hat{\mathcal{C}}^{\mu}_{\text{cl}}+\hbar\hat{\mathcal{C}}^{(n)\mu}_{\text{Q}},
	\end{eqnarray}
where
\begin{eqnarray}\label{C_cl}
\hat{\mathcal{C}}^{\mu}_{\text{cl}}
&=&\delta(q^2-m^2)\Big[
q^{\mu} q_{\alpha} \widehat{  \Sigma^{\text{cl}\alpha}_A f_V} -m^2\widehat{ \Sigma_A^{\text{cl}\mu} f_V }
- a^{\mu} (q_\alpha \widehat{\Sigma_{V}^{\text{cl}\alpha} f_A} + m \widehat{ \Sigma_{S}^{\text{cl}} f_A} ) 
+ m\frac{\epsilon^{\mu\nu\rho\sigma}}{2}q_{\nu} \widehat{\Sigma^{\text{cl}}_{T\rho\sigma} f_V}
\Big]
\end{eqnarray}
and
\begin{eqnarray}\nonumber
\hbar\hat{\mathcal{C}}^{(n)\mu}_{\text{Q}}&=&\hbar q^{\mu}\Big(\hat{\mathcal{C}}^{(n)}_{\mathfrak{q}2}+\delta(q^2-m^2)q_{\nu}\widehat{ \Sigma^{\text{Q}(n)\nu}_Af_V}\Big)
+\hbar m\Big[\hat{\mathcal{C}}^{(n)\mu}_{\mathfrak{m}2}
+\delta(q^2-m^2)\Big({\frac{1}{2}}\epsilon^{\mu\nu\rho\sigma}q_{\nu}\widehat{\Sigma^{\text{Q}(n)}_{T\rho\sigma}f_V}
\\
&&-m\widehat{\Sigma_A^{\text{Q}(n)\mu}f_V}
\Big)\Big]
. \label{eq:29}
\end{eqnarray} 
Analogously, in the rest frame, it is found
\begin{eqnarray}\label{C_Qnr}
\hbar\hat{\mathcal{C}}^{(n_r)\mu}_{\text{Q}}&=&\hbar\hat{\mathcal{C}}^{(n_r)\mu}_{2}
+\hbar \delta(q^2-m^2)\Big[
q^{\mu} q_{\alpha} \widehat{  \Sigma^{\text{Q}(n_r)\alpha}_A f_V} -m^2\widehat{ \Sigma_A^{\text{Q}(n_r)\mu} f_V }
+ m\frac{\epsilon^{\mu\nu\rho\sigma}}{2}q_{\nu} \widehat{\Sigma^{\text{Q}(n_r)}_{T\rho\sigma} f_V}
\Big].
\end{eqnarray}
Note that the classical part $\hat{\mathcal{C}}^{\mu}_{\text{cl}}$ is explicitly frame independent. Now, all the $\hbar$ terms are collected into $\hbar\hat{\mathcal{C}}^{(n)\mu}_{\text{Q}}$. Despite complication, one finds that $\hat{\mathcal{C}}^{\mu}_{\text{cl}}$ is proportional to $a^{\mu}f_A$; such a term hence results in the diffusion of spin \footnote{Here $\Sigma_{A}^{\text{cl}\alpha}$ and $\Sigma_{T\rho\sigma}^{\text{cl}}$ are in principle proportional to the axial-charge part of the Wigner functions for outgoing fermions. Therefore, the terms coupled to $f_V$ in $\hat{\mathcal{C}}^{\mu}_{\text{cl}}$ are also proportional to the spin four vector. One may see an explicit example for application to QGP in the following section. }. On the contrary, $\hat{\mathcal{C}}^{(n)\mu}_{\text{Q}}$ is instead proportional to $f_V$ and $\bar{f}_V$. Even when initial spin ($\sim a^{\mu}f_A$) is zero, such a term can lead to the spin polarization from the entangled vector-charge transport. Similarly, when applying Eqs.~(\ref{AKE_v1}), (\ref{AKE_v2}), and (\ref{AKE_vr}), one should recall these extra $\hbar$ corrections from $\hbar \bar{\Sigma}^{Q(n)}_{A\rho}$ and $\hbar\bar{\Sigma}^{\text{Q}(n)}_{T\rho\sigma}$ or from $\hbar \bar{\Sigma}^{Q(n_r)}_{A\rho}$ $\hbar\bar{\Sigma}^{\text{Q}(n_r)}_{T\rho\sigma}$. 

\subsection{Discussions for the collision terms} \label{sec:discussions}
In this subsection, we would like to make a comparison between the AKE in different frame choices and discuss about their physical interpretations. However, to avoid complications, we focus on the simplified versions in Eqs.~(\ref{AKE_v1}), (\ref{AKE_v2}), and (\ref{AKE_vr}), which are the primary results in connection to the spin polarization of quarks in QGP. For preciseness, we will hereafter dub the terms proportional to $\hbar$ in Eqs.~(\ref{AKE_v1}), (\ref{AKE_v2}), and (\ref{AKE_vr}) as explicit quantum ($\hbar$) corrections. In contrast, we refer the $\hbar$ corrections encoded in self-energies discussed in Sec. \ref{subsec_decompCQ} as implicit quantum ($\hbar$) corrections.  

First of all, Eqs.~(\ref{AKE_v1}) and (\ref{AKE_v2}) as mathematically equivalent expressions both work for an arbitrary mass of fermions. Nevertheless, the expression of Eq.~(\ref{AKE_v1}) could be more useful in the large-mass regime. When comparing Eq.~(\ref{AKE_v1}) with Eq.~(\ref{AKE_vr}) as an effective AKE with large-mass fermions, one finds they both incorporate the term $\frac{\hbar}{2}\epsilon^{\mu\nu\rho\sigma}
q_{\nu}(\partial_{\rho}\widehat{\Sigma_{V\sigma})f_V}$  as an explicit quantum correction in collisions. In the non-relativistic condition, such a term further dominates over the rest of explicit $\hbar$ corrections coupled with $S^{\mu\nu}_{m(n)}$ in Eq.~(\ref{AKE_v1}). When $m\rightarrow \infty$, one finds $q^{\mu}\rightarrow n^{\mu}m$ and accordingly $S^{\mu\nu}_{m(n)}\rightarrow 0$. It turns out that Eq.~(\ref{AKE_v1}) and Eq.~(\ref{AKE_nr}) coincide in the heavy-quark (fermion) limit. Consequently, as briefly mentioned in Sec.~\ref{subsec_AKE_vr}, Eq.~(\ref{AKE_v1}) further incorporates the  $\mathcal{O}(|{\bf q}|/m)$ corrections on top of Eq.~(\ref{AKE_vr}). It is thus more practical to utilize Eq.~(\ref{AKE_v1}) for exploring heavy-quark transport with the inclusion of non-relativistic corrections. On the contrary, Eq.~(\ref{AKE_v2}) has a more explicit connection to the CKT in the massless limit. The term $\hbar q^{\mu}S^{\rho\nu}_{m(n)} \widehat{ (\partial_{\rho} \Sigma_{V\nu})f_V}$ in explicit $\hbar$ corrections of collisions therein matches the $\hbar$ correction in the collision term of CKT up to a prefactor $q^{\mu}$. In fact, the prefactor $q^{\mu}$ further manifests that such a term is pertinent to the side-jump phenomena associated with the spin polarization enslaved by the momentum and chirality, which forces $a^{\mu}$ in the free-streaming part to align with $q^{\mu}$. On the other hand, the rest of $\hbar$ terms proportional to $m$ in Eq.~(\ref{AKE_v2}) stem from the finite-mass effect suppressed by $\mathcal{O}(m/|{\bf q}|)$ in the relativistic limit, which could modify the orientation of spin characterized by the direction of $a^{\mu}$. Therefore, the expression in Eq.~(\ref{AKE_v2}) could be more suitable for analyzing the axial-charge diffusion and spin polarization of light quarks in QGP.

In addition, similar to the case for CKT in the massless limit, the quantum corrections in the AKE now only come from collisions when choosing a constant frame vector in the absence of electromagnetic fields. As already mentioned in Sec.~\ref{subsec_decompCQ} for further separation of the classical and quantum parts in the collision term, the classical part in Eqs.~(\ref{AKE_v1}), (\ref{AKE_v2}), and (\ref{AKE_vr}) with the same expression will yield the spin diffusion. However, it is believe that the spin polarization in HIC is led by the local vorticity of QGP. It is obvious to see the explicit $\hbar$ corrections in the collision term now originate from the inhomogeneity of self-energies, from which the self-energies gradients could incorporate such vortical corrections in collisions. Also, these terms manifest the spin-orbit interactions through collisions, which entangle the dynamical evolution between  $a^{\mu}f_A$ and $f_V$. Nevertheless, the implicit $\hbar$ corrections encoded in self-energies should also be taken into account, where at least the $\hbar$ correction upon Wigner functions for outgoing quarks will contribute to the vortical corrections as well. 

In a recent study for the application of CKT on the chiral radiation transport theory for neutrinos in core-collapse supernovae, it is shown the similar $\hbar$ corrections of CKT give rise to the vorticity corrections associated with fluid helicity in the collision term for the neutrino absorption process \cite{Yamamoto:2020zrs}. One may expect a similar scenario when considering a strange quark probing the QGP near equilibrium yet with local vortical fields. Nonetheless, unlike the weak interaction governed the Weinberg-Salam model, the details of collisions in QGP are more sophisticated due to the interacting gluons. Even though the implicit $\hbar$ corrections encoded in self-energies from light quarks in equilibrium could be derived from equilibrium Wigner functions shown in e.g. Refs.~\cite{Hidaka:2017auj,Hattori:2019ahi}, how to include analogous corrections led by vorticity from polarized gluons is currently unknown. For the future application on the dynamical spin polarization of strange quarks traversing QGP, we will have to work out the Wigner functions for polarized gluons up to $\mathcal{O}(\hbar)$ with both classical and quantum contributions at least in equilibrium as one of essential ingredients. Such a development is beyond the scope of this work and left as the future research direction. It is however worthwhile to note that even the classical part of the collision term in AKE is an innovation. We will further apply such a theoretical framework to derive an explicit expression of the spin diffusion term for massive quarks traversing weakly-coupled QGP in the next section.

\section{Example : Spin diffusion of quarks in weakly coupled QGP}\label{sec_diffusion_QGP}
\subsection{Scattering between massive fermions and a medium}
We  now apply the formalism established in the previous sections to investigate the collision term for massive quarks traversing weakly-coupled QGP in relativistic heavy ion collisions. For simplicity, we will just focus on the spin-diffusion term such as $\hat{\mathcal{C}}^{\mu}_{\text{cl}}$ in the AKE and leave the $\hbar\hat{\mathcal{C}}^{(n)\mu}_{\text{Q}}$ for future study. Recently, a related study for spin diffusion has been presented in Ref.~\cite{Li:2019qkf} with a different approach. We will mostly follow the theoretical setup therein. In the following, 
we call fermions quarks and gauge bosons gluons interchangeably, and include the color-group factors. 
Here, the color degrees of freedom do not play crucial roles (like in the color conductivity), 
and the same computation holds for QED with simple replacements of the relevant degrees of freedom. 
Furthermore, we consider the massive quarks with quark mass much greater than the scale of thermal mass in QGP and accordingly neglect the Compton scattering with gluons as the subleading effects analogous to the study of heavy-quark transport in heavy ion collisions (See e.g. Ref.~\cite{Svetitsky:1987gq} and the same approximation in Ref.~\cite{Li:2019qkf}).  Note that the $\mathcal{O}(f_A^2)$ terms are not dropped  a priori in the calculations, whereas shall see that only the terms linear to $f_A$ remain in the final result, which thus agrees with our power-counting scheme.

The gluon-exchange processes between a massive fermion and the medium constitutes are written down as 
\begin{eqnarray}\label{Sigma_in_Pi}
	\Sigma^{>(<)}(q, X)=\lambda_c\int_{q'}\gamma^{\mu}S^{>(<)}(q', X)\gamma^{\nu}
 G^{>(<)}_{\mu\nu}(q-q', X) 
,
\end{eqnarray}
where $\lambda_c$ denotes an overall coefficient including the coupling and we drop $\mathcal{O}(\hbar^2)$ and higher-order correction in our order counting.
 The gluon propagator $ G^{>(<)}_{\mu\nu} $ contains the information of the spectral functions 
which depend on the scatterers. 
Here, we assume a dilute population of the massive fermions in the medium 
and neglect the contributions of the massive scatterers, 
and retain the massless-fermion and gluon scatterers. Having assumed the weakly coupled system, 
we focus on the lowest-order contributions in the coupling constant $ g_c $, i.e., 
the 2-to-2 scatterings between a massive quark and a massless quark/gluon.

Inserting Eq.~(\ref{spinor-decomp}), one can straightforwardly decompose 
the gamma structures in Eq.~(\ref{Sigma_in_Pi}) as 
\begin{eqnarray}\nonumber
	\chi^{>\mu\nu}_{q'}&\equiv&\gamma^{\mu}S^{>}(q')\gamma^{\nu}
	\\\nonumber
	&=&\bar{\mathcal{S}}\big(\eta^{\mu\nu}-i\spinGamma^{\mu\nu}\big)-i\bar{\mathcal{P}}\Big(\eta^{\mu\nu}\gamma^5+\frac{\epsilon^{\mu\nu\alpha\beta}}{2}\spinGamma_{\alpha\beta}\Big)
	+\bar{\mathcal{V}}_{\rho}\big(\eta^{\mu\rho}\gamma^{\nu}+\eta^{\rho\nu}\gamma^{\mu}-\eta^{\mu\nu}\gamma^{\rho}+i\epsilon^{\mu\nu\rho\sigma}\gamma^5\gamma_{\sigma}\big)
	\\\nonumber
	&&+\bar{\mathcal{A}}_{\rho}\big(\eta^{\mu\nu}\gamma^5\gamma^{\rho}-\eta^{\rho\nu}\gamma^5\gamma^{\mu}-\eta^{\mu\rho}\gamma^5\gamma^{\nu}+i\epsilon^{\mu\rho\nu\sigma}\gamma_{\sigma}\big)
	\\
	&&
		+\bar{\mathcal{S}}_{\alpha\beta}\Big(\eta^{\mu\alpha}\spinGamma^{\beta\nu}
	+\eta^{\nu\alpha}\spinGamma^{\beta\mu}+\frac{\eta^{\mu\nu}}{2}\spinGamma^{\alpha\beta}
	+i\eta^{\mu\alpha}\eta^{\beta\nu}+\frac{\epsilon^{\mu\nu\alpha\beta}}{2}\gamma^5
	\Big)
	.
\end{eqnarray}
Thus, contracted with the gluon propagator, we have 
\begin{eqnarray}\nonumber
	\chi^{>\mu\nu}_{q'} G^>_{\mu\nu}&=&
	\big(\bar{\mathcal{S}} G^>_{\mu\nu} + i\bar{\mathcal{S}}^{\mu\nu} G^>_{\mu\nu} \big)
	+i\gamma^5\Bigl(-\bar{\mathcal{P}} G^{>\mu}_{\mu} 
		-i\bar{\mathcal{S}}_{\alpha\beta} G^>_{\mu\nu} 
	\frac{\epsilon^{\mu\nu\alpha\beta}}{2}\Bigr) 
	\notag\\
	&&+\gamma^{\rho}\big(\bar{\mathcal{V}}^{\mu}  ( G^>_{(\mu\rho)} 
	-\bar{\mathcal{V}}_{\rho} G^{>\mu}_{\mu}-i\epsilon_{\mu\nu\sigma\rho}\bar{\mathcal{A}}^{\sigma} G^{>\mu\nu}\big)
	\notag\\
	&&
	+\gamma^5\gamma^{\rho}\big(
	-\bar{\mathcal{A}}^{\mu}  G^>_{(\mu\rho)}  
	+\bar{\mathcal{A}}_{\rho} G^{>\mu}_{\mu}
	+i\epsilon_{\mu\nu\sigma\rho}\bar{\mathcal{V}}^{\sigma} G^{>\mu\nu}\big)
	\notag\\
	&&+\frac{1}{2}\spinGamma^{\rho\sigma}\Big(
	2\bar{\mathcal{S}}^{\mu}_{\ \, \rho}  G^{>}_{(\mu\sigma)} 
	+\bar{\mathcal{S}}_{\rho\sigma} G^{>\mu}_{\mu}
	-2i\bar{\mathcal{S}} G^>_{\rho\sigma}
	-i\bar{\mathcal{P}}\epsilon_{\mu\nu\rho\sigma} G^{>\mu\nu}
	\Big),
\end{eqnarray}
where $A_{(\mu}B_{\nu)}=A_{\mu}B_{\nu} + A_{\nu}B_{\mu}$. 
We consider gluon propagators $ G^{<(>)}_{\mu\nu}$ symmetric in the Lorentz indices, 
and then find that  the imaginary terms in the above vanish in the contractions: 
\begin{eqnarray} \nonumber\label{sigma_decomp}
	\Sigma^>(q)&=&\lambda_c\int_{q'}\Bigg[\bar{\mathcal{S}} G^{>\mu}_{\mu}
	-i\bar{\mathcal{P}} G^{>\mu}_{\mu}\gamma^5+\gamma^{\rho}\big(2 \bar{\mathcal{V}}^{\mu}  G^>_{\mu\rho} 
	-\bar{\mathcal{V}}_{\rho} G^{>\mu}_{\mu}\big)
	\\
	&&
	+\gamma^5\gamma^{\rho}\big(
		- 2 \bar{\mathcal{A}}^{\mu} G^>_{\mu\rho} 
		+\bar{\mathcal{A}}_{\rho} G^{>\mu}_{\mu}\big)
	+\frac{\spinGamma^{\rho\sigma}}{2}
	\Big( 2 \bar{\mathcal{S}}_{\mu[\rho} G^{>\mu}_{\sigma]} 
	+\bar{\mathcal{S}}_{\rho\sigma} G^{>\mu}_{\mu}\Big)\Bigg],
	\label{eq:chiPi}
\end{eqnarray}
where $A_{[\mu}B_{\nu]}=A_{\mu}B_{\nu}-A_{\nu}B_{\mu}$. 
In general, $G^{<(>)}_{\mu\nu}$ possibly contains anti-symmetric components 
led by scatterings with spin-polarized scatterers in the medium.
For example, when considering the scattering with massless quarks, such anti-symmetric components can arise from the side-jump terms, whereas $\chi^{>\mu\nu}_{q'} G^>_{\mu\nu}$ should still remain real. Such quantum corrections from a polarized medium will not be considered in the present work.

One can now read out the corresponding terms up to $\mathcal{O}(\hbar)$ between Eqs.~(\ref{decomp_Sigma_general}) 
and (\ref{sigma_decomp}) as 
\begin{subequations}
\label{Sigma-G}
\begin{eqnarray}
\label{Sigma_S_hbar}
\bar \Sigma_S &=& \lambda_c \int_{q^\prime}   \bar \cS_{q^\prime}  G^{>\alpha}_\alpha
=   \frac{ \lambda_c }{m}  \int_{q^\prime}  ( q^\prime \cdot \bar \cV_{q^\prime})  G^{>\alpha}_\alpha 
,
\\
 \bar \Sigma_P &=&   \lambda_c \int_{q^\prime}   ( - \bar \cP_{q^\prime} )  G^{>\alpha}_\alpha,
\\
\bar\Sigma_{V\mu} &=& \lambda_c \int_{q^\prime} 
\bar \cV^\alpha_{q^\prime} ( 2 G ^{>}_{\alpha\mu} - G^{>\beta} _\beta \eta_{\alpha\mu}),  
\label{Sigma_V}
\\
\bar \Sigma_{A\mu} &=& \lambda_c \int_{q^\prime}
  \bar \cA^\alpha_{q^\prime}  ( - 2  G^> _{\alpha\mu} + G^{>\beta}_\beta  \eta_{\alpha\mu} ),  
  \label{Sigma_A}
\\
\bar \Sigma_{T\mu\nu} &=& \lambda_c \int_{q^\prime}
(  2 \bar \cS_{\alpha [\mu} G^{> \alpha}_{ \nu]}  
+   G^{>\alpha} _{\alpha}  \bar \cS_{\mu\nu}  
)
=
\frac{ \lambda_c}{m} \int_{q^\prime}
(  2 \bar S_{\alpha [\mu} G^{> \alpha}_{ \nu]}  
+  G^{>\alpha} _{\alpha}  \bar S_{\mu\nu}  
)
.
\label{Sigma_T_hbar}
\end{eqnarray}
\end{subequations}
The rightmost sides are obtain by using Eqs.~(\ref{eq:S-weak})-(\ref{replace_2}) 
up to the linear orders in $ \Sigma  $'s and $\hbar $. 
In the present case, we confirm that $\bar{\Sigma}_{P}$ is at $\mathcal{O}(\hbar^2)$ as anticipated earlier.

Now, given explicit forms of $G^>_{\mu\nu}$, $\bar{\mathcal{V}}^{\mu}$, and $\mathcal{\bar{A}}^{\mu}$, 
we can directly evaluate $\Sigma^>_{\mu}$ from Eq.~(\ref{Sigma-G}) and $\Sigma^<_{\mu}$ in the same fashion. 
Inserting these expressions into the collision terms in the SKE (\ref{SKE}), 
we have 
\begin{eqnarray}
(q\cdot\bar{\Sigma}_V+m\bar{\Sigma}_S)
=\lambda_c\int_{q^\prime} 2\pi\delta(q'^2-m^2)\Big(2q'^{\mu}G^>_{\mu\nu}q^{\nu}-p\cdot q'G^{>\mu}_{\mu}\Big)\bar{f}_{Vq'}
,
\end{eqnarray}
where $ p^\mu = q^\mu - q^{\prime \mu} $. 
The other term $(q\cdot\Sigma_V+m\Sigma_S)$ takes a similar form.
In addition, by making the decompositions 
$\bar{\Sigma}_{A\rho}=\bar{\Sigma}^{\text{cl}}_{A\rho}+\hbar \bar{\Sigma}^{Q(n)}_{A\rho}$ 
and $\bar{\Sigma}_{T\rho\sigma}=\bar{\Sigma}^{\text{cl}}_{T\rho\sigma}+\hbar\bar{\Sigma}^{\text{Q}}_{T\rho\sigma}$, 
we identify the classical and quantum parts. 
From the classical part in Eq.~(\ref{axial_sol}), 
Eqs.~(\ref{Sigma_A}) and (\ref{Sigma_T_hbar}) yield 
\begin{eqnarray}
&&
\bar{\Sigma}^{\text{cl}}_{A\rho}=\lambda_c\int_{q^\prime} 2\pi\delta(q'^2-m^2)\big(a_{q'\rho} G^{>\mu}_{\mu}-2a^{\mu}_{q'}G^>_{\mu\rho}
\big)\bar{f}_{A} (q'),
\\
&&
\label{Sigma_T_cl}
\bar{\Sigma}^{\text{cl}}_{T\rho\sigma}=-\frac{\lambda_c}{m}\int_{q^\prime}
2\pi\delta(q'^2-m^2) \Big(2 G^{>\mu}_{\quad[\sigma}\epsilon_{\mu\rho]\alpha\beta}+G^{>\mu}_{\mu}\epsilon_{\rho\sigma\alpha\beta}\Big)q'^{\alpha}a^{\beta}_{q'}\bar{f}_{A} (q').
\end{eqnarray} 
Those terms are further investigated with a specific gluon propagator 
provided by the hard-thermal loop approximation in the next section. 
The quantum parts are also identified in the same way. 
However, computation of those quantum corrections with specific gluon propagators are left as open issues. 
Note also that we have dropped possible antisymmetric parts of the gluon propagator in Eq.~(\ref{eq:chiPi}).

\subsection{Weakly coupled QGP and hard-thermal-loop approximation}

Although Eqs.~(\ref{Sigma_S_hbar})-(\ref{Sigma_T_hbar}) work for even non-equilibrium media, we now focus on the spin transport in equilibrium QGP as a concrete example. In such a case, we can make further simplification for the self-energy in Eq.~(\ref{Sigma_in_Pi}). For simplicity, we will only evaluate the spin-diffusion terms in the SKE and AKE, while the claculation for the spin-polarization term in the AKE is more involved, which will be presented in the followup work. 
We will implement the hard-thermal-loop (HTL) approximation, 
which allows us to derive the leading-logarithmic result in weakly coupled QCD as in the derivation shown in Ref.~\cite{Li:2019qkf} from a distinct approach. Recall that $p^{\mu}=(q-q')^{\mu}$ and $g_c$ denotes the coupling constant for strong interaction. We then apply the cut-gluon propagator in the Coulomb gauge and the fluid-rest frame, which gives rise to 
\begin{eqnarray}\label{HTLA_Pi}
G^{<(>)}_{\mu\nu}(q,q')\approx g^{<(>)}_{\text{eq}p}\Big[\rho_L(p)P^C_{\mu\nu}+\rho_T(p)P^T_{\mu\nu}\Big]
\end{eqnarray}
with
\begin{eqnarray}
P^C_{\mu\nu}\equiv u_{\mu}u_{\nu},\quad 
P^T_{\mu\nu}\equiv -\Theta_{\mu\alpha}\Theta_{\nu\beta}\left(\eta^{\alpha\beta}+\frac{p^{\alpha}p^{\beta}}{|{\bf p}|^2}\right)
=-\left(\Theta_{\mu\nu}+\frac{p_{\perp\mu}p_{\perp\nu}}{|{\bf p}|^2}\right)
,
\end{eqnarray}  
where $g^<_{\text{eq}p}=g_{0p}=1/(e^{\beta p\cdot u}-1)$, $g^>_{\text{eq}p}=1+g_{0p}$ 
and $  \Theta^{\mu\nu}\equiv\eta^{\mu\nu}-u^{\mu}u^{\nu}$. 
Here, $u^{\mu}$ and $\beta=1/T$ denote the fluid four velocity
 and the inverse of temperature in local equilibrium, respectively. 
We have introduced notations: 
\begin{subequations} \label{notations}
\begin{eqnarray}
&&
V^0 \equiv V \cdot u , \quad V_\perp^\mu \equiv V^\mu - V^0 u^\mu = \Theta^{\mu\nu} V_\nu ,
\\
&&
{\bf V}^i \equiv V_\perp^i , \quad \hat {\bf V}^i = \hat V_\perp^i \equiv V^i/ |{\bf V}| ,
\end{eqnarray}
\end{subequations}
for an arbitrary vector $ V^\mu $. 
Then, we have $ V_\perp \cdot  k_\perp = - {\bf V} \cdot \bk $ and, especially, $ V_\perp ^2 = -  |{\bf V}|^2 $ when $ k^\mu =V^\mu $.

In our setup, the HTL approximation is more precisely applied to $g_cT\ll |p^{\mu}|\ll T$.  On the other hand, $\rho_{L/T}(p)$ correspond to the HTL gluon spectral densities, which take explicit forms as (e.g., see Ref.~\cite{le2000thermal})
\begin{eqnarray}
\rho_L(p)\approx\frac{\pi m_D^2p_{0}}{ |\bp|^5 },\quad \rho_T(p)\approx\frac{\pi m_D^2p_{0}}{2  |\bp|^{5}\left(1-\left(\frac{{p_{0}}}{{|{\bf p}|}}\right)^2\right) },
\end{eqnarray}
where $m_D\sim g_cT$ corresponds to the Debye mass. The explicit form from the gluons in $ SU(N_c) $ color group 
and the $ N_f $-flavored massless quarks is given by 
\begin{eqnarray}
m_D^2=\frac{g_c^2T^2(2N_c+N_f)}{6},\quad \lambda_c=g_c^2C_2(F)=\frac{g_c^2(N_c^2-1)}{2N_c}.
\end{eqnarray}

Moreover, one should keep in mind the relation 
\begin{eqnarray}\label{Pi_HTL}
G^{>}_{\mu\nu}(p)=
(1 + g_{0p}^{-1}) G^{<}_{\mu\nu}(p)
\end{eqnarray}
from detailed balance. In light of the theoretical frameworks constructed in the previous section, we may now write down the SKE and AKE in the HTL approximation. It is interesting that one can linearize the kinetic equations in terms of the distribution functions by taking $g_{0p}^{-1}\rightarrow 0$ and $G^{>}_{\mu\nu}(p)\approx G^{<}_{\mu\nu}(p)$. However, in the practical calculation, we have to at least approximate 
$g_{0p}\approx {T}/{p_0}-{1}/{2}+{p_0}/{(12T)}+\mathcal{O}\left({p^3_0}/{T^3}\right)$ to keep all relevant terms contributing to the leading logarithmic order. In the following, we will append subindices to $f_{V/A}$ and $a^{\mu}$ for specifying their momentum dependence.

For the SKE in Eq.~(\ref{SKE}), it is found
\begin{eqnarray}\label{SKE_HTL}
0=\delta(q^2-m^2)\Big[q \cdot\partial f_{Vq}
+\lambda_c\int_p Q_1(q,p) \{ (1+ g_{0p}^{-1}) 
\bar{f}_{Vq'}f_{Vq}-\bar{f}_{Vq}f_{Vq'}\} \Big],
\end{eqnarray}
where 
\begin{eqnarray}\label{Q1}
Q_1(q,p)=2\pi\delta(q'^2-m^2)\Big(2q'^{\mu}G^<_{\mu\nu}q^{\nu}-p\cdot q'G^{<\mu}_{\mu}\Big).
\end{eqnarray}
By using $\bar{f}_{Vq}=1-f_{Vq}$, we may further rewrite Eq.~(\ref{SKE_HTL}) into 
\begin{eqnarray}\label{SKE_HTL_3}
0=\delta(q^2-m^2)\Big[
q\cdot\partial f_{Vq} +\lambda_c\int_p \big\{ 
Q_1(f_{Vq}-f_{Vq-p})+\tilde{Q}_1f_{Vq} (1- f_{Vq^\prime} )
\big\}
\Big],
\end{eqnarray}
where $\tilde{Q}_1=g_{0p}^{-1}Q_1$. One can then further approximate $f_{Vq}-f_{Vq-p}\approx p^{\mu}\partial_{q^{\mu}}f_{Vq}+\mathcal{O}(|p|/|q|)$ assuming the small-momentum transfer $|p|\ll |q|$.

Next, we can simplify the AKE with the same approximation. 
Inserting the rightmost sides of Eq.~(\ref{Sigma-G}) into Eq.~(\ref{C_cl}), we obtain 
\begin{subequations}\label{AKE_HTL3_non}
\begin{eqnarray}
&&
\delta(q^2-m^2) [\, q\cdot\partial\tilde{a}^{\mu}_q+F^{\nu\mu}\tilde{a}_{\nu q}
-  \hat{\mathcal{C}}^{\mu}_{\text{cl}} \, ] = 0,
\\
&&
\hat{\mathcal{C}}^{\mu}_{\text{cl}} = 
\lambda_c\int_p \Big(
-Q_1\tilde{a}^{\mu}_q
-\tilde{Q}_1(1-f_{Vq-p})\tilde{a}^{\mu}_q
+Q^{\mu\nu}_2\tilde{a}_{q-p\nu} 
+\tilde{Q}_2^{\mu\nu}f_{Vq}\tilde{a}_{q-p\nu}
\Big)
,
\end{eqnarray}
\end{subequations}
where $\tilde{a}^{\mu}_q\equiv a^{\mu}_qf_{Aq}$, $\tilde{Q}_2^{\mu\nu}=g_{0p}^{-1}Q_2^{\mu\nu}$, and 
\begin{eqnarray}\nonumber\label{Q2}
Q^{\mu\nu}_2
&=&-2\pi\delta(q'^2-m^2)\Big[\Big(p^{\mu}q^{\nu}-\eta^{\mu\nu}q\cdot p\Big) G^{<\rho}_{\rho}
-2\Big(p^{\mu}G^{<\nu}_{\rho}q^{\rho}-q\cdot pG^{<\nu\mu}\Big)
\\
&&+2\big(
q^{\nu}G^{<\rho\mu}q'_{\rho}-\eta^{\mu\nu}q'_{\sigma}G^{<\sigma\rho}q_{\rho}
\big)
\Big]
.
\end{eqnarray}
%
%
\if 0
derived from
\begin{eqnarray}\nonumber
&&\big(q^{\mu}q\cdot\bar{\Sigma}^{\text{cl}}_A-m^2\bar{\Sigma}_A^{\text{cl}\mu}+\frac{m}{2}\epsilon^{\mu\nu\rho\sigma}q_{\nu}\bar{\Sigma}^{\text{cl}}_{T\rho\sigma}\big)
\\\nonumber
&&=\lambda_c\int_p2\pi\delta(q'^2-m^2)\Big[\Big(q^{\mu}q\cdot a_{q'}-m^2a^{\mu}_{q'}-q'^{\mu}q\cdot a_{q'}+a^{\mu}_{q'}q\cdot q'\Big) {G}^{{>}\rho}_{\rho}
-2\Big(q^{\mu}a^{\nu}_{q'}{G}^{>}_{\nu\rho}q^{\rho}-m^2a_{q'\nu}{G}^{{>}\nu\mu}\Big)
\\
&&\quad-2\big(q\cdot q'a_{q'\nu}{G}^{{>}\nu\mu}-q'^{\mu}a^{\nu}_{q'}{G}^{>}_{\nu\rho}q^{\rho}
-q\cdot a_{q'}{G}^{{>}\nu\mu}q'_{\nu}+a^{\mu}_{q'}q'_{\nu}{G}^{{>}\nu\rho}q_{\rho}
\big)
\Big]\bar{f}_{Aq'}
\end{eqnarray}
with
\begin{eqnarray}
\epsilon^{\mu\nu\rho\sigma}\bar{\Sigma}^{\text{cl}}_{T\rho\sigma}=-\frac{4\lambda_c}{m}\int_p2\pi\delta(q'^2-m^2)\Big({G}^{{>}[\mu}_{\sigma}q'^{\nu]}a_{q'}^{\sigma}-G^{{>}[\mu}_{\sigma}a_{q'}^{\nu]}q'^{\sigma}+\frac{G^{{>}\alpha}_{\alpha}}{2}q'^{[\mu}a_{q'}^{\nu]}\Big)\bar{f}_{Aq'}
\end{eqnarray} 
from Eq.~(\ref{Sigma_T_cl}). 
\fi
%
%
Here we also use $\bar{f}_{Aq}=-f_{Aq}$. One can similarly approximate $\tilde{a}_{q-p\nu}\approx \tilde{a}_{q\nu}-p^{\beta}\partial_{q\beta}\tilde{a}_{q\nu}$. Such a diffusion term in collisions of the AKE has also been constructed in Ref.~\cite{Li:2019qkf} from a distinct approach, in which different parameterization of the spin vector is applied. Nonetheless, it may be more practical to adopt our parameterization for the spin vector, which has a direct connection to the axial-charge current equivalent to the spin polarization through the Wigner functions and the combination with the free-steaming part of the AKE. 
Note also that Eqs.~(\ref{SKE_HTL_3}) and (\ref{AKE_HTL3_non}) contain nonlinear terms in distribution functions. 
While those nonlinear terms are not included in Ref.~\cite{Li:2019qkf}, they are imperative to preserve the quantum statistics for fermions. For example, as will be shown, $f_{Vq}$ follows the Fermi-Dirac distribution instead of just the Boltzmann distribution in equilibrium with the vanishing collision term in the SKE. In addition, as shown in Eq.~(\ref{AKE_HTL3_non}), the nonlinear terms further reveal the entangled dynamics between the vector/axial charges and spin diffusion.

\subsection{SKE and AKE with diffusion effects in the leading-log approximation}

We now explicitly compute the collision terms in axial kinetic theory with the HTL approximation. 
Notations have been introduced in Eq.~(\ref{notations}).
The basic strategy is to collect all the terms up to $\mathcal{O}(|{\bf p}|^{-3})$ in the integrand. 
When combined with the integral measure, they give rise to 
the leading logarithmic results in $g_c$ 
with the cut-offs provided by the HTL resummation. 
Moreover, we will consider the onshell kinetic equations. We hence take $f_{Vq}=f_{Vq}({\bf q},X)$ as just a function of ${\bf q}$ and $X$ by using $q_0=E_q=\sqrt{|{\bf q}|^2+m^2}$ for fermions (here we neglect anti-fermions) in the Wigner functions. Similarly, we take $u\cdot \tilde{a}_q=-q_{\perp}\cdot \tilde{a}_q/q_0$ for $\tilde{a}_{q\mu}=\tilde{a}_{q\mu}({\bf q}, X)$.

\subsubsection{Results}

The computations for the diffusion terms in SKE and AKE are complicated yet straightforward. We hence present the details of computations for Eqs.~(\ref{SKE_HTL}) and (\ref{AKE_HTL3_non}) in Appendices~\ref{app_der_SKE} and \ref{app_der_AKE}, respectively. In the following, we just summarize the final results. 
Up to the leading logarithmic order in $g_c$, the SKE takes the form  
\begin{eqnarray}\label{SKE_HTL_4_non}\nonumber
0=\delta(q^2-m^2)\Bigg[ 
q\cdot\partial 
-\kappa_{\rm LL}
\Big\{
2(1-f_{Vq})
+
{\mathcal S}^{(1)} \hat{q}^{\beta}_{\perp}\partial_{q^{\beta}_{\perp}}
+
{\mathcal S}^{(2)} 
 \hat{q}_{\perp}^{\alpha}\hat{q}^{\beta}_{\perp}\partial_{q^{\alpha}_{\perp}}\partial_{q^{\beta}_{\perp}}
+{\mathcal S}^{(3)}  \eta^{\alpha\beta}\partial_{q^{\alpha}_{\perp}}
\partial_{q^{\beta}_{\perp}}\Big\}
\Bigg] f_{Vq},
\\
\end{eqnarray}
where we denote the coefficient of the leading log result 
$ \kappa_{\rm LL} \equiv [g_c^2C_2(F)m_D^2/(8\pi) ]  \ln (1/g_c)  $ and 
the Minkowski metric $\eta^{\alpha\beta}$. %and $ F_{Vq} \equiv  1 - 2 f_{Vq} $ 
We also introduce the four ``velocity'' $ v^\mu  = (v^0, \bv^i) \equiv q^\mu/m $, 
which has the normalization $ v^\mu v_\mu =1 $ under the delta function, 
and then the rapidity $\eta_q \equiv 2^{-1}\ln[{( E_q+|{\bf q}|)}/{(E_q-|{\bf q}|)}]
= 2^{-1}\ln[{( v^0+|{\bf v}|)}/{(v^0-|{\bf v}|)}]$. 
% $\eta_q \equiv 2^{-1}\ln[{( E_q+|{\bf q}|)}/{(E_q-|{\bf q}|)}]$. 
The coefficient in each term is given as 
\begin{align}
{\mathcal S}^{(1)}  =
\frac{mv_0^2 \theta_{-1}}{|{\bf v}|^2}(1-2f_{Vq})
,
\quad
{\mathcal S}^{(2)} =
\frac{mTv_0^2}{2|{\bf v}|^3} \left( \frac{|{\bf v}|^2\eta_q}{v_0^2}  + \frac{3\theta_1}{v_0} \right)
 ,
\quad
{\mathcal S}^{(3)}  
= \frac{mT}{2} \left(\frac{v_0^3\theta_{-3}}{|{\bf v}|^3}-3v_0 \right)
,
\end{align}
where $ \theta_n \equiv |\bv| - v_0^n \eta_q $. Note that Eq.~(\ref{SKE_HTL_4_non}) agrees with the result in Ref.~\cite{Li:2019qkf} except for additional nonlinear terms in $f_{Vq}$ coming from Fermi-Dirac statistics.

Carrying out  similar yet more sophisticated computations for Eq.~(\ref{AKE_HTL3_non}), we also derive the AKE with the spin-diffusion term up to leading-logarithmic order. Combining with the classical free-streaming part dictated by the BMT equation, the AKE reads
\begin{eqnarray}\nonumber\label{AKE_classical_QGP}
	0&=&\delta(q^2-m^2)\Bigg[q\cdot\partial\tilde{a}^{\mu}_q%+F^{\nu\mu}\tilde{a}_{\nu q}
	-\frac{ \kappa_{\rm LL}T}{E_q}
	\Big(\tilde{a}_q^{\mu}\grave{\mathcal{Q}}_\text{cl}^{(1)}+u^{\mu}\grave{\mathcal{Q}}_\text{cl}^{(2)}+\hat{q}^{\mu}_{\perp}\grave{\mathcal{Q}}_\text{cl}^{(3)}
	\\
	&&+\grave{\mathcal{Q}}_\text{cl}^{(4)}\hat{q}_{\perp}^{\nu}\partial_{q_{\perp\mu}}\tilde{a}_{q\nu}+\grave{\mathcal{Q}}_\text{cl}^{(5)}\hat{q}^{\nu}\partial_{q_{\perp}^{\nu}}\tilde{a}_{q}^{\mu}+\grave{\mathcal{Q}}_\text{cl}^{(6)}\eta^{\nu\rho}\partial_{q^{\nu}_{\perp}}\partial_{q^{\rho}_{\perp}}\tilde{a}_q^{\mu}+\grave{\mathcal{Q}}_\text{cl}^{(7)}\hat{q}_{\perp}^{\nu}\hat{q}_{\perp}^{\rho}\partial_{q^{\nu}_{\perp}}\partial_{q^{\rho}_{\perp}}\tilde{a}_q^{\mu}
	\Big)\Bigg],
\end{eqnarray} 
\begin{subequations}
\begin{eqnarray}\label{grave_Qcl_1}
\grave{\mathcal{Q}}_\text{cl}^{(1)}
&=&
\frac{2  m}{T} \Big[ \,
 \Big( v_0(1-2f_{Vq}) - \frac{T}{m}\Big)
- \frac{v_0^3 }{|{\bf v}|^2} m \, \theta_{-1} \, \hat{q}_{\perp}^{\rho}\partial_{q^{\rho}_{\perp}}f_{Vq}
\,\Big]
,\\
\nonumber\label{grave_Qcl_2}
\grave{\mathcal{Q}}_\text{cl}^{(2)}
&=&
m \frac{v_0}{ |{\bf v}|^3 }
\Big( \,
( \theta_1 -|{\bf v}|^3 ) \partial_{q^{\nu}_{\perp}}\tilde{a}_q^{\nu}
+ ( 3 \theta_1  +|{\bf v}|^3  ) \hat{q}^{\nu}_{\perp}\hat{q}^{\rho}_{\perp}\partial_{q^{\rho}}\tilde{a}_{q\nu}
\, \Big)
\nn
\\
&& 
+  \frac{m}{v_0 |{\bf v}|^2 T} \Big(
v_0^3(1-2f_{Vq})  \theta_{-1} -   \frac{2T}{m} (  \theta_{-1} + 2|\bv|^3 ) 
\Big)  \hat{q}_{\perp}\cdot \tilde{a}_q 
,\\\nonumber
\label{grave_Qcl_3}
\grave{\mathcal{Q}}_\text{cl}^{(3)}
&=&
\frac{m}{|{\bf v}|^2 }
\Big( \,  v_0^2 \theta_{-1} \, \partial_{q^{\nu}_{\perp}}\tilde{a}_q^{\nu}
+ ( 3 \theta_1 + |{\bf v}|^3 ) \hat{q}^{\nu}_{\perp}\hat{q}^{\rho}_{\perp}\partial_{q^{\rho}}\tilde{a}_{q\nu}
\, \Big)
\\
&&
+  \frac{m}{|\bv| T}  \theta_{-1} \Big(  v_0(1-2f_{Vq}) -  \frac{2T}{m}\Big) \hat{q}_{\perp}\cdot \tilde{a}_q
,\\
\grave{\mathcal{Q}}_\text{cl}^{(4)}
&=& -2m |{\bf v}|
,\\
\grave{\mathcal{Q}}_\text{cl}^{(5)}
&=& \frac{ m^2 v_0^3  }{|{\bf v}|^2T}  \theta_{-1}(1-2f_{Vq})
,\\
\grave{\mathcal{Q}}_\text{cl}^{(6)}
&=& -\frac{m^2 v_0^2}{2|\bv|^3}\left(3|\bv|^3 - v_0^2 \theta_{-3}\right)
,\\
\label{grave_Qcl_7}
\grave{\mathcal{Q}}_\text{cl}^{(7)}
&=& 
 \frac{m^2 v_0 }{2|{\bf v}|^3} \big(3 v_0\theta_1 +\eta_q|{\bf v}|^2\big)
.
\end{eqnarray} 
\end{subequations}

\subsubsection{Nonrelativistic limit}
In the non-relativistic limit ($m\gg T, |{\bf q}|$), 
%we have $ v_0 \sim 1+ |\bq|^2/(2m^2)$ and $ \theta_n \sim -(3n+2) |\bq|^3/(6m^3) $. 
we have $ v_0 \sim 1+ |\bv|^2/2$ and $ \theta_n \sim -(3n+2) |\bv|^3/6 $. 
We immediately find that the SKE and AKE reduce to similar forms, 
\begin{subequations}
\begin{eqnarray}
0&=&\Big(\partial_{0}+\frac{q^{\nu}_{\perp}\partial_{\perp\nu}}{m}\Big)f_{Vq}
+ \frac{2}{3} \kappa_{\rm LL} 
\Bigg[ \,
T\eta^{\nu\rho}\partial_{q^{\nu}_{\perp}}\partial_{q^{\rho}_{\perp}}-\frac{1}{m}\Big(3(1-f_{Vq})
+ (1-2f_{Vq}) q^{\nu}_{\perp}\partial_{q_{\perp}^{\nu}}\Big)
\, \Bigg]f_{Vq},
%\end{eqnarray}
%and
%\begin{eqnarray}\nonumber
\\
0&=&\Big(\partial_{0}+\frac{q^{\nu}_{\perp}\partial_{\perp\nu}}{m}\Big)\tilde{a}^{\mu}_q
+ \frac{2}{3} \kappa_{\rm LL} 
\Bigg[ \, 
T\eta^{\nu\rho}\partial_{q^{\nu}_{\perp}}\partial_{q^{\rho}_{\perp}}\tilde{a}_q^{\mu}
+\frac{1}{m}\Big\{ \, 
\Big(2(q_{\perp}^{\nu}\partial_{q^{\nu}_{\perp}}f_{Vq})-3 (1-2f_{Vq})\Big)\tilde{a}^{\mu}_q
\\
&&+2T(\partial_{q_{\perp}^{\nu}}\tilde{a}_q^{\nu})u^{\mu}
- (1-2f_{Vq})q^{\nu}_{\perp}\partial_{q^{\nu}_{\perp}}\tilde{a}^{\mu}_q
\, \Big\}
\, \Bigg],
\nn
\end{eqnarray}  
\end{subequations}
where we further retain the terms up to $\mathcal{O}(1/m)$. 
It turns out that the orientation of spin for heavy quarks is fixed yet the ``spin'' (axial-charge) density characterized by $f_{Aq}$ undergoes the diffusive process same as the vector-charge density led by $f_{Vq}$ when $m\rightarrow \infty$. Nonetheless, the modification upon the spin orientation by, e.g., the fluid velocity could emerge at higher orders suppressed by the mass of heavy quarks. 
Note that the Compton scattering, neglected in this work, also give rise to $ 1/m $ corrections, 
that, however, do not come with the logarithm enhancement $ \sim \log(1/g_c) $. 
Therefore, the above $ 1/m $ corrections provide the consistent results within the leading-log approximation. 

\subsubsection{Consistency checks}

In thermal equilibrium, the vector-charge distribution function takes 
the Fermi-Dirac form $f_{Vq}=1/(e^{(E_{q}-\mu)/T}+1)$ such that
\begin{eqnarray}\label{equil_rel_1}\nonumber
&&\partial_{q^{\beta}_{\perp}}f_{qV}=f_{Vq}(1-f_{Vq})\frac{q_{\perp\beta}}{E_qT},
\\
&&\partial_{q^{\alpha}_{\perp}}\partial_{q^{\beta}_{\perp}}f_{qV}=f_{Vq}(1-3f_{Vq}+2f_{Vq}^2)\frac{q_{\perp\alpha}q_{\perp\beta}}{E_q^2T^2}+\frac{f_{Vq}(1-f_{Vq})}{E_qT}\Bigg(\Theta_{\alpha\beta}+\frac{q_{\perp\alpha}q_{\perp\beta}}{E_q^2}\Bigg),
\end{eqnarray}
and hence
\begin{eqnarray}\nonumber\label{equil_rel_2}
&&\hat{q}^{\beta}_{\perp}\partial_{q^{\beta}_{\perp}}f_{qV}=-f_{Vq}(1-f_{Vq})\frac{|{\bf q}|}{E_qT},
\\\nonumber
&&\eta^{\alpha\beta}\partial_{q^{\alpha}_{\perp}}\partial_{q^{\beta}_{\perp}}f_{qV}=-f_{Vq}(1-3f_{Vq}+2f_{Vq}^2)\frac{|{\bf q}|^2}{E_q^2T^2}+\frac{f_{Vq}(1-f_{Vq})}{E_qT}\Bigg(3-\frac{|{\bf q}|^2}{E_q^2}\Bigg),
\\
&&\hat{q}^{\alpha}_{\perp}\hat{q}^{\beta}_{\perp}\partial_{q^{\alpha}_{\perp}}\partial_{q^{\beta}_{\perp}}f_{qV}=f_{Vq}(1-3f_{Vq}+2f_{Vq}^2)\frac{|{\bf q}|^2}{E_q^2T^2}-\frac{m^2f_{Vq}(1-f_{Vq})}{E_q^3T}.
\end{eqnarray}
Using Eqs.~(\ref{equil_rel_1}) and (\ref{equil_rel_2}), one can explicitly show that the Fermi-Dirac distribution satisfies the SKE in Eq.~(\ref{SKE_HTL_4_non}).

As for the AKE, although each coefficient $\grave{\mathcal Q}^{(i)}_\text{cl}$ takes a complicated form, we can make a cross check with the SKE in the massless limt. Generically, to consider the spin diffusion for massless or light quarks, it is inevitable to further incorporate the gluon Compton scattering. Nevertheless, taking the massless limit here is just to scrutinize the consistency of our results.
In the massless limit, $\tilde{a}^{\mu}_q=q^{\mu}f_A$ and thus
\begin{eqnarray}\nonumber
&&\partial_{q^{\beta}_{\perp}}\tilde{a}^{\mu}_q=(\Theta^{\mu}_{\beta}-u^{\mu}\hat{q}_{\perp\beta})f_A+q^{\mu}\partial_{q^{\beta}_{\perp}}f_A,
\quad
\partial_{q^{\beta}_{\perp}}\tilde{a}^{\beta}_q=3f_A+q^{\beta}_{\perp}\partial_{q^{\beta}_{\perp}}f_A,
\\\nonumber
&&\hat{q}_{\perp}^{\beta}\partial_{q^{\beta}_{\perp}}\tilde{a}^{\mu}_q=(\hat{q}^{\mu}_{\perp}+u^{\mu})f_A+q^{\mu}\hat{q}^{\beta}_{\perp}\partial_{q^{\beta}_{\perp}}f_A,
\\\nonumber
&&\partial_{q^{\alpha}_{\perp}}\partial_{q^{\beta}_{\perp}}\tilde{a}^{\mu}_q=(\Theta^{\mu}_{\beta}-u^{\mu}\hat{q}_{\perp\beta})\partial_{q^{\alpha}_{\perp}}f_A+
(\Theta^{\mu}_{\alpha}-u^{\mu}\hat{q}_{\perp\alpha})\partial_{q^{\beta}_{\perp}}f_A
+q^{\mu}\partial_{q^{\alpha}_{\perp}}\partial_{q^{\beta}_{\perp}}f_A
-\frac{u^{\mu}}{|{\bf q}|}\big(\Theta_{\alpha\beta}+\hat{q}_{\perp\alpha}\hat{q}_{\perp\beta}\big)f_A,
\\
&&\partial_{q_{\perp\beta}}\partial_{q^{\beta}_{\perp}}\tilde{a}^{\mu}_q=2(\partial_{q_{\perp\mu}}-u^{\mu}\hat{q}^{\beta}_{\perp}\partial_{q^{\beta}_{\perp}})f_A
+q^{\mu}\partial_{q_{\perp\beta}}\partial_{q^{\beta}_{\perp}}f_A
-\frac{2u^{\mu}}{|{\bf q}|}f_A,
\end{eqnarray}
where $\partial_{q^{\alpha}_{\perp}}\hat{q}_{\perp\beta}=\big(\Theta_{\alpha\beta}+\hat{q}_{\perp\alpha}\hat{q}_{\perp\beta}\big)/|{\bf q}|$. One can show that Eq.~(\ref{AKE_classical_QGP}) then reduces to 
	\begin{eqnarray}\nonumber\label{AKE_classical_massless}
	0&=&q^{\mu}\delta(q^2)\Big[ \, 
	q\cdot\partial f_{Aq}- \kappa_{\rm LL}\Big(2f_{Aq}\big(1-2f_{Vq}-q_{\perp}^{\rho}\partial_{q^{\rho}}f_{Vq}\big)
	\\
	&&	
	+\big(1-2f_{Vq}\big)q^{\rho}_{\perp}\partial_{q^{\rho}}f_{Aq}
	-|{\bf q}|T\eta^{\rho\nu}\partial_{q^{\rho}_{\perp}}\partial_{q^{\nu}_{\perp}}f_{Aq}
	\Big) \, \Big]
	.
	\end{eqnarray}
We may check the consistency with the SKE in the massless limit by dropping the nonlinear terms since the $f_A$ contributions are ignored in the SKE based on our power counting, which could cause discrepancies from the chirality-mixing terms.   
By further taking $f_{Vq}=(f_{Rq}+f_{Lq})/2$ and $f_{Aq}=f_{Rq}-f_{Lq}$, 
one finds that Eq.~(\ref{AKE_classical_massless}) is consistent with the SKE in Eq.~(\ref{SKE_HTL_4_non}) in the massless limit. 
Both the linearized Eqs.~(\ref{AKE_classical_massless}) and (\ref{SKE_HTL_4_non}) result in
\begin{eqnarray}\label{SKE_HTL_massless_non}
0=\delta(q^2-m^2)\Bigg[q\cdot\partial - \kappa_{\rm LL}\Bigg(2+q^{\beta}_{\perp}\partial_{q^{\beta}_{\perp}}
-|{\bf q}|T\eta^{\alpha\beta}\partial_{q^{\alpha}_{\perp}}\partial_{q^{\beta}_{\perp}}\Bigg)\Bigg]f_{R/Lq}.
\end{eqnarray}
Such a remarkable check should support the correctness of Eq.~(\ref{AKE_classical_QGP}). The check is also performed in Ref.~\cite{Li:2019qkf}.

\section{Concluding remarks and outlook}\label{sec_conclusion}
In this paper, we have derived the effective axial kinetic theory with background fields and collisions in the cases when the vector charge is more dominant than the axial charge (or more precisely the spin current) as natural conditions in most of physical systems. It is found that the AKE as a kinetic equation dictating the spin transport not only embraces the spin-diffusion term but also quantum corrections responsible for spin polarization, which reveals nontrivial entanglement of vector/axial-charge and spin transport through collisions. Our SKE and AKT work for an arbitrary mass and they reproduce the CKT in the massless limit. For the case of massive quarks,  we have shown how our formalism reproduces the spin-diffusion term up to the leading-logarithmic order in a weakly coupled QGP and the leading  of $\hbar$ expansion. 

%To give a quick view and simple explanation, here we summarize our findings in a short list:    
Here we further summarize the detailed achievements in a short list :
\begin{enumerate}
	\item In our power counting, the leading-order SKE remains the same as a classical Boltzmann equation,
	whereas the AKE governing the dynamics of spin polarization characterized by an axial-vector component $\mathcal{A}^{\mu}$ in Wigner functions can be written as $\Box^{(n)}\mathcal{A}^{\mu}=\hat{\mathcal{C}}^{\mu}_{\text{cl}}+\hbar\hat{\mathcal{C}}^{(n)\mu}_{\text{Q}}$. Here $\Box^{(n)}\mathcal{A}^{\mu}$ denotes the free-streaming AKE, while $\hat{\mathcal{C}}^{\mu}_{\text{cl}}$ and $\hbar\hat{\mathcal{C}}^{(n)\mu}_{\text{Q}}$ correspond to the ``classical'' and ``quantum'' parts of collisions. Such separation is explicitly shown in a generic spacetime-dependent frame and in the rest frame for massive fermions as well. 
	\item  It is found $\hat{\mathcal{C}}^{\mu}_{\text{cl}}$ and $\hbar\hat{\mathcal{C}}^{(n)\mu}_{\text{Q}}$ are proportional to $f_A$ and $f_V$, respectively. Consequently, $\hat{\mathcal{C}}^{\mu}_{\text{cl}}$ serves as a spin-diffusion term, which vanishes when $f_A=0$. On the contrary, the ``quantum'' correction $\hbar\hat{\mathcal{C}}^{(n)\mu}_{\text{Q}}$, which survives when $f_A=0$, is dubbed as the spin-polarization term and responsible for polarizing spin via the intertwined dynamics of vector-charge transport due to spin-orbit interaction. 
	\item Similar to the free-streaming case \cite{Hattori:2019ahi}, for a generic spacetime-dependent frame, $\hbar\hat{\mathcal{C}}^{(n)\mu}_{\text{Q}}$ can be separated into the term proportional to the four momentum and to the mass, which establishes a smooth connection to the CKT with collisions \cite{Hidaka:2016yjf,Hidaka:2017auj} and manifests spin enslavement by chirality in the massless limit. Also, we present the simplified versions of 
the AKE in Eq.~(\ref{AKE_v1}) [or Eq.~(\ref{AKE_vr})] and Eq.~(\ref{AKE_v2}) suitable for tracking the spin polarization of heavy and light quarks in QGP, respectively.
	\item The spin diffusion term $\hat{\mathcal{C}}^{\mu}_{\text{cl}}$ for massive quarks in weakly-coupled QGP is obtained up to the leading logarithmic order, which incorporates nonlinear terms in distribution functions as a consequence of quantum statistics for fermions. It turns out that even the spin diffusion is affected by entangled dynamics between $f_V$ and $f_A$ as opposed to the previous study with only the linearized collision terms in distribution functions \cite{Li:2019qkf}.
\end{enumerate}

Although we have explicitly evaluated the spin-diffusion term for massive quarks, which are sufficiently heavy for dropping the gluon Compton scattering, the quantum correction could be calculated in a similar fashion as the follow-up work. Moreover, when considering the spin transport for light quarks, it is inevitable to further incorporate the Compton scattering even for just spin diffusion. On the other hand, as already mentioned in the context, in rotating QGP, it is expected that both light quarks and gluons are polarized, which should be involved as quantum corrections in the self-energies. Recently, there have been some relevant studies for the quantum corrections upon polarized photons \cite{Yamamoto:2017uul,Huang:2018aly}.
Albeit the validity of our formalism is held even in the presence of such corrections, 
it could be challenging to systematically include the polarization of scattered gluons or even other quarks and to obtain an analytic form of the collision term. Nevertheless, to understand the dynamical evolution of the spin polarization for peculiarly strange quarks associated with the local polarization of $\Lambda$ hyperons, it will be essential to carry out the aforementioned studies in the future.      

On the other hand, our formalism is rather generic, which may have potential applications not only in heavy ion collisions but also other physical systems. For instance, it is proposed in Refs.~\cite{Yamamoto:2015gzz,Yamamoto:2016xtu,Masada:2018swb} that the electron and neutrino transport with anomalous effects led by chirality imbalance (axial charge) could influence the macroscopic hydrodynamic evolution of matter in core collapse supernovae. However, the chirality imbalance of electrons produced by the electron capture process may be compensated by elastic electron scattering with the effect of nonzero electron mass \cite{Ohnishi:2014uea,Grabowska:2014efa,Kaplan:2016drz}. Our formalism could be applied to track the axial-charge evolution in such a scenario. In such a case, Eq.~(\ref{AKE_v2}) will be useful for capturing the small-electron-mass effect.   

\acknowledgments
%\textit{Acknowledgments}.---
Y. H. was partially supported by Japan Society of Promotion of Science (JSPS), Grants-in-Aid for Scientific Research(KAKENHI) Grants No. 15H03652, 16K17716, and 17H06462.  Y. H. was also partially supported by RIKEN iTHES Project and iTHEMS Program. K.H. is supported in part by Japan Society of Promotion of Science (JSPS), Grants-in-Aid for Scientific Research (KAKENHI) under grant No. 20K03948 and also in part by Yukawa International Program for Quark-hadron Sciences (YIPQS). D.-L. Y. is supported by Keio Institute of Pure and Applied Sciences (KiPAS) project in Keio University and JSPS KAKENHI grant No. 20K14470. The authors also would like to thank participants of the molecular-type Yukawa Institute for Theoretical Physics (YITP) workshop "Quantum kinetic theories in magnetic and vortical fields (YITP-T-19-06)" for fruitful and useful discussions.

\appendix 

\section{Derivation of the master equations}\label{app_derivation_Meq}

\label{sec:master-eqs}

\subsection{Spinor decomposition}

\label{sec:spinor-decomp}

Based on the spinor decomposition of the propagators (\ref{spinor-decomp}) 
and the self-energies (\ref{decomp_Sigma_general}), 
we perform the decomposition of the Kadanoff-Baym equations (\ref{KB_eq}). 
One may use some useful relations such as  
\begin{eqnarray}
&&
 \gam^\mu \gam^\alpha \gam^\beta = {\eta}^{\mu\alpha} \gam^\beta - {\eta}^{\mu\beta} \gam^\alpha +  {\eta}^{\alpha\beta} \gam^\mu 
 - i \epsilon^{\mu\alpha\beta\lambda} \gam^5 \gam_\lambda
, \quad
 \gam^5\spinGamma_{\mu\nu} =  \frac{i}{2} \epsilon_{\mu\nu \kappa\sigma} \spinGamma^{\kappa \sigma}
 \nn
 .
\end{eqnarray}
Here, $\epsilon_{\mu\nu\rho\sigma}$ is the totally antisymmetric tensor with $\epsilon_{0123}=-1$.
The Kadanoff-Baym equations (\ref{KB_eq}) contain 
the commutators $  [\gamma^{\mu}, \nabla_{\mu}S^{<}] $ and $[(\slashed{\Pi}-m), S^{<} ]$ 
and their counterparts with the anticommutation relations. 
They can be decomposed with the following relations 
\begin{subequations}
\begin{eqnarray}
 \label{eq:comm0}
 [ \gam^\mu , S^{<} ] 
  &=&
 - 2\mathcal{A}^\mu \gamma^{5} + 2i  \mathcal{S}^{\mu}_{\ \, \alpha}  \gamma^{\alpha}
-2i \mathcal{P} \gamma^5\gamma^{\alpha}  -2i  \mathcal{V}_{\nu}  \spinGamma^{\mu\nu}
,
\\
\{ \gam^\mu , S^{<} \} 
  &=&
 2  \mathcal{V}^{\mu}  +   2  \mathcal{S}   \gamma^{\mu} 
+ \epsilon^\mu_{\ \, \nu\rho\alpha} \mathcal{S}^{\nu\rho} \gamma^5\gamma^\alpha
 - \epsilon^\mu_{\ \, \nu\alpha\beta} \mathcal{A}^{\nu}  \spinGamma^{\alpha\beta}
.
\label{eq:commf}
\end{eqnarray}
Based on the decompositions above, we now have
\begin{align}
%\label{eq:comm0}
\frac{i}{2}[\gamma^{\mu}, \nabla_{\mu}S^{<}]&=
-i\nabla_{\mu}\mathcal{A}^{\mu} \gamma^{5}
-\nabla^{\nu}\mathcal{S}_{\nu\mu}\gamma^{\mu}
+\nabla_{\mu}\mathcal{P}\gamma^5\gamma^{\mu}
+\nabla_{\mu}\mathcal{V}_{\nu}\sigma^{\mu\nu},\\
\frac{i}{2}\{\gamma^{\mu}, \nabla_{\mu}S^{<}\}&=
i\nabla^{\mu}\mathcal{V}_{\mu}
+i\nabla_{\mu}\mathcal{S}\gamma^{\mu} 
-\frac{i}{2}\epsilon^{\mu\nu\rho\sigma}\nabla_{\sigma}\mathcal{S}_{\nu\rho}\gamma^5\gamma_\mu
-\frac{i}{2}\epsilon_{\mu\nu\rho\sigma} \nabla^{\rho}\mathcal{A}^{\sigma} \sigma^{\mu\nu},\\
[(\slashed{\Pi}-m), S^{<} ]
&= -2\Pi^{\nu}\mathcal{A}_{\nu}\gamma^{5}
+2i\Pi^{\nu} \mathcal{S}_{\nu\mu}\gamma^{\mu}
-2i\Pi_{\mu}\mathcal{P}\gamma^5\gamma^{\mu}
-2i\Pi_{\mu}\mathcal{V}_{\nu}\sigma^{\mu\nu},\\
\{(\slashed{\Pi}-m), S^{<} \} 
&=2(\Pi^{\mu}\mathcal{V}_{\mu}-m\mathcal{S})
-2im \mathcal{P}\gamma^5
+  2( \Pi_{\mu}\mathcal{S}-m \mathcal{V}_{\mu})\gamma^\mu\notag\\
&\quad-(\epsilon^{\mu\nu\rho\sigma}\Pi_{\sigma}\mathcal{S}_{\nu\rho}
+2m\mathcal{A}^{\mu})\gamma^5\gamma_{\mu}
-(m \mathcal{S}_{\mu\nu}
+\epsilon_{\mu\nu\rho\sigma} \Pi^{\rho}\mathcal{A}^{\sigma}
)\sigma^{\mu\nu}.
%\label{eq:commf}
\end{align}
\end{subequations} 
Assuming the following decomposition $G(q,X)=\hat{K}^{\mu}G_{\mu}(q,X)$ and $F(q,X)=\hat{Q}^{\mu}F_{\mu}(q,X)$ with $\hat{K}^{\mu}$ and $\hat{Q}^{\mu}$ being arbitrary matrices, it is found
\begin{eqnarray}\nonumber
\{G,F\}_{\star}&=&\frac{1}{2}\{\hat{K}^{\mu},\hat{Q}^{\nu}\}\{G_{\mu}(q,X),F_{\nu}(q,X)\}_{\star}+\frac{1}{2}[\hat{K}^{\mu},\hat{Q}^{\nu}][G_{\mu}(q,X),F_{\nu}(q,X)]_{\star},
\\
\big[G,F\big]_{\star}&=&\frac{1}{2}[\hat{K}^{\mu},\hat{Q}^{\nu}]\{G_{\mu}(q,X),F_{\nu}(q,X)\}_{\star}+\frac{1}{2}\{\hat{K}^{\mu},\hat{Q}^{\nu}\}[G_{\mu}(q,X),F_{\nu}(q,X)]_{\star}.
\end{eqnarray} 
Accordingly, the self-energy parts can be decomposed in the same and straightforward way as 
\begin{equation}
\begin{split}
[\Sigma^> ,  S^<]_{\star}
&= i\bigl(
\{\bar{\Sigma}_{V\alpha},\mathcal{A}^{\alpha}\}_{\star}
- \{\bar{\Sigma}_{A\alpha},\mathcal{V}^{\alpha}\}_{\star}
\bigr)i\gamma^5\\
&
\quad+i\bigl(\{\bar{\Sigma}_P,\mathcal{A}_{\alpha}\}_{\star}
+ \{\bar{\Sigma}_{V}^{\mu},\mathcal{S}_{\mu\alpha}\}_{\star}
-\{\bar{\Sigma}_{A\alpha},\mathcal{P}\}_{\star}
+ \{\bar{\Sigma}_{T\alpha\beta},\mathcal{V}^{\beta}\}_{\star}
\bigr)\gamma^\alpha\\
&\quad+i\bigl( \{\bar{\Sigma}_P,\mathcal{V}_{\alpha}\}_{\star}- \{\bar{\Sigma}_{V\alpha},\mathcal{P}\}_{\star}
+ \{\bar{\Sigma}_{A}^{\mu},\mathcal{S}_{\mu\alpha}\}_{\star}
+\{\bar{\Sigma}_{T\alpha\beta},\mathcal{A}^{\beta}\}_{\star}
\bigr)\gamma^5\gamma^{\alpha}\\
&\quad+i\bigl( - \{\bar{\Sigma}_{V[\alpha},\mathcal{V}_{\beta]}\}_{\star}
+\{\bar{\Sigma}_{A[\alpha},\mathcal{A}_{\beta]}\}_{\star} 
- \{\bar{\Sigma}_{T\mu[\alpha},\mathcal{S}^{\mu}_{~\beta]}\}_{\star}
\bigr)\frac{\spinGamma^{\alpha\beta}}{2}\\
&\quad+[\bar{\Sigma}_S,\mathcal{S}]_{\star}-[\bar{\Sigma}_P,\mathcal{P}]_{\star}+ [\bar{\Sigma}_{V\mu},\mathcal{V}^{\mu}]_{\star}
-[\bar{\Sigma}_{A\mu},\mathcal{A}^{\mu}]_{\star}
+\frac{1}{2}[\bar{\Sigma}_{T\mu\nu},\mathcal{S}^{\mu\nu}]_{\star}
\\
&\quad+ \Bigl(
[\bar{\Sigma}_S,\mathcal{P}]_{\star}+ [\bar{\Sigma}_P,\mathcal{S}]_{\star} 
+\frac{1}{4}\epsilon_{\mu\nu\alpha\beta}[\bar{\Sigma}_{T}^{\mu\nu},\mathcal{S}^{\alpha\beta}]_{\star}
\Bigr)i\gamma^5\\
&\quad+ \Bigl([\bar{\Sigma}_S,\mathcal{V}_{\alpha}]_{\star}
+ [\bar{\Sigma}_{V\alpha},\mathcal{S}]_{\star}
+\frac{1}{2}\epsilon_{\mu\nu\lambda\alpha}(
 [\bar{\Sigma}_{A}^{\mu},\mathcal{S}^{\nu\lambda}]_{\star}
+[\bar{\Sigma}_{T}^{\mu\nu},\mathcal{A}^{\lambda}]_{\star}
)
\Bigr)\gamma^\alpha
\\
&\quad+\Bigl(
[\bar{\Sigma}_S,\mathcal{A}_{\alpha}]_{\star}
+[\bar{\Sigma}_{A\alpha},\mathcal{S}]_{\star}
+ \frac{1}{2}\epsilon_{\mu\nu\lambda\alpha}([\bar{\Sigma}_{V}^{\mu},\mathcal{S}^{\nu\lambda}]_{\star}
+ [\bar{\Sigma}_{T}^{\mu\nu},\mathcal{V}^{\lambda}]_{\star}) \Bigr)\gamma^5\gamma^{\alpha}\\
&\quad+ \bigl(
   [\bar{\Sigma}_S,\mathcal{S}_{\alpha\beta}]_{\star}
+ [\bar{\Sigma}_{T\alpha\beta},\mathcal{S}]_{\star}
 +\epsilon_{\mu\nu\alpha\beta}(-[\bar{\Sigma}_{V}^{\mu},\mathcal{A}^{\nu}]_{\star}
  +[\bar{\Sigma}_{A}^{\mu},\mathcal{V}^{\nu}]_{\star})\\
&\qquad-\frac{1}{2}  \epsilon_{\mu\nu\alpha\beta}([\bar{\Sigma}_P,\mathcal{S}^{\mu\nu}]_{\star} 
  +[\bar{\Sigma}_{T}^{\mu\nu},\mathcal{P}]_{\star})
  \bigr)\frac{\spinGamma^{\alpha\beta}}{2},
\end{split}
\end{equation}
and
\begin{equation}
\begin{split}
\{\Sigma^> ,  S^<\}_{\star}
&=
\{\bar{\Sigma}_S,\mathcal{S}\}_{\star}- \{\bar{\Sigma}_P,\mathcal{P}\}_{\star}
+ \{\bar{\Sigma}_{V\mu},\mathcal{V}^{\mu}\}_{\star}-\{\bar{\Sigma}_{A\mu},\mathcal{A}^{\mu}\}_{\star}
+ \frac{1}{2}\{{\bar{\Sigma}_{T\mu\nu}},{\mathcal{S}^{\mu\nu}}\}_{\star}
\\
&\quad+\Bigl( \{\bar{\Sigma}_S,\mathcal{P}\}_{\star}+ \{\bar{\Sigma}_P,\mathcal{S}\}_{\star}
+ \frac{1}{4}\epsilon_{\mu\nu\alpha\beta}\{{\bar{\Sigma}_{T}^{\mu\nu}},{\mathcal{S}^{\alpha\beta}}\}_{\star}
\Bigr)i\gamma^5
\\
&\quad+\Bigl( \{\bar{\Sigma}_S,\mathcal{V}_{\alpha}\}_{\star}
+ \{\bar{\Sigma}_{V\alpha},\mathcal{S}\}_{\star}
+\frac{1}{2}\epsilon_{\mu\nu\lambda\alpha}(\{\bar{\Sigma}_{A}^{\mu},\mathcal{S}^{\nu\lambda}\}_{\star}+\{{\bar{\Sigma}_{T}^{\mu\nu}},\mathcal{A}^{\lambda}\}_{\star} )
\Bigr)\gamma^\alpha\\
&\quad+\Bigl(\{\bar{\Sigma}_S,\mathcal{A}_{\alpha}\}_{\star}
+\{\bar{\Sigma}_{A\alpha},\mathcal{S}\}_{\star}
+\frac{1}{2}\epsilon_{\mu\nu\lambda\alpha}( \{\bar{\Sigma}_{V}^{\mu},\mathcal{S}^{\nu\lambda}\}_{\star}+\{\bar{\Sigma}_{T}^{\mu\nu},\mathcal{V}^{\lambda}\}_{\star} )
\Bigr)\gamma^5\gamma^\alpha \\
&\quad+\bigl( \{\bar{\Sigma}_S,\mathcal{S}_{\alpha\beta}\}_{\star}
+ \{\bar{\Sigma}_{T\alpha\beta},\mathcal{S}\}_{\star}
+\epsilon_{\mu\nu\alpha\beta}(-\{\bar{\Sigma}_{V}^{\mu},\mathcal{A}^{\nu}\}_{\star}
+\{\bar{\Sigma}_{A}^{\mu},\mathcal{V}^{\nu}\}_{\star})
\\
&\qquad-\frac{1}{2}\epsilon_{\mu\nu\alpha\beta} (
 \{\bar{\Sigma}_P,\mathcal{S}^{\mu\nu}\}_{\star}+\{{\Sigma}_{T}^{\mu\nu},\mathcal{P}\}_{\star})
\bigr)\frac{1}{2}\spinGamma^{\alpha\beta}\\
&\quad+i\bigl([\bar{\Sigma}_{V\alpha},\mathcal{A}^{\alpha}]_{\star}-[\bar{\Sigma}_{A\alpha},\mathcal{V}^{\alpha}]_{\star}
\bigr)i\gamma^{5}\\
&\quad+i\bigl([\bar{\Sigma}_P,\mathcal{A}_{\alpha}]_{\star}
+ [\bar{\Sigma}_{V}^{\mu},\mathcal{S}_{\mu\alpha}]_{\star}
- [\bar{\Sigma}_{A\alpha},\mathcal{P}]_{\star}
+[\bar{\Sigma}_{T\alpha\beta},\mathcal{V}^{\beta}]_{\star} 
\bigr)\gamma^\alpha\\
&\quad+i\bigl([\bar{\Sigma}_P,\mathcal{V}_{\alpha}]_{\star}
+ [\bar{\Sigma}_{A}^{\mu},\mathcal{S}_{\mu\alpha}]_{\star}
- [\bar{\Sigma}_{V\alpha},\mathcal{P}]_{\star}
 +[{\bar{\Sigma}_{T\alpha\beta}},\mathcal{A}^{\beta}]_{\star}
\bigr)\gamma^5\gamma^\alpha 
\\
&
\quad+i(- [\bar{\Sigma}_{V[\alpha},\mathcal{V}_{\beta]}]_{\star}
+[\bar{\Sigma}_{A[\alpha},\mathcal{A}_{\beta]}]_{\star} 
-[{\bar{\Sigma}_{T\mu[\alpha}},{\mathcal{S}^{\mu}_{~\beta]}}]_{\star}
 )\frac{1}{2}\spinGamma^{\alpha\beta},
\end{split}
\end{equation}
where we defined the antisymmetrization $ T_{[\mu\nu ]} = T_{\mu\nu} - T_{\nu\mu} $.
Plugging those decompositions back to the Kadanoff-Baym equations (\ref{KB_eq}), we find
\begin{eqnarray}\label{KB_eq1}
0&=&\{(\slashed{\Pi}-m), S^< \} +\frac{i\hbar}{2}\Big([\gamma^{\mu}, \nabla_{\mu}S^{<}]-[\Sigma^<, S^>]_{{\star}}+[\Sigma^>, S^<]_{{\star}}\Big)
\\\nonumber
&=&\mathscr{K}_S+\mathscr{K}_5{i}\gamma^5+\mathscr{K}_{V\mu}\gamma^{\mu}+\mathscr{K}_{5\mu}\gamma^{\mu}\gamma^5+\mathscr{K}_{T\mu\nu}{\frac{1}{2}}\sigma^{\mu\nu},\\
\label{KB_eq2}
0&=&[(\slashed{\Pi}-m), S^{<}]+\frac{i\hbar}{2}\Big(\{\gamma^{\mu}, \nabla_{\mu}S^{<}\}-\{\Sigma^<,  S^{>}\}_{{\star}}+\{\Sigma^>,  S^{<}\}_{{\star}}\Big)
\\\nonumber
&=&\mathscr{Q}_S+\mathscr{Q}_5{i}\gamma^5+\mathscr{Q}_{V\mu}\gamma^{\mu}+\mathscr{Q}_{5\mu}\gamma^{\mu}\gamma^5+\mathscr{Q}_{T\mu\nu}{\frac{1}{2}}\sigma^{\mu\nu},
\end{eqnarray}
where
\begin{align}
\mathscr{K}_{S}&=
2\Pi^{\mu}\mathcal{V}_{\mu}-2m\mathcal{S}\notag\\
&\quad+\frac{i\hbar}{2}\Bigl(\widehat{[{\Sigma}_S,\mathcal{S}]_{\star}}-\widehat{[{\Sigma}_P,\mathcal{P}]_{\star}}+ \widehat{[{\Sigma}_{V\mu},\mathcal{V}^{\mu}]_{\star}}
-\widehat{[{\Sigma}_{A\mu},\mathcal{A}^{\mu}]_{\star}}
+\frac{1}{2}\widehat{[{\Sigma}_{T\mu\nu},\mathcal{S}^{\mu\nu}]_{\star}}
\Bigr),\\
\mathscr{K}_{5}&=-2m \mathcal{P}-\hbar \tilde \cD\mathcal{A}^{\mu}
+\frac{\hbar}{2} \widehat{\{{\Sigma}_{A\mu},\mathcal{V}^{\mu}\}_{\star}}
+ \frac{i\hbar}{2}\Bigl(
\widehat{[{\Sigma}_S,\mathcal{P}]_{\star}}+ \widehat{[{\Sigma}_P,\mathcal{S}]_{\star} }
+\frac{1}{4}\epsilon_{\mu\nu\alpha\beta}\widehat{[{\Sigma}_{T}^{\mu\nu},\mathcal{S}^{\alpha\beta}]_{\star}}
\Bigr),\\
\mathscr{K}_{V\alpha}&=
  2 \Pi_{\alpha}\mathcal{S}- 2m \mathcal{V}_{\alpha}
-\hbar\tilde \cD^{\nu}\mathcal{S}_{\nu\alpha}
-\frac{\hbar}{2}\bigl(\widehat{\{{\Sigma}_P,\mathcal{A}_{\alpha}\}_{\star}}
-\widehat{\{{\Sigma}_{A\alpha},\mathcal{P}\}_{\star}}
+ \widehat{\{{\Sigma}_{T\alpha\mu},\mathcal{V}^{\mu}\}_{\star}}
\bigr)\notag\\
&\quad+\frac{i\hbar}{2} \Bigl(\widehat{[{\Sigma}_S,\mathcal{V}_{\alpha}]_{\star}}
+ \widehat{[{\Sigma}_{V\alpha},\mathcal{S}]_{\star}}
+\frac{1}{2}\epsilon_{\mu\nu\lambda\alpha}(
 \widehat{[{\Sigma}_{A}^{\mu},\mathcal{S}^{\nu\lambda}]_{\star}}
+\widehat{[{\Sigma}_{T}^{\mu\nu},\mathcal{A}^{\lambda}]_{\star}}
)
\Bigr),\\
\mathscr{K}_{5\alpha}&=
-\epsilon_{\alpha\nu\rho\sigma}\Pi^{\sigma}\mathcal{S}^{\nu\rho}-2m\mathcal{A}_{\alpha}
+\hbar\tilde \cD_{\alpha}\mathcal{P}
-\frac{\hbar}{2}\bigl( \widehat{\{{\Sigma}_P,\mathcal{V}_{\alpha}\}_{\star}}
+ \widehat{\{{\Sigma}_{A}^{\mu},\mathcal{S}_{\mu\alpha}\}_{\star}}
+\widehat{\{{\Sigma}_{T\alpha\mu},\mathcal{A}^{\mu}\}_{\star}}
\bigr)\notag\\
&\quad+\frac{i\hbar}{2}\Bigl(
\widehat{[{\Sigma}_S,\mathcal{A}_{\alpha}]_{\star}}
+\widehat{[{\Sigma}_{A\alpha},\mathcal{S}]_{\star}}
+ \frac{1}{2}\epsilon_{\mu\nu\lambda\alpha}(\widehat{[{\Sigma}_{V}^{\mu},\mathcal{S}^{\nu\lambda}]_{\star}}
+ \widehat{[{\Sigma}_{T}^{\mu\nu},\mathcal{V}^{\lambda}]_{\star}}) \Bigr),\\
\mathscr{K}_{T\alpha\beta}&=
-2m \mathcal{S}_{\alpha\beta}
-2\epsilon_{\alpha\beta\rho\sigma} \Pi^{\rho}\mathcal{A}^{\sigma}
+\hbar\tilde \cD_{[\alpha}\mathcal{V}_{\beta]}
-\frac{\hbar}{2}\bigl( 
\widehat{\{{\Sigma}_{A[\alpha},\mathcal{A}_{\beta]}\}_{\star}}
- \widehat{\{{\Sigma}_{T\mu[\alpha},\mathcal{S}^{\mu}_{~\beta]}\}_{\star}}
\bigr)\notag\\
&\quad+ \frac{i\hbar}{2}\bigl(
   \widehat{[{\Sigma}_S,\mathcal{S}_{\alpha\beta}]_{\star}}
+ \widehat{[{\Sigma}_{T\alpha\beta},\mathcal{S}]_{\star}}
 +\epsilon_{\mu\nu\alpha\beta}(-\widehat{[{\Sigma}_{V}^{\mu},\mathcal{A}^{\nu}]_{\star}}
  +\widehat{[{\Sigma}_{A}^{\mu},\mathcal{V}^{\nu}]_{\star}})\notag\\
&\qquad-\frac{1}{2}  \epsilon_{\mu\nu\alpha\beta}(\widehat{[{\Sigma}_P,\mathcal{S}^{\mu\nu}]_{\star} }
  +\widehat{[{\Sigma}_{T}^{\mu\nu},\mathcal{P}]_{\star}})
  \bigr),
\end{align}
and
\begin{align} 
\mathscr{Q}_{S}&=i\hbar\tilde \cD_{\mu}\mathcal{V}^{\mu}
+\frac{i\hbar}{2}\Bigl(\widehat{\{{\Sigma}_S,\mathcal{S}\}_{\star}}- \widehat{\{{\Sigma}_P,\mathcal{P}\}_{\star}}
-\widehat{\{{\Sigma}_{A\mu},\mathcal{A}^{\mu}\}_{\star}}
+ \frac{1}{2}\widehat{\{{{\Sigma}_{T\mu\nu}},{\mathcal{S}^{\mu\nu}}\}_{\star}}\Bigr),\\
\mathscr{Q}_{5}&=2i\Pi^{\nu}\mathcal{A}_{\nu}
+\frac{i\hbar}{2}\Bigl( \widehat{\{{\Sigma}_S,\mathcal{P}\}_{\star}}+ \widehat{\{{\Sigma}_P,\mathcal{S}\}_{\star}}
+ \frac{1}{4}\epsilon_{\mu\nu\alpha\beta}\{\widehat{{{\Sigma}_{T}^{\mu\nu}},{\mathcal{S}^{\alpha\beta}}\}_{\star}}
\Bigr)\notag\\
&\quad-\frac{\hbar}{2}\bigl(\widehat{[{\Sigma}_{V\alpha},\mathcal{A}^{\alpha}]_{\star}}-\widehat{[{\Sigma}_{A\alpha},\mathcal{V}^{\alpha}]_{\star}}
\bigr),\\
\mathscr{Q}_{V\alpha}&=2i\Pi^{\nu} \mathcal{S}_{\nu\alpha}
+i\hbar\tilde \cD_{\alpha}\mathcal{S}
+\frac{i\hbar}{2}\Bigl( \widehat{\{{\Sigma}_S,\mathcal{V}_{\alpha}\}_{\star}}
+\frac{1}{2}\epsilon_{\mu\nu\lambda\alpha}(\widehat{\{{\Sigma}_{A}^{\mu},\mathcal{S}^{\nu\lambda}\}_{\star}}
+\widehat{\{{{\Sigma}_{T}^{\mu\nu}},\mathcal{A}^{\lambda}\}_{\star}} )
\Bigr)\notag\\
&
\quad-\frac{\hbar}{2}\bigl(\widehat{[{\Sigma}_P,\mathcal{A}_{\alpha}]_{\star}}
+ \widehat{[{\Sigma}_{V}^{\mu},\mathcal{S}_{\mu\alpha}]_{\star}}
- \widehat{[{\Sigma}_{A\alpha},\mathcal{P}]_{\star}}
+\widehat{[{\Sigma}_{T\alpha\beta},\mathcal{V}^{\beta}]_{\star} }
\bigr),\\
\mathscr{Q}_{5\alpha}&=-2i\Pi_{\alpha}\mathcal{P}
-\frac{i\hbar}{2}\epsilon_{\alpha\nu\rho\sigma}\tilde \cD^{\nu}\mathcal{S}^{\rho\sigma}
+\frac{i\hbar}{2}\Bigl(\widehat{\{{\Sigma}_S,\mathcal{A}_{\alpha}\}_{\star}}
+\widehat{\{{\Sigma}_{A\alpha},\mathcal{S}\}_{\star}}
-\frac{1}{2}\epsilon_{\alpha\nu\rho\sigma}\widehat{\{{\Sigma}_{T}^{\nu\rho},\mathcal{V}^{\sigma}\}_{\star}} 
)
\Bigr)\notag\\
&\quad-\frac{\hbar}{2}\bigl(\widehat{[{\Sigma}_P,\mathcal{V}_{\alpha}]_{\star}}
+ \widehat{[{\Sigma}_{A}^{\mu},\mathcal{S}_{\mu\alpha}]_{\star}}
- \widehat{[{\Sigma}_{V\alpha},\mathcal{P}]_{\star}}
 +\widehat{[{{\Sigma}_{T\alpha\mu}},\mathcal{A}^{\mu}]_{\star}}
\bigr),\\
\mathscr{Q}_{T\alpha\beta}&=-2i\Pi_{[\alpha}\mathcal{V}_{\beta]}
-i\hbar\epsilon_{\alpha\beta\mu\nu} \tilde \cD^{\mu}\mathcal{A}^{\nu}
+\frac{i\hbar}{2}\bigl( \widehat{\{{\Sigma}_S,\mathcal{S}_{\alpha\beta}\}_{\star}}
+ \widehat{\{{\Sigma}_{T\alpha\beta},\mathcal{S}\}_{\star}}
+\epsilon_{\alpha\beta\mu\nu}\widehat{\{{\Sigma}_{A}^{\mu},\mathcal{V}^{\nu}\}_{\star}}
\notag\\
&\quad
-\frac{1}{2}\epsilon_{\alpha\beta\mu\nu} (
 \widehat{\{{\Sigma}_P,\mathcal{S}^{\mu\nu}\}_{\star}}
 +\widehat{\{{\Sigma}_{T}^{\mu\nu},\mathcal{P}\}_{\star}})
\bigr)
-\frac{\hbar}{2}(- \widehat{[{\Sigma}_{V[\alpha},\mathcal{V}_{\beta]}]_{\star}}
+\widehat{[{\Sigma}_{A[\alpha},\mathcal{A}_{\beta]}]_{\star} }
-\widehat{[{{\Sigma}_{T\mu[\alpha}},{\mathcal{S}^{\mu}_{~\beta]}}]_{\star}}
 ).
 \label{eq:A16}
\end{align}
Here we introduce a shorthand notation $   \widehat { X Y  }=  \bar X Y - X \bar Y   $, where $  X$ and $Y  $ are the coefficients of the Clifford decomposition of the propagators and self-energies, e.g., $ \widehat{ \{\Sigma_{V\mu}, \cV_\nu\}_{\star}} = \{\bar \Sigma_{V\mu}, \cV_\nu\}_{\star} -  \{\Sigma_{V\mu}, \bar \cV_\nu\}_{\star}$.  We also introduced $\tilde \cD_\mu \M = \nabla_\mu \M + \widehat { \{\Sigma_{V\mu}, \M\}_{\star} }/2$.
$\mathscr{K}_{S}=\mathscr{K}_{5}=\mathscr{K}_{V\alpha}=\mathscr{K}_{5\alpha}=\mathscr{Q}_{T\alpha\beta}=\mathscr{Q}_{S}=\mathscr{Q}_{5}=\mathscr{Q}_{V\alpha}=\mathscr{Q}_{5\alpha}=\mathscr{Q}_{T\alpha\beta}=0$ give the set of full quantum master equations. 
Explicitly, they read 
\begin{subequations}
	\label{eq:Eq-sum0}
	\begin{eqnarray}
	&&\label{M2} 
	m \cS = \Pi^{\mu} \mathcal{V}_{\mu} 
	{-\frac{\hbar^2}{4}[ \widehat{ \Sigma_{S} \cS} - \widehat{\Sigma_{P}  \cP}  +  \widehat{\Sigma_{V\mu}  \cV^\mu} -   \widehat{\Sigma_{A\mu}  \cA^\mu} 
		+ \frac{1}{2} \widehat{\Sigma_{T\mu\nu} \cS^{\mu\nu}}   ]_\text{P.B.}+\mathcal{O}(\hbar^3)}
	,
	\\
	&&\label{M5} 
	m\cP =  - \frac{ \hbar}{2} ( \tilde \cD_{\mu}\mathcal{A}^{\mu} - \widehat {  \Sigma_{A\mu} \cV^\mu}) 
	{-\frac{\hbar^2}{4}[  \widehat{\Sigma_{S}   \cP} +   \widehat{\Sigma_P  \cS} 
		+  \frac{1}{4} \epsilon_{\mu\nu\alpha\beta} \widehat{ \Sigma_{T}^{\mu\nu} \cS^{\alpha\beta}} ]_\text{P.B.}+\mathcal{O}(\hbar^3)}
	,
	\\\nonumber
	&&\label{M7}
	2 \Pi_{\alpha}\mathcal{S}  - \hbar \tilde \cD^{\nu}\mathcal{S}_{\nu\alpha} - 2m \cV_\alpha
	- \hbar (  \widehat { \Sigma_P  \cA_\alpha } -\widehat { \Sigma_{A\alpha}  \cP } -  \widehat {\Sigma_{T\mu\alpha} \cV^\mu } ) 
	\\
	&&={\frac{\hbar^2}{2}[\widehat{ \Sigma_{S} \cV_\alpha}  + \widehat{ \Sigma_{V \alpha} \cS} 
		+  \frac{1}{2}  \epsilon_{\mu\nu\lambda\alpha}
		( \widehat{ \Sigma_{A}^{\mu}  \cS^{\nu\lambda}}  + \widehat{ \Sigma_{T}^{\mu\nu}  \cA^{\lambda}} )
		]_\text{P.B.}+\mathcal{O}(\hbar^3)}
	,
	\\
	&&\label{M9}
	\hbar \tilde \cD_{\alpha}\mathcal{P}  - \epsilon_{\alpha\nu\rho\sigma}\Pi^{\sigma}\mathcal{S}^{\nu\rho} 
	-  2m \cA_\alpha
	-\hbar (   \widehat { \Sigma_P\cV_\alpha }   +\widehat {\Sigma_{A}^\mu \cS_{\mu\alpha} }
	- \widehat { \Sigma_{T\mu\alpha} \cA^{\mu} } )\notag
	\\
	&&={\frac{\hbar^2}{2}[ \widehat{ \Sigma_{S}  \cA_\alpha}   + \widehat{ \Sigma_{A\alpha}   \cS}
		+ \frac{1}{2}  \epsilon_{\mu\nu\lambda\alpha}
		( \widehat{ \Sigma_{V}^{\mu}  \cS^{\nu\lambda}}  + \widehat{ \Sigma_{T}^{\mu\nu}  \cV^{\lambda}} )
		]_\text{P.B.}+\mathcal{O}(\hbar^3)}
	,
	\\
	&&\label{M4} \label{replace_3}\nonumber
	m \cS_{\alpha\beta} 
	+ \epsilon_{\alpha\beta\rho\sigma} \Pi^{\rho}\mathcal{A}^{\sigma} 
	-\frac{ \hbar }{2} ( \tilde \cD_{[\alpha}\mathcal{V}_{\beta]}
	- \widehat { \Sigma_{A[\alpha}   \cA_{\beta]} } +  \widehat {\Sigma_{T\mu[\alpha} \cS^\mu_{\ \,\beta]} } )
	\\
	&&=
	{-\frac{\hbar^2}{4}
		[ \widehat{ \Sigma_{S}   \cS_{\alpha\beta}} +  \widehat{ \Sigma_{T\alpha\beta}   \cS}
		-  \epsilon_{\mu\nu\alpha\beta} \widehat{ \Sigma_{V}^{\mu} \cA^\nu}
		+   \epsilon_{\mu\nu\alpha\beta} \widehat{ \Sigma_{A}^{\mu}   \cV^\nu} 
		-  \frac{1}{2}  \epsilon_{\mu\nu\alpha\beta}  (\widehat{ \Sigma_{P}  \cS^{\mu\nu}}  
		+  \widehat{ \Sigma_{T}^{\mu\nu}  \cP} )
		]_\text{P.B.}} 
			+\mathcal{O}(\hbar^3)
	,\notag\\
	\end{eqnarray}
\end{subequations}
and 
\begin{subequations}
	\label{eq:Eq-diff0}
	\begin{eqnarray}
	&&\label{M1} 
	\tilde \cD^{\mu} \cV_\mu =
	-  \widehat{ \Sigma_{S}  \cS} + \widehat{ \Sigma_{P} \cP } 
	+  \widehat{ \Sigma_{A\mu} \cA^\mu }
	- \frac{1}{2} \widehat{ \Sigma_{T\mu\nu} \cS^{\mu\nu}  } +\mathcal{O}(\hbar^3)
	,
	\\
	&&\label{M6}
	2\Pi^{\nu}\mathcal{A}_{\nu} =
	- \hbar(  \widehat{ \Sigma_{S}  \cP} + \widehat{ \Sigma_P \cS }
	+  \frac{1}{ 4 } \epsilon_{\mu\nu\alpha\beta} \widehat{\Sigma_{T}^{\mu\nu} \cS^{\alpha\beta}} ) 
	{-\frac{\hbar^2}{2}[ -  \widehat{ \Sigma_{V\alpha} \cA^\alpha} +  \widehat{ \Sigma_{A\alpha}  \cV^\alpha}]_\text{P.B.}+\mathcal{O}(\hbar^3)}
	,
	\\
	&&\label{M8}\nonumber
	2 \Pi^{\nu} \mathcal{S}_{\nu\alpha}+  \hbar \tilde \cD_{\alpha}\mathcal{S}
	+\hbar (  \widehat{ \bar \Sigma_{S} \cV_\alpha}
	+  \frac{1}{2}  \epsilon_{\mu\nu\lambda\alpha}
	( \widehat{ \Sigma_{A}^{\mu} \cS^{\nu\lambda} } + \widehat{ \Sigma_{T}^{\mu\nu} \cA^{\lambda} } )
	)
	\\
	&&=
	{\frac{\hbar^2}{2}[   \widehat{ \Sigma_P  \cA_\alpha} + \widehat{ \Sigma_{V}^{\mu}  \cS_{\mu\alpha}} - \widehat{ \Sigma_{A{\alpha}}  \cP} +  \widehat{ \Sigma_{T\alpha\beta}  \cV^\beta}  ]_\text{P.B.}+\mathcal{O}(\hbar^3)}
	,
	\\
	&& \label{M10}\nonumber
	2 \Pi_{\alpha}\mathcal{P}  
	+ \frac{\hbar}{2}\epsilon_{\alpha\nu\rho\sigma} 
	( \tilde \cD^{\nu}\mathcal{S}^{\rho\sigma} +   \widehat{ \Sigma_{T}^{\nu\rho} \cV^{\sigma}} )
	-\hbar ( \widehat{ \Sigma_{S} \cA_\alpha }  + \widehat{ \Sigma_{A\alpha}  \cS} )
	\\
	&&=1-{\frac{\hbar^2}{2}[  \widehat{ \Sigma_P \cV_\alpha} -   \widehat{ \Sigma_{V{\alpha}}   \cP} 
		+  \widehat{\Sigma_{A}^\mu   \cS_{\mu\alpha}} 
		+  \widehat{\Sigma_{T\alpha\beta}  \cA^{\beta}}  ]_\text{P.B.}+\mathcal{O}(\hbar^3)}
	,
	\\
	&&\label{M3}\nonumber
	\Pi_{[\alpha}\mathcal{V}_{\beta]} 
	+ \frac{\hbar}{2}\epsilon_{\alpha\beta\mu\nu} 
	( \tilde \cD^{\mu}\mathcal{A}^{\nu}    -   \widehat{ \Sigma_{A}^{\mu}  \cV^\nu} )
	- \frac{\hbar}{2} ( \widehat{ \Sigma_{S} \cS_{\alpha\beta} } +  \widehat{ \Sigma_{T\alpha\beta}  \cS} )
	+ \frac{\hbar}{4}  \epsilon_{\mu\nu\alpha\beta} (  \widehat{ \Sigma_{P}  \cS^{\mu\nu}  }  + \widehat{ \Sigma_{T}^{\mu\nu} \cP} )
	\\
	&&={\frac{\hbar^2}{4}[  \widehat{ \Sigma_{V{[}\alpha}    \cV_{\beta{]}}} {-}  \widehat{ \Sigma_{A{[}\alpha}    \cA_{\beta{]}}} 
		{+}  \widehat{ \Sigma_{T\mu{[}\alpha}  \cS^\mu_{\ \,\beta{]}}}]_\text{P.B.}+\mathcal{O}(\hbar^3)}
	,
	\end{eqnarray}
\end{subequations}
where $[AB]_\text{P.B.}=\{A(q,X),B(q,X)\}_\text{P.B.}$ is a shorthand notation for the Poisson bracket.

\subsection{Eliminating $\mathcal{S}$, $\mathcal{P}$ and $\mathcal{S}_{\mu\nu}$}

\label{sec:elimination}

Equations~(\ref{M2}), (\ref{M5}) and (\ref{M4}) can be used 
to eliminate $\mathcal{S}$, $\mathcal{P}$, and $\mathcal{S}_{\mu\nu}$ from the other master equations. 
Since $ \cS^{\mu\nu} $ is contracted with $\Sigma_{T\mu\nu}$ and $\bar{\Sigma}_{T\mu\nu}$ on the right-hand side, 
one may not express $\mathcal{S}_{\mu\nu}$ as an explicit function of $\mathcal{V}_{\mu}$ and $\mathcal{A}_{\mu}$ only. 
However, assuming that the interaction is sufficiently weak, 
we may drop the nonlinear terms in the self-energy. 
Within this assumption, we may rewrite Eqs.~(\ref{M2}), (\ref{M5})  and (\ref{replace_3}) as
\begin{subequations} \label{eq:SPS}
\begin{eqnarray}
\label{eq:S-weak}
\cS_{\alpha\beta} 
&\approx &
- \frac{1}{m} \epsilon_{\alpha\beta\rho\sigma} \Pi^{\rho}\mathcal{A}^{\sigma} 
+ \frac{ \hbar }{2m} \{  \cD_{[\alpha}\mathcal{V}_{\beta]}
- \widehat { \Sigma_{A[\alpha}   \cA_{\beta]} } 
+  {\frac{q_{\mu}}{m}}  \epsilon_{\ \ \ \, [\alpha}^{\mu \rho\sigma} \widehat{ \Sigma_{T\beta] \rho}  \mathcal{A}_{\sigma} }  \,\}
\nn
\\\nonumber
&&
+ \frac{ \hbar^2 }{4m^{{2}}}  ( \widehat{ \Sigma_{T\mu[\alpha}  \nabla^{\mu} \mathcal{V}_{\beta]} }
- \widehat{ \Sigma_{T\mu[\alpha}  \nabla_{\beta]} \mathcal{V}^{\mu} }) 
{-\frac{\hbar^2}{4m}
	[ \frac{\epsilon_{\alpha\beta\mu\nu}}{m}\widehat{ \Sigma_{S}(q^{\nu}\mathcal{{A}}^{\mu})} +  \frac{1}{m}\widehat{ \Sigma_{T\alpha\beta}(q\cdot\mathcal{V})}}
\\
&&{	-  \epsilon_{\mu\nu\alpha\beta} \widehat{ \Sigma_{V}^{\mu} \cA^\nu}
	+   \epsilon_{\mu\nu\alpha\beta} \widehat{ \Sigma_{A}^{\mu}   \cV^\nu} 
	-  \frac{1}{m}\widehat{ \Sigma_{P} (q_{[\alpha}\mathcal{A}_{\beta]})}  
	]_\text{P.B.}+\mathcal{O}(\hbar^3)}, 
\\
\label{replace_1}
\cS &=& \frac{\Pi^{\mu}}{m} \mathcal{V}_{\mu} 
{-\frac{\hbar^2}{4m}[ \frac{\widehat{\Sigma_{S} \c(q\cdot \mathcal{V})}}{m}  +  \widehat{\Sigma_{V\mu}  \cV^\mu} -   \widehat{\Sigma_{A\mu}  \cA^\mu} 
	- \frac{1}{2m} \epsilon^{\mu\nu\alpha\beta}\widehat{\Sigma_{T\mu\nu} (q_{\alpha}\cA_{\beta})}   ]_\text{P.B.}+\mathcal{O}(\hbar^3)},
	\\
	\label{replace_2}
\cP &=&  - \frac{ \hbar}{2m} ( \tilde \cD_{\mu}\mathcal{A}^{\mu} - \widehat {  \Sigma_{A\mu} \cV^\mu}) 
{-\frac{\hbar^2}{4m}[ \frac{\widehat{\Sigma_P  (q\cdot\mathcal{V})}}{m} 
	+  \frac{1}{ 2m }\widehat{ \Sigma_{T}^{\mu\nu} (q_{[\mu}\mathcal{A}_{\nu]})} ]_\text{P.B.}+\mathcal{O}(\hbar^3)}.
\end{eqnarray}
\end{subequations}

Inserting Eqs.~(\ref{replace_1}), (\ref{replace_2}) and (\ref{eq:S-weak}) 
and maintaining the linear terms in the self-energies 
and the explicit $  \hbar $ dependence up to $ {\mathcal O}(\hbar) $ 
(or $\mathcal{O}(\hbar^2)$ for involving at least the next-leading-order corrections), 
the master equations (\ref{M7}) and (\ref{M9}) up to $\mathcal{O}(\hbar)$ are reduced as 
\begin{subequations}
	\label{eq:Eq-sum}
	\begin{eqnarray}
	&&q^{\mu}  ( q \cdot \mathcal{V})    - m ^2 \cV^\mu
	\approx \frac{ \hbar }{2}  [ m (  \widehat { \Sigma_P  \cA^\mu }  + \widehat {\Sigma_{T}^{\mu\alpha} \cV_\alpha } ) 
	- 2 ( \tilde F^{\mu\beta} {\mathcal{A}}_\beta 
	+ \frac{1}{2}  \epsilon^{\mu \alpha\beta\gamma}  q_{\alpha} \Delta_\beta \mathcal{A}_{\gam} ) 
	-   \epsilon^{\mu \alpha \beta\gam} q_\alpha \widehat{ \Sigma_{V\beta} \cA_\gam }
	]
	,
	\nn
	\\
%	\label{M7_N}
	\\
	&&
%	\label{M9_N}
	  q^2 \mathcal{A}^\mu -  q^\mu q \cdot \mathcal{A}  -  m^2 \cA^\mu
	\\
	&& \quad \quad
	\approx  
	\frac{1}{2}\hbar \Bigl[ m (   \widehat { \Sigma_P\cV^\mu }   + \widehat { \Sigma_{T}^{\mu\alpha} \cA_{\alpha} } ) 
	+ \epsilon^{\mu\alpha\beta\gam} q_{\alpha} \cD_{\beta}\mathcal{V}_\gam
	- 2  \epsilon^{\mu\alpha\beta\gam} q_{\alpha} \widehat {\Sigma_{A\beta}  \mathcal{A}_\gam }
	-{\frac{1}{m}} \epsilon^{\mu}_{\ \, \alpha\beta\gamma}    \epsilon^{ \beta\lambda\rho\sigma } q^{\alpha}  q_{\lambda}
	\widehat{   \Sigma_{T \rho}^{\gam}   \mathcal{A}_{\sigma} }
	\Bigr]
	.
	\nn
	\end{eqnarray}
\end{subequations}
Similarly, the master equations (\ref{M1})-(\ref{M3}) up to $\mathcal{O}(\hbar)$ are reduced as 
\begin{subequations}
	\label{eq:Eq-diff}
	\begin{eqnarray}
	&&%\label{M1_N} 
	\cD^{\mu} \cV_\mu \approx
	\widehat{ \Sigma_{A\mu} \cA^\mu }
	-  \frac{1}{2m} [ 2 q_\mu \widehat{ \Sigma_{S}   \mathcal{V}  ^\mu } 
	-  \epsilon_{\mu\nu\alpha\beta} q^{\alpha} \widehat{ \Sigma_{T}^{\mu\nu}  \mathcal{A}^{\beta}   } 
	+ \hbar \widehat{ \Sigma_{P}  (\nabla \cdot  \mathcal{A} )} 
	+ \frac{\hbar}{2} \widehat{ \Sigma_{T}^{\mu\nu}  \nabla_{[\mu}\mathcal{V}_{\nu]}  } 
	],
	\\
	&&%\label{M6_N}
	q \cdot \mathcal{A}  \approx
	- \frac{\hbar}{2m}  q_\mu ( \widehat{ \Sigma_P  \mathcal{V}^\mu }
	+  \widehat{\Sigma_{T}^{\mu\nu}  \mathcal{A}_{\nu}  } ),
	\\
	&&%\label{M8_N}
	\cD_{\mu}  (q \cdot \cV) - q^{\nu}  D_{[\mu}\mathcal{V}_{\nu]} 
	\nn
	\\\nonumber
	&&%\hspace{5cm}
	\approx -  m  \widehat{ \Sigma_{S} \cV_\mu}
	+ q_{[\mu}  \widehat { \Sigma_{A \alpha]}   \cA^{\alpha} } 
	+ {\frac{1}{m}} q^{\nu}   q_{\alpha}  \epsilon_{\ \ \ \, [\mu}^{\alpha \beta\gam} \widehat{ \Sigma_{T\nu] \beta}  \mathcal{A}_{\gam} } 
	+ \frac{1 }{2}  \epsilon_{\mu\nu\rho\sigma} 
	\{
	m \widehat{ \Sigma_{T}^{\nu\rho} \cA^{\sigma} } 
	- \hbar (\partial^\nu F^{ \rho\beta}) \partial_{q\beta} \mathcal{A}^{\sigma} 
	+  \hbar\widehat{ \Sigma_{A}^{\nu} ( \nabla^{\rho}\mathcal{V}^{\sigma}) }
	\}
	\\\nonumber
	&&\quad
	{-\frac{ \hbar  q^{\nu}}{2m} (  
\widehat {\Sigma_{T\gamma[\nu}\Delta^{\gamma}\mathcal{V}_{\mu]}}
-\widehat {\Sigma_{T\gamma[\nu}\Delta_{\mu]}\mathcal{V}^{\gamma}
} )}
	+{\frac{\hbar q^{\nu}}{2}
		[   \frac{1}{m}\widehat{ \Sigma_{T\nu\mu}(q\cdot\mathcal{V})}}
	{	-  \epsilon_{\nu\mu\rho\sigma} \widehat{ \Sigma_{V}^{\rho} \cA^\sigma}
		+   \epsilon_{\nu\mu\rho\sigma} \widehat{ \Sigma_{A}^{\rho}   \cV^\sigma} 
		-  \frac{1}{m}\widehat{ \Sigma_{P} (q_{[\nu}\mathcal{A}_{\mu]})}  
		]_\text{P.B.}}
	\\
	&&\quad+{\frac{\hbar m}{2}[   \widehat{ \Sigma_P  \cA_\mu} -\frac{\epsilon_{\nu\mu\rho\sigma}}{m} \widehat{ \Sigma_{V}^{\nu}(q^{\rho}\mathcal{A}^{\sigma})} +  \widehat{ \Sigma_{T\mu\beta}  \cV^\beta}  ]_\text{P.B.}},
	\\
	&& %\label{M10_N}
	F_{\mu\nu} \cA^\nu - q\cdot \cD \cA_\mu
	+ \frac{ \hbar }{4}  \epsilon_{\mu\nu\rho\sigma}  [ \cD^{\nu} ,   \cD^{\rho} ] \mathcal{V}^{\sigma}
	-{\frac{\hbar}{2m}q_{\mu}[ \widehat{\Sigma_P (q\cdot\mathcal{V})} 
		+  \frac{1}{2} \widehat{ \Sigma_{T}^{\rho\sigma} (q_{[\rho}\mathcal{A}_{\sigma]})} ]_\text{P.B.}} 
	\\\nonumber
	&&
	= m [ \widehat{ \Sigma_{S} \cA_{\mu} } 
	- \frac{1}{2}  \epsilon_{{\mu}\nu\rho\sigma}   \widehat{ \Sigma_{T}^{\nu\rho} \cV^{\sigma}} ]
	+  q_\alpha\widehat{ \Sigma_{A{\mu}} \mathcal{V}^\alpha } - q_\mu\widehat {  \Sigma_{A\alpha} \cV^\alpha} 
	+ \frac{\hbar}{2}\epsilon_{\mu\nu\rho\sigma}  \Delta^{\nu}
	[ 
	\widehat { \Sigma_{A}^{\rho}   \cA^{\sigma} } 
	-  {\frac{1}{m}}q^{\alpha}  \epsilon^{\ \ \ \ \rho}_{\alpha\beta\gamma} \widehat{ \Sigma_{T}^{{\sigma\beta}}  \mathcal{A}^{\gamma} }  
	]
	\\\nonumber
	&&\quad{-}{\frac{\hbar m}{2}[  \widehat{ \Sigma_P \cV_{\mu}} 
		{+}\frac{1}{m}\epsilon_{\mu{\nu}\rho\sigma}  \widehat{\Sigma_{A}^{\nu}  (q^{{\rho}}\mathcal{A}^{{\sigma}})} 
		+  \widehat{\Sigma_{T{\mu\nu}}  \cA^{\nu}}  ]_\text{P.B.}},
	\\
	&&%\label{M3_N}
	q^{[\mu}\mathcal{V}^{\nu]} 
	+ \frac{\hbar}{2}\epsilon^{\mu\nu\alpha\beta} 
	( \cD_{\alpha}\mathcal{A}_{\beta}    -   \widehat{ \Sigma_{A\alpha}  \cV_\beta} )
	\approx - \frac{\hbar}{2m} [  \epsilon^{\mu\nu\alpha\beta} q_{\alpha} \widehat{ \Sigma_{S} \mathcal{A}_{\beta} } 
	- q_\alpha  \widehat{ \Sigma_{T}^{\mu\nu}    \mathcal{V}^\alpha } 
	+ q^{[\mu}  \widehat{ \Sigma_{P}  \mathcal{A}^{\nu]}  
		% - \frac{ \hbar }{4}  \widehat{ \Sigma_{T}^{\alpha\beta} \nabla \cdot  \mathcal{A} }
	}
	], 
	\end{eqnarray}
\end{subequations}
where  we defined $\cD_\mu \M = \nabla_\mu \M + \widehat {\Sigma_{V\mu} \M}$, and 
we used the following relation 
\begin{eqnarray}
2\epsilon_{\nu\mu\rho\sigma}(\Pi^{\nu}\Pi^{\rho})\mathcal{A}^{\sigma}
= \frac{\hbar^2}{6}\epsilon_{\nu\mu\rho\sigma}\big(\partial^{\rho}F^{\beta\nu}+\partial^{\beta}F^{\rho\nu}\big)\partial_{q\beta}\mathcal{A}^{\sigma}
=\frac{\hbar^2}{2}\epsilon_{\nu\mu\rho\sigma}(\partial^{\rho}F^{\beta\nu})\partial_{q\beta}\mathcal{A}^{\sigma}.
\end{eqnarray}
The master equations (\ref{eq:Eq-sum}) and (\ref{eq:Eq-diff}) are in general 
the full master equations with collisional effects up to $\mathcal{O}(\hbar^1)$. 
In our power counting $\mathcal{V}^{\mu}\sim\mathcal{O}(\hbar^0)$ and $\mathcal{A}^{\mu}\sim\mathcal{O}(\hbar^1)$, 
these equations reduce to Eqs.~(\ref{MA_1})-(\ref{MA_red}).

\section{Angular-momentum decomposition for fermions and spin polarization}\label{app_AM_decomp}
We briefly review the derivation of the gauge-invariant angular-momentum decomposition for fermions proposed in Ref.~\cite{Wakamatsu:2010cb} as a covariant version of the Ji's decomposition \cite{Ji:1996ek}. Such a derivation is more explicitly shown in the review paper \cite{Leader:2013jra} (See also Ref.~\cite{Fukushima:2020qta}). Here we just summarize the derivation therein. The starting point is the Belinfante angular momentum for ``on-shell'' fermions,
\begin{eqnarray}
M^{\mu\nu\rho}_B
=\frac{x^{\nu}}{4}\bar{\psi}\big(\gamma^{\mu}i\overleftrightarrow{D}^{\rho}+\gamma^{\rho}i\overleftrightarrow{D}^{\mu}\big)\psi-(\nu\leftrightarrow\rho),
\end{eqnarray} 
where $\bar{\psi}\overleftrightarrow{D}_{\mu}\psi=\bar{\psi}\big(\overrightarrow{D}_{\mu}-\overleftarrow{D}^{\dagger}_{\mu}\big)\psi$ and $D_{\mu}=\partial_{\mu}+ieA_{\mu}$ denotes the covariant derivative. One may rewrite $M^{\mu\nu\rho}_B$ as
\begin{eqnarray}
M^{\mu\nu\rho}_B=\frac{x^{\nu}}{2}\bar{\psi}\gamma^{\mu}i\overleftrightarrow{D}^{\rho}\psi-
\frac{x^{\nu}}{4}\bar{\psi}\big(\gamma^{\mu}i\overleftrightarrow{D}^{\rho}-\gamma^{\rho}i\overleftrightarrow{D}^{\mu}\big)\psi-(\nu\leftrightarrow\rho).
\end{eqnarray} 
Next, one has to employ the relations,
\begin{eqnarray}\nonumber
\sigma^{\mu\nu}i\overrightarrow{\slashed{D}}&=&\gamma^{\nu}\overrightarrow{D}^{\mu}-\gamma^{\mu}\overrightarrow{D}^{\nu}{-}i\epsilon^{\mu\nu\alpha\beta}\gamma_{\beta}\gamma_5\overrightarrow{D}_{\alpha},
\\
i\overleftarrow{\slashed{D}}\sigma^{\mu\nu}&=&\overleftarrow{D}^{\mu}\gamma^{\nu}-\overleftarrow{D}^{\nu}\gamma^{\mu}{-}i\epsilon^{\mu\nu\alpha\beta}\gamma_{\beta}\gamma_5\overleftarrow{D}_{\alpha},
\end{eqnarray}
%{YH: I put the negative sign in front of $\epsilon$. For example, $\sigma^{01} \gamma^{2} = i\gamma^{0}\gamma^{1}\gamma^{2}=i\gamma^{0}\gamma^{1}\gamma^{2}\gamma^{3}\gamma_{3}=\gamma_{5}\gamma_{3}=-\gamma_{3}\gamma_{5}=-\epsilon^{0123}\gamma_{3}\gamma_{5}.$}
with $\sigma^{\mu\nu}=i[\gamma^{\mu},\gamma^{\nu}]/{2}$ and the equations of motion $i\overrightarrow{\slashed{D}}\psi=m\psi$ and $i\bar{\psi}\overleftarrow{\slashed{D}}=-m\bar{\psi}$, to obtain a useful identity,
\begin{eqnarray}
\bar{\psi}\big(\gamma^{\mu}i\overleftrightarrow{D}^{\rho}-\gamma^{\rho}i\overleftrightarrow{D}^{\mu}\big)\psi={-}\epsilon^{\rho\mu\alpha\beta}\partial_{\alpha}\big(\bar{\psi}\gamma_{\beta}\gamma_5\psi\big).
\end{eqnarray} 
By using the identity, it is found
\begin{eqnarray}
M^{\mu\nu\rho}_B=\frac{i}{2}\bar{\psi}\gamma^{\mu}\big(x^{\nu}\overleftrightarrow{D}^{\rho}-x^{\rho}\overleftrightarrow{D}^{\nu}\big)\psi
{-}\frac{1}{2}\epsilon^{\mu\nu\rho\beta}\bar{\psi}\gamma_{\beta}\gamma_5\psi
{-}\frac{1}{4}\partial_{\alpha}\Big(\big(x^{\nu}\epsilon^{\mu\rho\alpha\beta}-x^{\rho}\epsilon^{\mu\nu\alpha\beta}\big)\bar{\psi}\gamma_{\beta}\gamma_5\psi\Big),
\end{eqnarray}
and thus yields the gauge-invariant angular-momentum decomposition by dropping the surface term above,
\begin{eqnarray}\label{can_AM_tensor}
M^{\mu\nu\rho}_C=\frac{i}{2}\bar{\psi}\gamma^{\mu}\big(x^{\nu}\overleftrightarrow{D}^{\rho}-x^{\rho}\overleftrightarrow{D}^{\nu}\big)\psi
-\frac{1}{2}\epsilon^{\mu\nu\rho\beta}\bar{\psi}\gamma_{\beta}\gamma_5\psi,
\end{eqnarray}
where the first term above is regarded as the orbital angular momentum and the second term proportional to an axial-charge current is responsible for spin. 
This gauge-invariant decomposition introduced in Ref.~\cite{Wakamatsu:2010cb} is defined as the ``canonical angular momentum'' in Ref.~\cite{Yang:2018lew} though it actually agrees with the usual canonical angular momentum obtained from the Noether's theorem only in the absence of gauge fields.

\section{Decomposition of the collision term in AKE}\label{app_decomp_AKE}
The expression of Eq.~(\ref{C_nQ_2}) can be equivalently written as
\begin{eqnarray}\nonumber
\hat{\mathcal{C}}^{(n)\mu}_{2}&=&\frac{\delta(q^2-m^2)}{2}\Bigg\{\epsilon^{\mu\nu\rho\sigma}
q_{\nu}\big((\Delta_{\rho}\bar{\Sigma}_{V\sigma})f_V-(\Delta_{\rho}\Sigma_{V\sigma})\bar{f}_V\big)
+2S^{\mu\nu}_{m(n)}\Big(m\big(\Sigma_S\Delta_{\nu}\bar{f}_V-\bar{\Sigma}_S\Delta_{\nu}f_V\big)
\\\nonumber
&&-F^{\rho}_{\,\,\,\nu}\big(\Sigma_{V\rho}\bar{f}_V-\bar{\Sigma}_{V\rho}f_V\big)
-q^{\rho}\big(\bar{f}_V\Delta_{\nu}\Sigma_{V\rho}-f_V\Delta_{\nu}\bar{\Sigma}_{V\rho}\big)
+\bar{f}_V(q\cdot\Delta\Sigma_{V\nu})-f_V(q\cdot\Delta\bar{\Sigma}_{V\nu})\Big)
\\\nonumber
&&+2(\Sigma_{V\nu}\bar{f}_V-\bar{\Sigma}_{V\nu}f_V)
\Big(q\cdot\Delta S^{\mu\nu}_{m(n)}-F^{\mu}_{\,\,\,\lambda}S^{\lambda\nu}_{m(n)}\Big)\Bigg\}
-m\delta(q^2-m^2)S^{\mu\nu}_{m(n)}\Delta_{\nu}C_S[f_V]
\\\nonumber
&&-2S^{\mu\nu}_{m(n)}q^{\lambda}F_{\lambda\nu}\delta'(q^2-m^2)\big(C_{V}[f_V]+mC_S[f_V]\big)
-\delta(q^2-m^2)\big(\Delta_{\nu}S^{\mu\nu}_{m(n)}\big)
\\\nonumber
&&\times\big(C_{V}[f_V]+mC_S[f_V]\big)
+\tilde{F}^{\mu\nu}q_{\nu}\delta'(q^2-m^2)\big(C_{V}[f_V]+mC_S[f_V]\big)
\\
&&+\frac{\delta(q^2-m^2)}{2m}\Big[q^{\mu}\big(f_Vq\cdot\Delta\bar{\Sigma}_{P}-\bar{f}_Vq\cdot\Delta\Sigma_P\big)
-m^2\bigl( (\partial^{\mu}\bar{\Sigma}_P)f_{V}-(\partial^{\mu}{\Sigma}_P)\bar{f}_{V}\bigr)\Big]
.
\end{eqnarray}
By analogy with the collisionless case, we would like to decompose $\hat{\mathcal{C}}^{(n)\mu}_2$ above into the piece proportional to $q^{\mu}$, which survives in the massless limit and reproduces the collision term in CKT, and another piece proportional to $m$, which stems from the purely finite-mass correction.

The Schouton identity gives rise to
\begin{eqnarray}
S^{\mu\nu}_{m(n)}q^{\alpha}=q^{\mu}S^{\alpha\nu}_{m(n)}+q^{\nu}S^{\mu\alpha}_{m(n)}+\frac{\epsilon^{\mu\nu\rho\alpha}q_{\rho}}{2}-\frac{\epsilon^{\mu\nu\rho\alpha}}{2(q\cdot n+m)}\big(mq_{\rho}+q^2n_{\rho}\big).
\end{eqnarray}
We thus find
\begin{eqnarray}\nonumber
&&2S^{\mu\nu}_{m(n)}\Big(\bar{f}_V(q\cdot\Delta\Sigma_{V\nu})-f_V(q\cdot\Delta\bar{\Sigma}_{V\nu})\Big)
\\\nonumber
&&=2\Big[\big(q^{\mu}S^{\rho\nu}_{m(n)}+q^{\nu}S^{\mu\rho}_{m(n)}\big)\big(\bar{f}_V\Delta_{\rho}\Sigma_{V\nu}-f_V\Delta_{\rho}\bar{\Sigma}_{V\nu}\big)\Big]
-\epsilon^{\mu\nu\rho\sigma}
q_{\nu}\big((\Delta_{\rho}\bar{\Sigma}_{V\sigma})f_V-(\Delta_{\rho}\Sigma_{V\sigma})\bar{f}_V\big)
\\
&&\quad -\frac{\epsilon^{\mu\nu\rho\sigma}(mq_{\rho}+q^2n_{\rho})}{(q\cdot n+m)}\big(\bar{f}_V\Delta_{\sigma}\Sigma_{V\nu}-f_V\Delta_{\sigma}\bar{\Sigma}_{V\nu}\big)
\end{eqnarray}
and $\hat{\mathcal{C}}^{(n)\mu}_{2}$ becomes
\begin{eqnarray}\nonumber\label{C_nQ_3}
\hat{\mathcal{C}}^{(n)\mu}_{2}&=&\frac{\delta(q^2-m^2)}{2}\Bigg\{2q^{\mu}S^{\rho\nu}_{m(n)}\big(\bar{f}_V\Delta_{\rho}\Sigma_{V\nu}-f_V\Delta_{\rho}\bar{\Sigma}_{V\nu}\big)
+2S^{\mu\nu}_{m(n)}\Big(m\big(\Sigma_S\Delta_{\nu}\bar{f}_V-\bar{\Sigma}_S\Delta_{\nu}f_V\big)
\\\nonumber
&&-F^{\rho}_{\,\,\,\nu}\big(\Sigma_{V\rho}\bar{f}_V-\bar{\Sigma}_{V\rho}f_V\big)\Big)
-\frac{\epsilon^{\mu\nu\rho\sigma}m(q_{\rho}+mn_{\rho})}{(q\cdot n+m)}\big(\bar{f}_V\Delta_{\sigma}\Sigma_{V\nu}-f_V\Delta_{\sigma}\bar{\Sigma}_{V\nu}\big)
+2(\Sigma_{V\nu}\bar{f}_V-\bar{\Sigma}_{V\nu}f_V)
\\\nonumber
&&\times \Big(q\cdot\Delta S^{\mu\nu}_{m(n)}-F^{\mu}_{\,\,\,\lambda}S^{\lambda\nu}_{m(n)}\Big)\Bigg\}
-2S^{\mu\nu}_{m(n)}q^{\lambda}F_{\lambda\nu}\delta'(q^2-m^2)\big(C_{V}[f_V]+mC_S[f_V]\big)
-\delta(q^2-m^2)
\\\nonumber
&&\times\Big[\big(\Delta_{\nu}S^{\mu\nu}_{m(n)}\big)\big(C_{V}[f_V]+mC_S[f_V]\big)
+mS^{\mu\nu}_{m(n)}\Delta_{\nu}C_S[f_V]\Big]
\\\nonumber
&&+\tilde{F}^{\mu\nu}q_{\nu}\delta'(q^2-m^2)\big(C_{V}[f_V]+mC_S[f_V]\big)
+\frac{\delta(q^2-m^2)}{2m}\Big[q^{\mu}\big(f_Vq\cdot\Delta\bar{\Sigma}_{P}-\bar{f}_Vq\cdot\Delta\Sigma_P\big)
\\
&&-m^2\bigl( (\partial^{\mu}\bar{\Sigma}_P)f_{V}-(\partial^{\mu}{\Sigma}_P)\bar{f}_{V}\bigr)\Big]
.
\end{eqnarray}

Next, applying the Schouton identity again, it is found
\begin{eqnarray}\nonumber
&&\delta(q^2-m^2)q\cdot\Delta S^{\mu\nu}_{m(n)}
\\\nonumber
&&=\delta(q^2-m^2)\Bigg[q^{\mu}\Delta_{\alpha}S^{\alpha\nu}_{m(n)}+q^{\nu}\Delta_{\alpha}S^{\mu\alpha}_{m(n)}
+q^{\rho}\epsilon^{\mu\nu\alpha\sigma}\Delta_{\alpha}\frac{q_{\rho}n_{\sigma}}{2(q\cdot n+m)}
+q^{\sigma}\epsilon^{\mu\nu\rho\alpha}\Delta_{\alpha}\frac{q_{\rho}n_{\sigma}}{2(q\cdot n+m)}\Bigg]
\\\nonumber
&&=\delta(q^2-m^2)\Bigg[q^{\mu}\Delta_{\alpha}S^{\alpha\nu}_{m(n)}+q^{\nu}\Delta_{\alpha}S^{\mu\alpha}_{m(n)}
+m^2\Delta_{\alpha}\frac{\epsilon^{\mu\nu\alpha\sigma}n_{\sigma}}{2(q\cdot n+m)}
+\frac{q^{\rho}F_{\rho\alpha}\epsilon^{\mu\nu\alpha\sigma}n_{\sigma}}{2(q\cdot n+m)}
+\tilde{F}^{\mu\nu}
\\
&&\quad-m\Delta_{\alpha}\frac{\epsilon^{\mu\nu\rho\alpha}q_{\rho}}{2(q\cdot n+m)}
-\frac{\epsilon^{\mu\nu\rho\alpha}F_{\sigma\alpha}n^{\sigma}q_{\rho}}{2(q\cdot n+m)}
\Bigg]
\end{eqnarray}
and
\begin{eqnarray}
-S^{\lambda\nu}_{m(n)}F^{\mu}_{\,\,\,\lambda}=-\frac{q_{\rho}n_{\sigma}}{2(q\cdot n+m)}\big(
\epsilon^{\lambda\mu\rho\sigma}F^{\nu}_{\,\,\,\lambda}+\epsilon^{\lambda\nu\mu\sigma}F^{\rho}_{\,\,\,\lambda}+\epsilon^{\lambda\nu\rho\mu}F^{\sigma}_{\,\,\,\lambda}
\big),
\end{eqnarray}
which lead to
\begin{eqnarray}\nonumber
&&\delta(q^2-m^2)\Big(q\cdot\Delta S^{\mu\nu}_{m(n)}-S^{\lambda\nu}_{m(n)}F^{\mu}_{\,\,\,\lambda}\Big)
\\\nonumber
&&=\delta(q^2-m^2)\Bigg[q^{\mu}\Delta_{\alpha}S^{\alpha\nu}_{m(n)}+\tilde{F}^{\mu\nu}
+q^{\nu}\Delta_{\alpha}S^{\mu\alpha}_{m(n)}
-m\Delta_{\alpha}\frac{\epsilon^{\mu\nu\rho\alpha}(q_{\rho}+mn_{\rho})}{2(q\cdot n+m)}
+S^{\mu\rho}_{m(n)}F^{\nu}_{\,\,\,\rho}
\\
&&\quad+\frac{q^{\rho}F_{\rho\alpha}\epsilon^{\mu\nu\alpha\sigma}n_{\sigma}}{(q\cdot n+m)}
\Bigg].
\end{eqnarray}
We hence obtain
\begin{eqnarray}\nonumber\label{C_nQ_4}
\hat{\mathcal{C}}^{(n)\mu}_{2}&=&\frac{\delta(q^2-m^2)}{2}\Bigg\{2q^{\mu}S^{\rho\nu}_{m(n)}\big(\bar{f}_V\Delta_{\rho}\Sigma_{V\nu}-f_V\Delta_{\rho}\bar{\Sigma}_{V\nu}\big)
+2S^{\mu\nu}_{m(n)}m\big(\Sigma_S\Delta_{\nu}\bar{f}_V-\bar{\Sigma}_S\Delta_{\nu}f_V\big)
\\\nonumber
&&-\frac{\epsilon^{\mu\nu\rho\sigma}m(q_{\rho}+mn_{\rho})}{(q\cdot n+m)}\big(\bar{f}_V\Delta_{\sigma}\Sigma_{V\nu}-f_V\Delta_{\sigma}\bar{\Sigma}_{V\nu}\big)
+2\big(\bar{f}_V\Sigma_{V\nu}-f_V\bar{\Sigma}_{V\nu}\big)\Bigg[q^{\mu}\Delta_{\alpha}S^{\alpha\nu}_{m(n)}+\tilde{F}^{\mu\nu}
\\\nonumber
&&+\frac{q^{\rho}F_{\rho\alpha}\epsilon^{\mu\nu\alpha\sigma}n_{\sigma}}{(q\cdot n+m)}
-m\Delta_{\alpha}\frac{\epsilon^{\mu\nu\rho\alpha}(q_{\rho}+mn_{\rho})}{2(q\cdot n+m)}
\Bigg]\Bigg\}
-2S^{\mu\nu}_{m(n)}q^{\lambda}F_{\lambda\nu}\delta'(q^2-m^2)C_{V}[f_V]
+\delta'(q^2-m^2)
\\\nonumber
&&
\times\tilde{F}^{\mu\nu}q_{\nu}\big(C_{V}[f_V]+mC_S[f_V]\big)\\\notag
&&-m\Big(2S^{\mu\nu}_{m(n)}q^{\lambda}F_{\lambda\nu}\delta'(q^2-m^2)
+\delta(q^2-m^2)\big( (\Delta_{\nu}S^{\mu\nu}_{m(n)})
+S^{\mu\nu}_{m(n)}\Delta_{\nu}
\big)
\Big)C_{S}[f_{V}]
\\
&&
+\frac{\delta(q^2-m^2)}{2m}\Big[q^{\mu}\big(f_Vq\cdot\Delta\bar{\Sigma}_{P}-\bar{f}_Vq\cdot\Delta\Sigma_P\big)
-m^2\bigl( (\partial^{\mu}\bar{\Sigma}_P)f_{V}-(\partial^{\mu}{\Sigma}_P)\bar{f}_{V}\bigr)\Big]
.
\end{eqnarray}
On the other hand, one finds
\begin{eqnarray}\nonumber
-2S^{\mu\nu}_{m(n)}\delta'(q^2-m^2)q^{\lambda}F_{\lambda\nu}
&=&-2\delta'(q^2-m^2)\Big(q^{\mu}S^{\rho\nu}_{m(n)}+q^{\nu}S^{\mu\rho}_{m(n)}\Big)F_{\rho\nu}
-2\tilde{F}^{\mu\rho}q_{\rho}\delta'(q^2-m^2)
\\
&&+\frac{2\delta'(q^2-m^2)\tilde{F}^{\mu\rho}}{(q\cdot n+m)}m\big(q_{\rho}+mn_{\rho}\big)-\frac{2\delta(q^2-m^2)}{q\cdot n+m}\tilde{F}^{\mu\rho}n_{\rho},
\end{eqnarray}
which yields
\begin{eqnarray}\nonumber
-2S^{\mu\nu}_{m(n)}\delta'(q^2-m^2)q^{\lambda}F_{\lambda\nu}
&=&-\delta'(q^2-m^2)\Big(q^{\mu}S^{\rho\nu}_{m(n)}F_{\rho\nu}
+\tilde{F}^{\mu\rho}q_{\rho}-\frac{m(q_{\rho}+mn_{\rho})\tilde{F}^{\mu\rho}}{(q\cdot n+m)}\Big)
\\
&&-\frac{\delta(q^2-m^2)}{q\cdot n+m}\tilde{F}^{\mu\rho}n_{\rho}.
\end{eqnarray}
In addition, one can show
\begin{eqnarray}
\frac{q^{\rho}F_{\rho\alpha}\epsilon^{\mu\nu\alpha\sigma}n_{\sigma}}{(q\cdot n+m)}
=\frac{1}{q\cdot n+m}\big({-}q^{\mu}\tilde{F}^{\nu\sigma}n_{\sigma}+\tilde{F}^{\mu\sigma}n_{\sigma}q^{\nu}+m\tilde{F}^{\mu\nu}\big)-\tilde{F}^{\mu\nu}.
\end{eqnarray}
Accordingly, $\hat{\mathcal{C}}^{(n)}_{2}$ takes the form
\begin{eqnarray}\nonumber\label{C_nQ_5}
\hat{\mathcal{C}}^{(n)\mu}_{2}&=&\frac{\delta(q^2-m^2)}{2}\Bigg\{2q^{\mu}S^{\rho\nu}_{m(n)}\big(\bar{f}_V\Delta_{\rho}\Sigma_{V\nu}-f_V\Delta_{\rho}\bar{\Sigma}_{V\nu}\big)
+2S^{\mu\nu}_{m(n)}m\big(\Sigma_S\Delta_{\nu}\bar{f}_V-\bar{\Sigma}_S\Delta_{\nu}f_V\big)
\\\nonumber
&&-\frac{\epsilon^{\mu\nu\rho\sigma}m(q_{\rho}+mn_{\rho})}{(q\cdot n+m)}\big(\bar{f}_V\Delta_{\sigma}\Sigma_{V\nu}-f_V\Delta_{\sigma}\bar{\Sigma}_{V\nu}\big)
+2\big(\bar{f}_V\Sigma_{V\nu}-f_V\bar{\Sigma}_{V\nu}\big)\Bigg[q^{\mu}\Delta_{\alpha}S^{\alpha\nu}_{m(n)}
\\\nonumber
&&{-}\frac{q^{\mu}\tilde{F}^{\nu\sigma}n_{\sigma}}{(q\cdot n+m)}
+\frac{m\tilde{F}^{\mu\nu}}{(q\cdot n+m)}
-m\Delta_{\alpha}\frac{\epsilon^{\mu\nu\rho\alpha}(q_{\rho}+mn_{\rho})}{2(q\cdot n+m)}
\Bigg]\Bigg\}
\\\nonumber
&&+\frac{\delta(q^2-m^2)}{2m}\Big[q^{\mu}\big(f_Vq\cdot\Delta\bar{\Sigma}_{P}-\bar{f}_Vq\cdot\Delta\Sigma_P\big)
-m^2\bigl( (\partial^{\mu}\bar{\Sigma}_P)f_{V}-(\partial^{\mu}{\Sigma}_P)\bar{f}_{V}\bigr)\Big]
\\\nonumber
&&-\delta'(q^2-m^2)\Big(q^{\mu}S^{\rho\nu}_{m(n)}F_{\rho\nu}
-\frac{m(q_{\rho}+mn_{\rho})\tilde{F}^{\mu\rho}}{(q\cdot n+m)}\Big)C_{V}[f_V]
-m\Big(\big(2S^{\mu\nu}_{m(n)}q^{\lambda}F_{\lambda\nu}-\tilde{F}^{\mu\nu}q_{\nu}\big)
\\
&&\times \delta'(q^2-m^2)
+\delta(q^2-m^2)\big(\Delta_{\nu}S^{\mu\nu}_{m(n)}\big)
+\delta(q^2-m^2)S^{\mu\nu}_{m(n)}\Delta_{\nu}\Big)C_S[f_V]
,
\end{eqnarray}
which can be further rearranged as
\begin{eqnarray}\nonumber\label{C_nQ_6}
\hat{\mathcal{C}}^{(n)\mu}_{2}&=&q^{\mu}\Bigg\{\delta(q^2-m^2)\Bigg[S^{\rho\nu}_{m(n)}\big(\bar{f}_V\Delta_{\rho}\Sigma_{V\nu}-f_V\Delta_{\rho}\bar{\Sigma}_{V\nu}\big)
+\big(\bar{f}_V\Sigma_{V\nu}-f_V\bar{\Sigma}_{V\nu}\big)
\Bigg(\partial_{\alpha}S^{\alpha\nu}_{m(n)}+\frac{S^{\rho\nu}_{m(n)}F_{\rho\sigma}n^{\sigma}}{q\cdot n+m}
\Bigg)
\\\nonumber
&&
+\frac{1}{2m}\big(f_Vq\cdot\Delta\bar{\Sigma}_{P}-\bar{f}_Vq\cdot\Delta\Sigma_P\big)
\Bigg]
-\delta'(q^2-m^2)S^{\rho\nu}_{m(n)}F_{\rho\nu}C_{V}[f_V]
\Bigg\}
\\\nonumber
&&+
m\Bigg\{\delta(q^2-m^2)\Bigg[
S^{\mu\nu}_{m(n)}\big(\Sigma_S\Delta_{\nu}\bar{f}_V-\bar{\Sigma}_S\Delta_{\nu}f_V\big)
-\Delta_{\nu}\big(S^{\mu\nu}_{m(n)}C_S[f_V]\big)
-\frac{\epsilon^{\mu\nu\rho\sigma}(q_{\rho}+mn_{\rho})}{{2}(q\cdot n+m)}
\\\nonumber
&&\times\big(\bar{f}_V\Delta_{\sigma}\Sigma_{V\nu}-f_V\Delta_{\sigma}\bar{\Sigma}_{V\nu}\big)
+\big(\bar{f}_V\Sigma_{V\nu}-f_V\bar{\Sigma}_{V\nu}\big)\Bigg(\frac{\tilde{F}^{\mu\nu}}{(q\cdot n+m)}
-\epsilon^{\mu\nu\rho\alpha}\Delta_{\alpha}\frac{(q_{\rho}+mn_{\rho})}{2(q\cdot n+m)}
\Bigg)\\\notag
&&
-\frac{1}{2}\bigl( (\partial^{\mu}\bar{\Sigma}_P)f_{V}-(\partial^{\mu}{\Sigma}_P)\bar{f}_{V}\bigr)
\Bigg]
\\
&&+\delta'(q^2-m^2)\Bigg[\frac{(q_{\rho}+mn_{\rho})\tilde{F}^{\mu\rho}}{(q\cdot n+m)}C_{V}[f_V]
-(2S^{\mu\nu}_{m(n)}q^{\lambda}F_{\lambda\nu}-\tilde{F}^{\mu\nu}q_{\nu})C_S[f_V]\Bigg]
\Bigg\},
\end{eqnarray}
where we utilize
\begin{eqnarray}
\Delta_{\alpha}S^{\alpha\nu}_{m(n)}{-}\frac{\tilde{F}^{\nu\sigma}n_{\sigma}}{(q\cdot n+m)}=\partial_{\alpha}S^{\alpha\nu}_{m(n)}+\frac{S^{\rho\nu}_{m(n)}F_{\rho\sigma}n^{\sigma}}{q\cdot n+m}.
\end{eqnarray}

We can now write $\hat{\mathcal{C}}^{(n)\mu}_2$ in a more succinct form as 
\begin{eqnarray}\label{C_2n_decomp}
\hat{\mathcal{C}}^{(n)\mu}_{2}=q^{\mu}\hat{\mathcal{C}}^{(n)}_{\mathfrak{q}2}+m\hat{\mathcal{C}}^{(n)\mu}_{\mathfrak{m}2},
\end{eqnarray}
where 
\begin{eqnarray}\nonumber
\hat{\mathcal{C}}^{(n)}_{\mathfrak{q}2}&=&
\delta(q^2-m^2)\Bigg[
- S^{\rho\nu}_{m(n)} \widehat{ (\Delta_{\rho} \Sigma_{V\nu})f_V}
- \widehat{ \Sigma_{V\nu} f_V}
\left(\partial_{\alpha}S^{\alpha\nu}_{m(n)}
+\frac{S^{\rho\nu}_{m(n)}F_{\rho\sigma}n^{\sigma}}{q\cdot n+m}
\right)
\\
&&
+ \frac{1}{2m} (q\cdot\Delta \widehat{ \Sigma_P) f_V} \Bigg]
-\delta'(q^2-m^2)S^{\rho\nu}_{m(n)}F_{\rho\nu}C_{V}[f_V],
\end{eqnarray}
and
\begin{eqnarray}\nonumber
\hat{\mathcal{C}}^{(n)\mu}_{\mathfrak{m}2}&=&
\delta(q^2-m^2)\Bigg[\big(\Delta_{\nu}S^{\nu\mu}_{m(n)}\big)C_S[f_V]
- S^{\nu\mu}_{m(n)}  (\Delta_{\nu} \widehat{ \Sigma_S) f_V} 
+ \frac{\epsilon^{\mu\nu\rho\sigma}(q_{\rho}+mn_{\rho})}{2(q\cdot n+m)}
( \Delta_{\sigma} \widehat{  \Sigma_{V\nu}) f_V }
\\\nonumber
&&
- \frac{1}{2}  (\partial^{\mu} \widehat{ \Sigma_P)f_{V} } 
- \frac{\epsilon^{\mu\nu\alpha\beta} \widehat{ \Sigma_{V\nu} f_V}  }{2(q\cdot n+m)}
\Bigg(m\partial_{\alpha}n_{\beta}
-(q_{\beta}+mn_{\beta})\frac{(\partial_{\alpha}q\cdot n+n^{\rho}F_{\rho\alpha})}{q\cdot n+m}
\Bigg)\Bigg]
\\
&&
+\delta'(q^2-m^2)\Bigg[\frac{(q_{\rho}+mn_{\rho})\tilde{F}^{\mu\rho}}{(q\cdot n+m)}C_{V}[f_V]
-2S^{\mu\nu}_{m(n)}q^{\lambda}F_{\lambda\nu}C_S[f_V]+\tilde{F}^{\mu\nu}q_{\nu}C_S[f_V]
\Bigg].
\end{eqnarray}
Here, $C_{V}[f_V]\equiv -q_\mu \widehat{ \Sigma_V^\mu  f_{V} } $ 
and $C_S[f_V]\equiv - \widehat{ \Sigma_S  f_{V} }  $. 
It is clear to see that $\hat{\mathcal{C}}^{(n)}_{\mathfrak{q}2}$ reduces to the $\mathcal{O}(\hbar)$ correction in the collision term of the CKT found in Refs.~\cite{Hidaka:2016yjf,Hidaka:2017auj} by taking $m=0$ {\footnote{In Refs.~\cite{Hidaka:2016yjf,Hidaka:2017auj}, $\Sigma_P$ and $\bar{\Sigma}_P$ are set to zero. As argued previously, such terms are actually expected to be at higher order and vanishing in the massless limit.}}.
Combining with $\hat{\mathcal{C}}^{(n)\mu}_1$ and collisionless part, one finds that the CKT with collisions is smoothly reproduced by AKE in the massless limit when $m=0$ and $a^{\mu}=q^{\mu}$.
\section{Derivation of the spin diffusion for SKE}\label{app_der_SKE}

From Eqs.~(\ref{HTLA_Pi}) and (\ref{Q1}), we find    
\begin{eqnarray}
Q_1(p)&=&2\pi\delta(q'^2-m^2)\Big(2q'^{\mu}G^<_{\mu\nu}q^{\nu}-p\cdot q'G^{<\mu}_{\mu}\Big)\notag\\
&=&2\pi\delta(q'^2-m^2)g_{0p}\frac{p_0}{T}\Big(2\hat{\rho}_T(|{\bf q}|^2(1-z^2)- p_{0}q_{0}+ |{\bf q}||{\bf p}|z)+\hat{\rho}_L(2q_0^2-p_0q_0-|{\bf q}||{\bf p}|z)\Big),\notag\\
\label{app_Q1}
\end{eqnarray}
where we defined $\hat{\rho}_{T/L}\equiv\rho_{T/L}T/p_0$ and $  z\equiv \hat \bq \cdot \hat \bp $. 
We also used the transversality $ p_\mu P_T^{\mu\nu} =0 $ 
and the on-shell conditions $ m^2 = q^{2} = p^2 + q^{\prime2} + 2 p\cdot q'$ 
and $ m^2 = q^{\prime2} =  p^2 + q^{2} - 2 p\cdot q$ 
that lead to relations $   2 p\cdot q = p^2 =- 2 p\cdot q'  $.

The on-shell condition can be also arranged as 
\begin{eqnarray}
q^{\prime2} -m^2= p^2 -  2q \cdot p 
= (p^0 -  q^0 )^2  + (2 |\bq ||\bp| z -\bp^2  -  q_0^2).
\end{eqnarray}
Therefore, the delta function is nonzero when the momenta satisfy the conditions 
\begin{eqnarray}
p^0 = q^0 \pm \sqrt{  \bp^2 - 2 |\bq ||\bp| z +  q_0^2 } 
\sim q^0 \pm \left( q^0 -\frac{ |\bq ||\bp| z}{ q_0} 
+ \frac{ \bp^2 (q_0^2 - \bq^2 z^2 ) }{ 2q_0^3}  
\right),
\end{eqnarray}
where the square root is expanded for a small momentum transfer $ |\bp| $. 
We take the positive-energy solution $ q^{\prime 0 } = q^0 - p^0 >0 $ for particles from the lower sign. 
Defining $ \bar p \equiv   |\bq ||\bp| z/ q_0 -  \bp^2 (q_0^2 - \bq^2 z^2 ) / (2q_0^3) $, 
we have\footnote{It turns out that the $\mathcal{O}(|{\bf p}|^2)$ term in the Delta function does not contribute to the leading logarithmic order.}
\begin{eqnarray}
\delta(q^{\prime2}-m^2) 
\sim 
\frac{1}{2 q^0} \left( 1 + \frac{ |\bq ||\bp| z}{ q_0^2} 
+ \order( |\bp|^2)  
\right) 
\, \delta (p^0 - \bar p)
.
\end{eqnarray}
Based on the dispersion relation above, we can approximate the spectral function as 
\begin{eqnarray}\label{rhoT_exp}
\hat{\rho}_T(p) 
\rightarrow
\frac{\pi Tm_D^2}{2|{\bf p}|^5}I_T(z)
\Bigg(1-\frac{|{\bf q}||{\bf p}|z}{q_0^2} 
 + \order( |\bp|^2 ) 
\Bigg),
\end{eqnarray}
where $I_T(z)\equiv [1- ( |{\bf q}|z/q_0)^2 ]^{-1}$. 
Here, the presence of the delta function $ \delta (p^0 - \bar p) $ is assumed on the right-hand side of arrow. 
On the other hand, we have $\hat{\rho}_L(p)=\pi Tm_D^2/|{\bf p}|^5$.

Note that the difference between the distribution functions provides positive powers 
in the small momentum transfer limit: 
\begin{eqnarray} \label{f-f}
f_{Vq}-f_{Vq-p}\approx p^{\beta}\partial_{q^{\beta}}f_{Vq}-\frac{p^{\alpha}p^{\beta}}{2}\partial_{q^{\alpha}}\partial_{q^{\beta}}f_{Vq}
,
\end{eqnarray}
where $ p^0 \sim \order(|\bp|^1) $. 
Therefore, we have found the momentum dependences 
\begin{eqnarray} \label{order-p}
\hat \rho_{L,T} \sim |\bp|^{-5}  ,
\quad f_{Vq} -f_{V q-p} \sim |\bp| ,
\quad d^3\bp \sim d|\bp| |\bp|^2 .
\end{eqnarray}
This order counting suggests that one should maintain the other factors up to $  \order (|\bp|) $ to get the leading-log result. 
Therefore, we expand the remaining factor as 
\begin{eqnarray}%\nonumber
\label{delta_exp}
&&\delta( q^{\prime2}-m^2)g_{0p}\frac{p_0}{T}
\approx 
\frac{1}{2|q_0|}\Bigg[ 1+\frac{|{\bf p}||{\bf q}|z}{q_0^2}\Big(1-\frac{q_0}{2T}\Big)
+ \order(|\bp|^2) 
\Bigg] 
\delta (p^0 - \bar p) 
,
\end{eqnarray}
where we used $g_{0p}\approx {T}/{p_0}-{1}/{2} $. 
Combining all pieces together, it is found
\begin{eqnarray}\nonumber\label{Q1_int}
\int_pQ_1(f_{Vq}-f_{Vq-p})
&\approx& \frac{\pi Tm_D^2}{2|q_0|}\int_p\frac{1}{|{\bf p}|^5}
(2\pi)\delta\left(p_0-\frac{\bf p\cdot q}{q_0} \right)
\\
&&\times 
\Bigg(1-\frac{|{\bf p}||{\bf q}|z}{2q_0T}\Bigg)
\Bigg[2q_0^2
+|{\bf q}|^2 (1-z^2) I_T(z)
\Bigg](f_{Vq}-f_{Vq-p}).
\end{eqnarray}

In Eq.~(\ref{f-f}), we apply the following decomposition of $p^{\beta}_{\perp}$ 
in terms of the transverse and longitudinal components with respect to $q_{\perp}^{\mu}$: 
\begin{eqnarray}\label{rel_1}
p^{\mu}_{\perp}=\hat{q}^{\mu}_{\perp}|{\bf p}|z+\hat{\Theta}^{\mu\nu}_qp_{\nu},\quad
\hat{\Theta}^{\alpha\beta}_q=\Theta^{\alpha\beta}+\hat{q}^{\alpha}_{\perp}\hat{q}^{\beta}_{\perp},
\end{eqnarray}
where $q_{\alpha}\hat{\Theta}_q^{\alpha\beta}=0$.
Moreover, The distribution function $f_{Vq}$ in general can be a function of $q_0$ before implementing the on-shell condition. It is in fact more convenient to keep the off-shell form when $F^{\mu\nu}\neq 0$ due to the presence of $\delta'(q^2-m^2)$ term in the AKE. Nonetheless, in the absence of background fields, it is more practical and convenient to write down the on-shell kinetic equations. We hence take $f_{Vq}=f_{Vq}({\bf q},X)$ as just a function of ${\bf q}$ and $X$ by using $q_0=E_q=\sqrt{|{\bf q}|^2+m^2}$ for fermions (here we neglect anti-fermions) in the Wigner functions. Accordingly, all the terms proportional to $\partial_{q_0}f_{Vq}$ can be dropped.

Applying the replacement 
\begin{eqnarray}\label{rel_2}
p^{\alpha}p^{\beta}\rightarrow p_0^2u^{\alpha}u^{\beta}
+p_0|{\bf p}|z\big(u^{\alpha}\hat{q}^{\beta}_{\perp}+u^{\beta}\hat{q}^{\alpha}_{\perp}\big)
+|{\bf p}|^2\Big(\hat{q}^{\alpha}_{\perp}\hat{q}^{\beta}_{\perp}z^2-\frac{\hat{\Theta}^{\alpha\beta}_q}{2}\big(1-z^2\big)\Big)
\end{eqnarray}
for the integrand and carrying out the integration, Eq.~(\ref{Q1_int}) results in 
\begin{eqnarray}\label{Q1_int_2}
&&\delta(q^2-m^2)\int_pQ_1(f_{Vq}-f_{Vq-p})
\\\nonumber
&&\approx -\frac{\pi m_D^2\delta(q^2-m^2)}{2q_0}\int^{T}_{m_D}\frac{d|{\bf p}|}{(2\pi)^2|{\bf p}|}\int^1_{-1} dz \Bigg(\frac{|{\bf q}|^2 (1-z^2) I_T(z) }{2}+q_0^2\Bigg)\Bigg[\frac{|{\bf q}|z^2}{q_0}\hat{q}^{\beta}_{\perp}
+\Big(\hat{q}^{\alpha}_{\perp}\hat{q}^{\beta}_{\perp}z^2
\\\nonumber
&&\quad-\frac{\hat{\Theta}^{\alpha\beta}_q}{2}\big(1-z^2\big)\Big)T\partial_{q^{\alpha}}
\Bigg]\partial_{q^{\beta}}f_{Vq}
\\\nonumber
&&=-\frac{m_D^2\delta(q^2-m^2)}{8\pi q_{0}}\ln (1/g_c)
\Bigg[
q_{0}\Bigg(j^T_1-\frac{q_{0}^2}{|{\bf q}|^2}j_2^T
+j_1^L\Bigg)q^{\beta}_{\perp}\partial_{q^{\beta}_{\perp}}
-T\Bigg(J^T_0-\frac{4q_{0}^2}{|{\bf q}|^2}j^T_1+\frac{3q_{0}^4}{|{\bf q}|^4}j^T_2
+\frac{q_{0}^2}{|{\bf q}|^2}
\\\nonumber
&&\quad
\times(j^L_0-3j^L_1)\Bigg)
\frac{q^{\beta}_{\perp}q^{\alpha}_{\perp}}{2}\partial_{q^{\alpha}_{\perp}}\partial_{q^{\beta}_{\perp}}
-T\Bigg(|{\bf q}|^2\left(J^T_0-\frac{2q_{0}^2}{|{\bf q}|^2}j^T_1+\frac{q_{0}^4}{|{\bf q}|^4}j^T_2\right)+q_{0}^2(j^L_0-j^L_1)\Bigg)\frac{\eta^{\alpha\beta}}{2}\partial_{q^{\alpha}_{\perp}}\partial_{q^{\beta}_{\perp}}
\Bigg]f_{Vq}
,
\end{eqnarray}
where we take $u^{\beta}\partial_{q^{\beta}}f_{Vq}=0$ and
\begin{eqnarray}\nonumber\label{jT}
&&j_0^T\equiv \int^1_{-1} dz\frac{1}{2\left(1-\frac{|{\bf q}|^2z^2}{q_0^2}\right)}=\frac{q_0}{|{\bf q}|}\eta_q,\quad 
j_1^T\equiv \int^1_{-1} dz\frac{\left(\frac{|{\bf q}|z}{q_0}\right)^2}{2\left(1-\frac{|{\bf q}|^2z^2}{q_0^2}\right)}=\frac{q_0}{|{\bf q}|}\eta_q-1,
\\
&&
j_2^T\equiv \int^1_{-1} dz\frac{\left(\frac{|{\bf q}|z}{q_0}\right)^4}{2\left(1-\frac{|{\bf q}|^2z^2}{q_0^2}\right)}=\frac{q_0}{|{\bf q}|}\eta_q-\frac{|{\bf q}|^2}{3q_0^2}-1,
\end{eqnarray}
where $  \eta_q=2^{-1}\ln\left[({q_0+|{\bf q}|)}/({q_0-|{\bf q}|})\right]$, 
and
\begin{eqnarray}\label{jL}
&&j_0^L\equiv \int^1_{-1} dz=2,\quad j_1^L\equiv \int^1_{-1} z^2dz=\frac{2}{3}.%\quad j_2^L\equiv \int^1_{-1} z^4dz=\frac{2}{5}.
\end{eqnarray}
By inputting explicit forms of $j^{L/T}_i$ into (\ref{Q1_int_2}), we eventually obtain
\begin{eqnarray}\nonumber\label{Q1_int_3}
&&\delta(q^2-m^2)\int_pQ_1(f_{Vq}-f_{Vq-p})
\\\nonumber
&&=-\frac{m_D^2\delta(q^2-m^2)}{8\pi}\ln (1/g_c)
\Bigg[\frac{E_q^2}{|{\bf q}|^2}\Bigg(1-\frac{m^2\eta_q}{q_0|{\bf q}|}\Bigg)q^{\beta}_{\perp}
+\frac{m^2T}{2|{\bf q}|^3}\Bigg(\left(1-\frac{3E^2_q}{|{\bf q}|^2}\right)\eta_q+\frac{3E_q}{|{\bf q}|}\Bigg)q_{\perp}^{\alpha}q^{\beta}_{\perp}\partial_{q^{\alpha}_{\perp}}
\\
&&\quad-\frac{E_qT}{2}\Bigg(\left(3-\frac{E_q^2}{|{\bf q}|^2}\right)+\frac{m^4}{E_q|{\bf q}|^3}\Bigg)\eta^{\alpha\beta}\partial_{q^{\alpha}_{\perp}}
\Bigg]\partial_{q^{\beta}_{\perp}}f_{Vq},
\end{eqnarray}
where we take $q_{0}=E_{q}$.

Subsequently, we should also calculate the $\tilde{Q}_1f_{Vq}$ term in Eq.~(\ref{SKE_HTL_3}). 
While the same counting (\ref{order-p}) is applied to this case, 
there is not the order-one suppression from the difference $ f_{Vq} - f_{V q-p} $. 
Nevertheless, a factor of $ g_{0p}^{-1} $ provides the same order of suppression instead. 
Therefore, maintaining the linear terms from the other factors as in the computation above, 
we find 
\begin{eqnarray}\nonumber
&&\delta(q^2-m^2)\int_p\tilde{Q}_1f_{Vq}
\\\nonumber
&&\approx -\frac{\pi m_D^2\delta(q^2-m^2)}{2q_0}\int^{T}_{m_D}\frac{d|{\bf p}|}{(2\pi)^2|{\bf p}|}\int^1_{-1} dz\Bigg(\frac{|{\bf q}|^2I_T(1-z^2)}{2}+q_0^2\Bigg)
\frac{1}{q_0}\left(1-\frac{|{\bf q}|^2z^2}{q_0^2}\right)f_{Vq}
\\
&&=-\frac{m_D^2\delta(q^2-m^2)}{8\pi q_0}\ln (1/g_c)\Bigg[q_0\big(j^L_0-j^T_1+j^T_2\big)-\frac{|{\bf q}|^2}{q_0}\big(j^L_1-j^T_0+j^T_1\big)\Bigg]f_{Vq},
\end{eqnarray}
which yields
\begin{eqnarray}\label{Q1t_int_non}\nonumber
&&\delta(q^2-m^2)\int_p\tilde{Q}_1f_{Vq}(1-f_{Vq-p})
\\
&&\approx-\frac{m_D^2\delta(q^2-m^2)}{4\pi}\ln (1/g_c)\Bigg(f_{Vq}(1-f_{Vq})-\frac{q_0^2}{|{\bf q}|^2}\left(1-\frac{\eta_qm^2}{q_0|{\bf q}|}\right)f_{Vq}q^{\beta}_{\perp}\partial_{q^{\beta}_{\perp}}f_{Vq}\Bigg).
\end{eqnarray}
Accordingly, the SKE results in the form given in Eq.~(\ref{SKE_HTL_4_non}).

\section{Derivation of the spin diffusion for AKE}\label{app_der_AKE}

To evaluate the collision term in AKE, we implement the same strategy as in the SKE. 

\subsection{Half of the terms: $ [Q_2^{\mu\nu}\tilde{a}_{q'\nu}-Q_1\tilde{a}^{\mu}_q ]$}

For convenience, we first apply the following decompositions to two of the four terms in Eq.~(\ref{AKE_HTL3_non}): 
\begin{eqnarray}
Q_2^{\mu\nu}\tilde{a}_{q'\nu}-Q_1\tilde{a}^{\mu}_q
=2\pi\delta(q'^2-m^2)\big(Q^{\mu}_{\text{cl}(a)}+Q^{\mu}_{\text{cl}(b)}+Q^{\mu}_{\text{cl}(c)}\big),
\end{eqnarray}
where 
\begin{subequations}
\begin{eqnarray}
&&
Q^{\mu}_{\text{cl}(a)}\equiv 2(q'^{\rho}G^<_{\rho\nu}q^{\nu})(\tilde{a}^{\mu}_{q'}-\tilde{a}^{\mu}_{q})+G^{<\rho}_{\rho}(p\cdot q'\tilde{a}^{\mu}_q+p\cdot q\tilde{a}^{\mu}_{q'})
=g_{0p}\frac{p_0}{T}\Big(\hat{Q}^{T\mu}_{\text{cl}(a)}+\hat{Q}^{L\mu}_{\text{cl}(a)}\Big),
\\
&&
Q^{\mu}_{\text{cl}(b)}\equiv-\big(G^{<\rho}_{\rho}p^{\mu}+2q'_{\rho}G^{<\rho\mu}\big)q\cdot\tilde{a}_{q'}
=g_{0p}\frac{p_0}{T}\Big(\hat{Q}^{T\mu}_{\text{cl}(b)}+\hat{Q}^{L\mu}_{\text{cl}(b)}\Big),
\\
&&
Q^{\mu}_{\text{cl}(c)} 
\equiv 
2\big(p^{\mu}G^{<\nu}_{\rho}q^{\rho}-q\cdot pG^{<\nu\mu}\big)\tilde{a}_{q'\nu}=g_{0p}\frac{p_0}{T}\big(\hat{Q}^{T\mu}_{\text{cl}(c)}+\hat{Q}^{L\mu}_{\text{cl}(c)}\big)
.
\end{eqnarray}
\end{subequations}
The remaining two terms will be addressed in Sec.~\ref{sec:d}. 
We utilized $q'\cdot a_{q'}= (q'^2-m^2) f_{Aq'} =0$ from the on-shell condition. 
On the rightmost side, the terms proportional to the transverse and longitudinal spectral functions 
are denoted as $\hat{Q}^{T\mu}_{\text{cl}}$ and $\hat{Q}^{L\mu}_{\text{cl}}$, respectively. 
Note that, similar to $ f_{V q-p} $ in the SKE, the expansion of $ \tilde a_{q-p}^\mu  $ 
give rise to the factors of $ p $ for the small momentum transfer limit. 
Recalling the order counting in Eq.~(\ref{order-p}),  
we thus maintain the terms in $ \hat{Q}^{T,L \mu}_{\text{cl}} $ up to $ \order(|\bp|^2) $ in the following. 
For $  Q^{\mu}_{\text{cl}(a)} $, we have  
\begin{subequations}
\begin{eqnarray}\nonumber
\hat{Q}^{T\mu}_{\text{cl}(a)}&=&2\hat{\rho}_T\Big[|{\bf q}|^2(1-z^2)(\tilde{a}^{\mu}_{q-p}-\tilde{a}^{\mu}_{q})
-\Big(\big(p_0q_0-|{\bf p}||{\bf q}|z\big)\big(\tilde{a}^{\mu}_{q-p}+\tilde{a}^{\mu}_{q}\big)
-(p_0^2-|{\bf p}|^2)\tilde{a}^{\mu}_{q-p}\Big)\Big]
\\\nonumber
&\approx&2\hat{\rho}_T\Big[
\Big(2(|{\bf p}||{\bf q}|z-p_0q_0\big)+p_0^2-|{\bf p}|^2\Big)\tilde{a}_{q}^{\mu}
-\Big(|{\bf q}|^2(1-z^2)-q_0p_0+|{\bf p}||{\bf q}|z\Big)p^{\beta}\partial_{q^{\beta}}\tilde{a}^{\mu}_q
\\
&&+|{\bf q}|^2
(1-z^2)\frac{p^{\alpha}p^{\beta}}{2}\partial_{q^{\beta}}\partial_{q^{\alpha}}\tilde{a}^{\mu}_q
+\mathcal{O}(|{\bf p}|^3)\Big],
\\
\nonumber
\hat{Q}^{L\mu}_{\text{cl}(a)}&=&\hat{\rho}_L\Big[2(q_0^2-q_0p_0)(\tilde{a}^{\mu}_{q-p}-\tilde{a}^{\mu}_{q})+\Big(\big(p_0q_0-|{\bf p}||{\bf q}|z\big)\big(\tilde{a}^{\mu}_{q-p}+\tilde{a}^{\mu}_{q}\big)
-(p_0^2-|{\bf p}|^2)\tilde{a}^{\mu}_{q}\Big)\Big]
\\\nonumber
&\approx&\hat{\rho}_L\Big[\Big(|{\bf p}|^2-p_0^2-2(|{\bf p}||{\bf q}|z-p_0q_0\big)\Big)\tilde{a}_{q}^{\mu}
-\Big(2q_0^2-q_0p_0-|{\bf p}||{\bf q}|z\Big)p^{\beta}\partial_{q^{\beta}}\tilde{a}^{\mu}_q
\\
&&
+q_0^2p^{\alpha}p^{\beta}\partial_{q^{\alpha}}\partial_{q^{\beta}}\tilde{a}^{\mu}_q
+\mathcal{O}(|{\bf p}|^3)\Big].
\end{eqnarray}
\end{subequations}
As for $ Q^{\mu}_{\text{cl}(b)}  $, we have 
\begin{subequations}
\begin{eqnarray}\nonumber
\hat{Q}^{T\mu}_{\text{cl}(b)}&=&2\hat{\rho}_T(p^{\mu}+q_{\perp}^{\mu}-|{\bf q}|\hat{p}^{\mu}_{\perp}z)p\cdot\tilde{a}_{q-p}
\\\nonumber
&\approx&2\hat{\rho}_T\Big[(p^{\mu}+q_{\perp}^{\mu}-|{\bf q}|\hat{p}^{\mu}_{\perp}z)p^{\nu}\tilde{a}_{\nu q}
+\big(|{\bf q}|\hat{p}^{\mu}_{\perp}z-q^{\mu}_{\perp}\big)p^{\nu}p^{\rho}\partial_{q^{\rho}}\tilde{a}_{\nu q}
+\mathcal{O}(|{\bf p}|^3)\Big],
\\
\nonumber
\hat{Q}^{L\mu}_{\text{cl}(b)}&=&\hat{\rho}_L\Big(2(p_0-q_0)u^{\mu}-p^{\mu}\Big)p\cdot\tilde{a}_{q-p}
\\
&\approx& \hat{\rho}_L\Big[(p_0u^{\mu}-p_{\perp}^{\mu}-2q_0u^{\mu})p^{\nu}\tilde{a}_{q\nu}
+2q_0u^{\mu}p^{\nu}p^{\rho}\partial_{q^{\rho}}\tilde{a}_{q\nu}
+\mathcal{O}(|{\bf p}|^3)\Big].
\end{eqnarray}
\end{subequations}
Here, we drop the terms proportional to $q\cdot a_{q}=(q^2-m^2)f_{Aq}$ based on the on-shell condition. 
Finally, for $Q^{\mu}_{\text{cl}(c)}   $, we have 
\begin{subequations}
\begin{eqnarray}\nonumber
\hat{Q}^{T\mu}_{\text{cl}(c)}&=&2\hat{\rho}_T\Big(p^{\mu}\big(|{\bf q}|z\hat{p}^{\nu}_{\perp}-q^{\nu}_{\perp}\big)+(q_0p_0-|{\bf q}||{\bf p}|z)(\Theta^{\mu\nu}+\hat{p}^{\mu}_{\perp}\hat{p}^{\nu}_{\perp})\Big)\tilde{a}_{q-p\nu}
\\\nonumber
&\approx&2\hat{\rho}_T\Big((p_0u^{\mu}+p_{\perp}^{\mu})\big(|{\bf q}|z\hat{p}^{\nu}_{\perp}-q^{\nu}_{\perp}\big)+(q_0p_0-|{\bf q}||{\bf p}|z)(\Theta^{\mu\nu}+\hat{p}^{\mu}_{\perp}\hat{p}^{\nu}_{\perp})\Big)
\\
&&\times\Big(\tilde{a}_{q\nu}-p^{\beta}\partial_{q^{\beta}}\tilde{a}_{q\nu}+\mathcal{O}(|{\bf p}|^2)\Big)
,
\\
\nonumber
\hat{Q}^{L\mu}_{\text{cl}(c)}&=&\hat{\rho}_L\Big(2q_0u^{\nu}p^{\mu}-2(q_0p_0-|{\bf q}||{\bf p}|z)u^{\mu}u^{\nu}\Big)\tilde{a}_{q-p\nu}
\\
&\approx&\hat{\rho}_L\big(2q_0u^{\nu}p_{\perp}^{\mu}+2|{\bf q}||{\bf p}|zu^{\mu}u^{\nu}\big)\big(\tilde{a}_{q\nu}-p^{\beta}\partial_{q^{\beta}}\tilde{a}_{q\nu}+\mathcal{O}(|{\bf p}|^2)\big).
\end{eqnarray}
\end{subequations}
In the following, we evaluate those terms one by one. 
As in the case of SKE, we first retrieve the integrals for $ |\bp| $ that give rise to the logarithm 
in the following subsections. 
The angle integrals, with respect to $  z$, will be performed in Sec.~\ref{sec:AKE-results} afterwards.

\subsubsection{Evaluating $ {Q}^{\mu}_{\text{cl}(a)} $}

Combining with Eqs.~(\ref{delta_exp}) and (\ref{rhoT_exp}), we find 
\begin{eqnarray}\nonumber\label{int_Qcla_2}
&&\delta(q^2-m^2)\int_p2\pi\delta((q-p)^2-m^2)Q^{\mu}_{\text{cl}(a)}
\\\nonumber
&&
\approx
\frac{\pi Tm_D^2\delta(q^2-m^2)}{2q_{0}}\int^{T}_{m_D}\frac{d|{\bf p}|}{(2\pi)^2|{\bf p}|}\int^1_{-1} dz\Bigg[\Bigg(\frac{q_0z^2}{T}+\frac{I_Tz^2(1-z^2)|{\bf q}|^2}{2q_0T}\Bigg)q^{\beta}_{\perp}\partial_{q^{\beta}_{\perp}}
\\
&&\quad-\left(q_0^2+\frac{I_T|{\bf q}|^2(1-z^2)}{2}\right)\Big((1-z^2)\eta^{\alpha\beta}+(1-3z^2)\hat{q}^{\alpha}_{\perp}\hat{q}^{\beta}_{\perp}\Big)\frac{\partial_{q^{\alpha}_{\perp}}\partial_{q^{\beta}_{\perp}}}{2}
\Bigg]\tilde{a}_{q}^{\mu}
,
\end{eqnarray}
where we take $u^{\beta}\partial_{q^{\beta}}\tilde{a}_{q}^{\mu}=0$ by imposing $q_0=E_q=\sqrt{|{\bf q}|^2+m^2}$ upon $\tilde{a}^{\mu}_q$ in advance and utilize
\begin{eqnarray}
\Big(\hat{q}^{\alpha}_{\perp}\hat{q}^{\beta}_{\perp}z^2-\frac{\hat{\Theta}^{\alpha\beta}_q}{2}\big(1-z^2\big)\Big)=-\frac{\Theta^{\alpha\beta}}{2}(1-z^2)-\frac{(1-3z^2)}{2}\hat{q}^{\alpha}_{\perp}\hat{q}^{\beta}_{\perp}.
\end{eqnarray}

\subsubsection{Evaluating $ {Q}^{\mu}_{\text{cl}(b)} $}

Next, we evaluate the term proportional to $  Q^{\mu}_{\text{cl}(b)}$: 
\begin{eqnarray}
\delta(q^2-m^2)\int_p2\pi\delta((q-p)^2-m^2)Q^{\mu}_{\text{cl}(b)}
\approx
\frac{\pi Tm_D^2\delta(q^2-m^2)}{2q_{0}}\int\frac{d^3{\bf p}}{(2\pi)^3}\frac{1}{|{\bf p}|^{5}}
\big(\tilde{\mathcal{K}}^{T\mu}_{\text{cl}(b)}+\tilde{\mathcal{K}}^{L\mu}_{\text{cl}(b)}\big)
,
\notag\\ \label{int_Qclb_1}
\end{eqnarray}
where 
\begin{subequations}
\begin{eqnarray}\nonumber
\tilde{\mathcal{K}}^{T\mu}_{\text{cl}(b)}&=&I_T\Bigg[
|{\bf p}|(q^{\mu}_{\perp}-\hat{p}^{\mu}_{\perp}|{\bf q}| z) \left(\frac{|{\bf q}|}{q_0}u^{\nu} z+\hat{p}_{\perp}^{\nu}\right)
+\frac{|{\bf p}|^2}{2q_0^3T}\Big(|{\bf q}|^2 z^2 \big(u^{\nu} ((T-q_0)q_{\perp}^{\mu}+2 q_0 T u^{\mu})+\hat{p}_{\perp}^{\mu} \hat{p}_{\perp}^{\nu} q_0^2\big)
\\\nonumber
&&
+q_0^2 T \big(2 \hat{p}_{\perp}^{\mu} \hat{p}^{\nu}_{\perp} q_0-q^{\mu}_{\perp} u^{\nu}\big)
+|{\bf q}| q_0^2 z (3 \hat{p}^{\mu}_{\perp} T u^{\nu}-\hat{p}_{\perp}^{\nu} q_{\perp}^{\mu}+2 \hat{p}_{\perp}^{\nu} T u^{\mu})+|{\bf q}|^3\hat{p}_{\perp}^{\mu}  u^{\nu} z^3 (q_0-T)\Big)
\\
&&-\big(q_{\perp}^{\mu}-\hat{p}_{\perp}^{\mu}|{\bf q}|z\big)\left(\frac{|{\bf q}|}{q_0}u^{\nu} z+\hat{p}_{\perp}^{\nu}\right)
\left(\frac{|{\bf q}|}{q_{0}}u^{\rho}z+\hat{p}_{\perp}^{\rho}\right)|{\bf p}|^{2}
\partial_{q^{\rho}}
+\mathcal{O}(|{\bf p}|^3)
\Bigg]\tilde{a}_{q\nu },
\\
\nonumber
\tilde{\mathcal{K}}^{L\mu}_{\text{cl}(b)}&=&-2 \tilde{a}_{q\nu} u^{\mu} |{\bf p}|(\hat{p}_{\perp}^{\nu} q_0 + |{\bf q}|u^{\nu} z)
+\frac{|{\bf p}|^2}{q_0^2T}\Big( u^{\mu} \left( u^{\nu} \left(|{\bf q}|^2 z^2 (q_0-2 T)+q_0^2 T\right)+\hat{p}_{\perp}^{\nu} |{\bf q}| q_0 z (q_0-T)\right)
\\
&&
-\hat{p}_{\perp}^{\mu}q_0 T (|{\bf q}|u^{\nu} z+\hat{p}_{\perp}^{\nu} q_0)\Big)\tilde{a}_{q\nu}
+2q_0u^{\mu}\left(\frac{|{\bf q}|}{q_0}u^{\nu} z+\hat{p}_{\perp}^{\nu}\right)\left(\frac{|{\bf q}|}{q_{0}}u^{\rho}z+\hat{p}_{\perp}^{\rho}\right)|{\bf p}|^{2}\partial_{q^{\rho}}\tilde{a}_{q\nu}+\mathcal{O}(|{\bf p}|^3).\notag\\
\end{eqnarray}
\end{subequations}
It is clear that $\mathcal{O}(|{\bf p}|)$ terms in $\tilde{\mathcal{K}}^{T/L\mu}_{\text{cl}(b)}$ can be dropped by symmetry of the integration. We should further employ the useful decomposition and replacements in Eqs.~(\ref{rel_1}), (\ref{rel_2}), and the following relation in the integral,
\begin{eqnarray}\nonumber\label{rel_3}
zp^{\nu}_{\perp}p^{\mu}_{\perp}p^{\rho}_{\perp}&=&z|{\bf p}|^3(\hat{q}_{\perp}^{\nu}z+\hat{p}^{\nu}_{T})(\hat{q}_{\perp}^{\mu}z+\hat{p}^{\mu}_{T})(\hat{q}_{\perp}^{\rho}z+\hat{p}^{\rho}_{T})
\\
&\rightarrow &|{\bf p}|^3\Bigg(z^4\hat{q}^{\nu}_{\perp}\hat{q}^{\mu}_{\perp}\hat{q}^{\rho}_{\perp}-\frac{z^2(1-z^2)}{2}\Big(\hat{q}^{\nu}_{\perp}\hat{\Theta}_q^{\mu\rho}+\hat{q}^{\mu}_{\perp}\hat{\Theta}_q^{\nu\rho}+\hat{q}^{\rho}_{\perp}\hat{\Theta}_q^{\mu\nu}\Big)\Bigg),
\end{eqnarray}
where $\hat{p}^{\mu}_{T}=\hat{\Theta}^{\mu\nu}_q\hat{p}_{\perp\nu}$ and we drop terms with odd powers of $z$.
Dropping the vanishing terms by symmetry in Eq.~(\ref{int_Qclb_1}), we obtain
\begin{eqnarray}\label{int_Qclb_2}\nonumber
&&\delta(q^2-m^2)\int_p2\pi\delta((q-p)^2-m^2)Q^{\mu}_{\text{cl}(b)}
\\
&&\approx
\frac{\pi Tm_D^2\delta(q^2-m^2)}{2q_{0}}\int^{T}_{m_D}\frac{d|{\bf p}|}{(2\pi)^2|{\bf p}|}\int^1_{-1} dz
\Big(\mathcal{K}^{T\mu}_{\text{cl}(b)}+\mathcal{K}^{L\mu}_{\text{cl}(b)}\Big),
\end{eqnarray}
where
\begin{eqnarray}\nonumber
\mathcal{K}^{T\mu}_{\text{cl}(b)}&=&I_T\Bigg\{\Bigg[u^{\mu}z^2\Bigg(\frac{q^{\nu}_{\perp}}{q_0}
+\frac{|{\bf q}|^2u^{\nu}}{q_0^2}\Bigg)
-\frac{\hat{q}_{\perp}^{\mu}}{4q_0^3T}\Bigg(q_{0}^{2}\hat{q}^{\nu}_{\perp}\Big(3|{\bf q}|^2z^2(1-z^2)+2q_0T(1-3z^2)\Big)
\\\nonumber
&&+2u^{\nu}|{\bf q}|\Big(z^2(1-z^2)|{\bf q}|^2(q_0-T)+q_0^2T(1-3z^2)\Big)\Bigg)
-\frac{\Theta^{\mu\nu}}{4}(1-z^2)\Bigg(2+\frac{|{\bf q}|^2z^2}{q_0T}\Bigg)\Bigg]\tilde{a}_{q\nu}
\\\nonumber
&&+\frac{1}{2}\Bigg[q_{\perp}^{\mu}(1-z^2)\Big(\hat{q}^{\nu}_{\perp}\hat{q}^{\rho}_{\perp}(1-5z^2)+(1-z^2)\Theta^{\rho\nu}
-3\frac{q^{\rho}_{\perp}}{q_0}u^{\nu}z^2\Big)
\\
&&-z^2(1-z^2)\Bigg(\Theta^{\mu\nu}q_{\perp}^{\rho}+\Theta^{\rho\mu}\Big(q_{\perp}^{\nu}+\frac{|{\bf q}|^2}{q_0}u^{\nu}\Big)\Bigg)
\Bigg]\partial_{q^{\rho}_{\perp}}\tilde{a}_{q\nu}\Bigg\}
\end{eqnarray}
and
\begin{eqnarray}
\mathcal{K}^{L\mu}_{\text{cl}(b)}&=&u^{\mu}\Bigg[u\cdot\tilde{a}_q\Bigg(1+\frac{|{\bf q}|^2z^2(q_0-2T)}{q_0^2T}\Bigg)+\hat{q}_{\perp}\cdot\tilde{a}_q\frac{|{\bf q}|z^2(q_0-T)}{q_0T}
\Bigg]
+\hat{q}^{\mu}_{\perp}\Bigg[\hat{q}_{\perp}\cdot\tilde{a}_q\frac{(1-3z^2)}{2}
\\\nonumber
&&-u\cdot\tilde{a}_q\frac{|{\bf q}|z^2}{q_0}\Bigg]
+\Theta^{\mu\nu}\tilde{a}_{\nu}\frac{(1-z^2)}{2}
+u^{\mu}\Big(2z^2q^{\rho}_{\perp}u^{\nu}-\hat{q}_{\perp}^{\rho}\hat{q}_{\perp}^{\nu}q_0(1-3z^2)-\Theta^{\rho\nu}q_0(1-z^2)\Big)\partial_{q^{\rho}_{\perp}}\tilde{a}_{q\nu}.
\end{eqnarray}
Since $\delta(q^2-m^2)q\cdot \tilde{a}_q=\delta(q^2-m^2)(E_qu\cdot \tilde{a}_q+q_{\perp}\cdot \tilde{a}_q)=0$, we can use $u\cdot \tilde{a}_q=-q_{\perp}\cdot \tilde{a}_q/q_0$ to further simplify $\mathcal{K}^{T/L\mu)}_{\text{cl}(b)}$, which results in
\begin{eqnarray}\nonumber
\mathcal{K}^{T\mu}_{\text{cl}(b)}&=&I_T\Bigg\{\hat{q}_{\perp}\cdot \tilde{a}_q\Bigg[u^{\mu}\frac{m^2|{\bf q}|z^2}{q_0^3}
-\frac{\hat{q}^{\mu}_{\perp}}{2}\Bigg(z^2(1-z^2)\Big(1+\frac{q_0}{2T}+\frac{m^4}{q_0^4}\Big(1-\frac{q_0}{T}\Big)\Big)+\frac{m^2}{q_0^2}\Big(1-5z^2+2z^4
\\\nonumber
&&+\frac{q_0}{2T}z^2(1-z^2)\Big)\Bigg)\Bigg]
-\frac{\Theta^{\mu\nu}\tilde{a}_{q\nu}}{2}(1-z^2)\left(1+\frac{|{\bf q}|^2z^2}{2q_0T}\right)
+\frac{1}{2}(1-z^2)\Bigg[\hat{q}_{\perp}^{\mu}\Bigg(|{\bf q}|\hat{q}_{\perp}^{\rho}\hat{q}^{\nu}_{\perp}(1-5z^2)
\\
&&+|{\bf q}|\Theta^{\rho\nu}(1-z^2)-\frac{3|{\bf q}|^2z^2}{q_0}\hat{q}_{\perp}^{\rho}u^{\nu}\Bigg)
-|{\bf q}|z^2\Theta^{\mu\nu}\hat{q}^{\rho}_{\perp}-\Big(|{\bf q}|\hat{q}_{\perp}^{\nu}+\frac{|{\bf q}|^2}{q_0}u^{\nu}\Big)z^2\Theta^{\rho\mu}
\Bigg]\partial_{q^{\rho}_{\perp}}\tilde{a}_{q\nu}\Bigg\},\notag\\
\end{eqnarray}
and
\begin{eqnarray}\nonumber
\mathcal{K}^{L\mu}_{\text{cl}(b)}&=&u^{\mu}\Bigg[\frac{m^2|{\bf q}|z^2}{q_0^2T}
-\frac{|{\bf q}|}{q_0}\Bigg(1-\frac{|{\bf q}|^2z^2}{q_0^2}+\frac{m^2z^2}{q_0^2}\Bigg)
\Bigg]\hat{q}_{\perp}\cdot\tilde{a}_q
+\hat{q}^{\mu}_{\perp}\Bigg[\frac{(1-z^2)}{2}-\frac{m^{2}z^2}{q_0^{2}}\Bigg]\hat{q}_{\perp}\cdot\tilde{a}_q
\\
&&
+\Theta^{\mu\nu}\tilde{a}_{\nu}\frac{(1-z^2)}{2}
+u^{\mu}\Big(2z^2q^{\rho}_{\perp}u^{\nu}-\hat{q}_{\perp}^{\rho}\hat{q}_{\perp}^{\nu}q_0(1-3z^2)-\Theta^{\rho\nu}q_0(1-z^2)\Big)\partial_{q^{\rho}_{\perp}}\tilde{a}_{q\nu}.
\end{eqnarray}

\subsubsection{Evaluating $ {Q}^{\mu}_{\text{cl}(c)} $}

In the same way, we should cope with the other term %$\int_pQ^{\mu}_{\text{cl}(c)}$. It is found
\begin{eqnarray}
\delta(q^2-m^2)\int_p2\pi\delta((q-p)^2-m^2)Q^{\mu}_{\text{cl}(c)}
\approx
\frac{\pi Tm_D^2\delta(q^2-m^2)}{2q_{0}}\int\frac{d^3{\bf p}}{(2\pi)^3}\frac{1}{|{\bf p}|^{5}}
\Big(\tilde{\mathcal{K}}^{T\mu}_{\text{cl}(c)}+\tilde{\mathcal{K}}^{L\mu}_{\text{cl}(c)}\Big),\notag \label{int_Qclc_1}\\
\end{eqnarray}
where
\begin{eqnarray}\nonumber
\tilde{\mathcal{K}}^{T\mu}_{\text{cl}(c)}&=&I_T\Bigg[
-|{\bf p}|(q^{\nu}_{\perp}-\hat{p}^{\nu}_{\perp}|{\bf q}| z) \left(\frac{|{\bf q}|}{q_0}u^{\mu} z+\hat{p}_{\perp}^{\mu}\right)
+\frac{|{\bf p}|^2}{2q_0^3T}\Big(|{\bf q}|q_0 z (q_{\perp}^{\nu}-\hat{p}_{\perp}^{\nu} |{\bf q}| z) (\hat{p}_{\perp}^{\mu} q_0+|{\bf q}|u^{\mu} z)
\\\nonumber
&&-T (q_0^2-|{\bf q}|^2 z^2)
(\hat{p}_{\perp}^{\mu} \hat{p}_{\perp}^{\nu} q_0+\hat{p}_{\perp}^{\nu}|{\bf q}|u^{\mu} z+\Theta^{\mu\nu} q_0-q_{{\perp}}^{\nu}u^{\mu})\Big)
\\
&&+|{\bf p}|^{2}(q^{\nu}_{\perp}-\hat{p}^{\nu}_{\perp}|{\bf q}| z) \left(\frac{|{\bf q}|}{q_0}u^{\mu} z+\hat{p}_{\perp}^{\mu}\right)\left(\frac{|{\bf q}| }{q_{0}}u^{\rho}z+\hat{p}^{\rho}_{\perp}\right)\partial_{q^{\rho}}
+\mathcal{O}(|{\bf p}|^3)
\Bigg]\tilde{a}_{q\nu},
\end{eqnarray}
and
\begin{eqnarray}\nonumber
\tilde{\mathcal{K}}^{L\mu}_{\text{cl}(c)}&=&2|{\bf p}|(\hat{p}^{\mu}_{\perp}q_0+|{\bf q}|u^{\mu}z)u\cdot\tilde{a}_q
-\frac{|{\bf p}|^2}{q_0^2T}|{\bf q}|z\big(q_0-2T\big)\big(\hat{p}_{\perp}^{\mu}q_0+|{\bf q}|u^{\mu}z\big)u\cdot\tilde{a}_q
\\
&&-2|{\bf p}|^{2}\big(q_0\hat{p}_{\perp}^{\mu}+|{\bf q}|zu^{\mu}\big)\left(\frac{|{\bf q}| }{q_{0}}u^{\rho}z+\hat{p}^{\rho}_{\perp}\right)\partial_{q^{\rho}}u\cdot\tilde{a}_q+\mathcal{O}(|{\bf p}|^3).
\end{eqnarray}
As above, the $\mathcal{O}(|{\bf p}|)$ terms in $\tilde{\mathcal{K}}^{T/L\mu}_{\text{cl}(c)}$ do not contribute to the integral. 
Now, by employing Eqs.~(\ref{rel_1}), (\ref{rel_2}), and (\ref{rel_3}), we find 
\begin{eqnarray}\nonumber
\delta(q^2-m^2)\int_p2\pi\delta((q-p)^2-m^2)Q^{\mu}_{\text{cl}(c)}
\approx
\frac{\pi Tm_D^2\delta(q^2-m^2)}{2q_{0}}\int^{T}_{m_D}\frac{d|{\bf p}|}{(2\pi)^2|{\bf p}|}\int^1_{-1} dz
\Big(\mathcal{K}^{T\mu}_{\text{cl}(c)}+\mathcal{K}^{L\mu}_{\text{cl}(c)}\Big),
\\
\end{eqnarray}
where
\begin{subequations}
\begin{eqnarray}\nonumber\label{int_Qclc_2}
\mathcal{K}^{T\mu}_{\text{cl}(c)}&=&I_T\Bigg[
\frac{u^{\mu}q^{\nu}_{\perp}}{2q_0}(1-z^2)\Bigg(1+\frac{|{\bf q}|^2z^2}{q_0^2T}(q_0-T)\Bigg)
-\frac{\Theta^{\mu\nu}}{4}\Bigg((1+z^2)\left(1-\frac{|{\bf q}|^2z^2}{q_0^2}\right)-\frac{|{\bf q}|^2}{q_0T}z^{2}(1-z^2)\Bigg)
\\\nonumber
&&+\frac{\hat{q}^{\mu}_{\perp}\hat{q}^{\nu}_{\perp}}{4}\Bigg((1-3z^2)\left(1-\frac{|{\bf q}|^2z^2}{q_0^2}\right)+\frac{3|{\bf q}|^2}{q_0T}z^{2}(1-z^2)\Bigg)
+\frac{u^{\mu}|{\bf q}|z^2(1-z^2)}{2q_0}\big(3\hat{q}^{\rho}_{\perp}q^{\nu}_{\perp}+|{\bf q}|\Theta^{\rho\nu}\big)\partial_{q^{\rho}_{\perp}}
\\
&&
+\frac{(1-z^2)}{2}\Big(z^2\Theta^{\mu\nu}q^{\rho}_{\perp}+q^{\mu}_{\perp}\big((5z^2-1)\hat{q}^{\nu}_{\perp}\hat{q}^{\rho}_{\perp}+z^2\Theta^{\rho\nu}\big)-(1-z^2)q_{\perp}^{\nu}\Theta^{\rho\mu}
\Big)\partial_{q^{\rho}_{\perp}}
\Bigg]\tilde{a}_{q\nu},
\\
\nonumber
\mathcal{K}^{L\mu}_{\text{cl}(c)}&=&
-u\cdot\tilde{a}_q\frac{(q_0-2T)|{\bf q}|z^2}{q_0T}\Bigg(\frac{|{\bf q}|}{q_0}u^{\mu}+\hat{q}_{\perp}^{\mu}
\Bigg)
+q_0\Big(\hat{q}^{\mu}_{\perp}\hat{q}^{\rho}_{\perp}(1-3z^2)+\Theta^{\mu\rho}(1-z^2)
\\
&&-2q_{\perp}^{\rho}q_0^{-1}u^{\mu}z^2\Big)\partial_{q^{\rho}_{\perp}}(u\cdot\tilde{a}_{q}).
\end{eqnarray}
\end{subequations}

\subsection{Rest of the terms: $  [Q_2^{\mu\nu}f_{Vq}\tilde{a}_{q'\nu}-Q_1(1-f_{Vq'})\tilde{a}^{\mu}_q]$}

\label{sec:d}

As we have evaluated two of the four terms in Eq.~(\ref{AKE_HTL3_non}), 
the remaining two terms are given as 
\begin{eqnarray}
{\mathcal C}_2 \equiv \delta(q^2-m^2)\int_p g_{0p}^{-1}\Big[Q_2^{\mu\nu}f_{Vq}\tilde{a}_{q'\nu}-Q_1(1-f_{Vq'})\tilde{a}^{\mu}_q\Big].
\end{eqnarray}
While $ Q_1 $ has been given in the computation of the SKE, 
we now have the tensor part 
\begin{eqnarray}
Q^{\mu\nu}_2\tilde{a}_{q'\nu}
&=&-2\pi\delta(q'^2-m^2)g_{0p}\Big\{p^{\mu}\Big[\rho_L\Big(p^{\nu}-2q_0u^{\nu}\Big)+2\rho_T\Big(q_{\perp}^{\nu}-|{\bf q}|z\hat{p}_{\perp}^{\nu}-p^{\nu}\Big)\Big]\tilde{a}_{q'\nu}
\nn
\\\nonumber
&&-\tilde{a}^{\mu}_{q'}\Big[\rho_L\Big(q\cdot p+2(q_0^2-q_0p_0)\Big)
+2\rho_T\Big(|{\bf q}|^2(1-z^2)-q\cdot p\Big)\Big]
\\\nonumber
&&+2u^{\mu}\rho_L\Big(q\cdot pu^{\nu}+(q_0-p_0)p^{\nu}\Big)\tilde{a}_{q'\nu}
-2\tilde{a}^{\mu}_{q'\perp}q\cdot p\rho_T-2q_{\perp}^{\mu}p\cdot \tilde{a}_{q'}\rho_T
\\
&&-2\hat{p}_{\perp}^{\mu}\Big(q\cdot p\hat{p}^{\nu}_{\perp}-|{\bf q}|zp^{\nu}\Big)\rho_{T}\tilde{a}_{q'\nu}\Bigg\}
.
\end{eqnarray}
Combining the two terms, it is found
\begin{eqnarray}
{\mathcal C}_2 \approx \frac{\pi m_D^2\delta(q^2-m^2)}{2q_{0}}\int\frac{d^3{\bf p}}{(2\pi)^3}
\frac{1}{|{\bf p}|^{5}}\Big(\tilde{\mathcal{K}}^{T\mu}_{\text{cl}(d)}+\tilde{\mathcal{K}}^{L\mu}_{\text{cl}(d)}\Big),
\end{eqnarray}
where the integrand is given as 
\begin{subequations}
\begin{eqnarray}\nn
\tilde{\mathcal{K}}^{T\mu}_{\text{cl}(d)}&=&-I_T\Bigg[\frac{|{\bf q}|^3|{\bf p}|}{q_0T}z(1-z^2)(1-2f_{Vq})\tilde{a}_{q}^{\mu}+\frac{|{\bf q}|^2|{\bf p}|^2}{2q_0^3T}\Bigg(\tilde{a}_q^{\mu} \left(z^2-1\right) \Big(q_0^2 \big(1-2z |{\bf q}|\hat{p}_{\perp}^{\rho}(\partial_{q^{\rho}}f_{Vq})-2f_{Vq}\big)
\\\nonumber
&&+(2f_{Vq}-1) |{\bf q}|^2 z^2\Big)+2 f_{Vq}q_0 z \Big(\tilde{a}_{q\nu} \big(\hat{p}^{\mu}_{\perp} \hat{q}^{\nu}_{\perp}q_0+\hat{p}^{\mu}_{\perp} |{\bf q}|u^{\nu} z^2-\hat{p}^{\nu}_{\perp} (\hat{q}^{\mu}_{\perp}q_0+|{\bf q}|u^{\mu} z^2)
\\
&&-\hat{q}^{\mu}_{\perp} |{\bf q}|u^{\nu} z+\hat{q}_{\perp}^{\nu} |{\bf q}|u^{\mu} z\big)
-|{\bf q}|\hat{p}^{\rho}_{\perp}(\partial_{q^{\rho}}\tilde{a}_q^{\mu}) q_0 \left(z^2-1\right)\Big)\Bigg)\Bigg]
+ \mathcal{O}(|{\bf p}|^3)
,
\\
\tilde{\mathcal{K}}^{L\mu}_{\text{cl}(d)}&=&-\Bigg[\frac{2|{\bf q}||{\bf p}|q_0z}{T}(1-2f_{Vq})\tilde{a}_{q}^{\mu}
+\frac{|{\bf p}|^2}{q_0^2T}\Bigg(\tilde{a}_q^{\mu} \Big(q_0^3 (2z|{\bf q}| \hat{p}_{\perp}^{\rho}(\partial_{q^{\rho}}f_{Vq}) +2f_{Vq}-1)+(1-2f_{Vq}) |{\bf q}|^2 q_0 z^2\Big)
\nn
\\
&&+2 f_{Vq} |{\bf q}|q_0^{{2}} z \Big( \hat{p}^{\nu}_{\perp}\tilde{a}_{q\nu}u^{\mu}+q_0\hat{p}^{\rho}_{\perp}(\partial_{q^{\rho}_{\perp}}\tilde{a}_q^{\mu})-u^{\nu}\tilde{a}_{q\nu} \hat{p}^{\mu}_{\perp}\Big)\Bigg)
\Bigg] + \mathcal{O}(|{\bf p}|^3).
\end{eqnarray}
\end{subequations}
Here, we utilize $u\cdot \tilde{a}_q=-q_{\perp}\cdot \tilde{a}_q/q_0$ in computations as above.
Carrying out the integration, we obtain
\begin{eqnarray}
{\mathcal C}_2 
&\approx &
\frac{m_D^2T\delta(q^2-m^2)}{8\pi E_q}\ln (1/g_c)
\Big(\tilde{a}_q^{\mu}\tilde{\mathcal{Q}}_\text{cl}^{(1)}+u^{\mu}\tilde{\mathcal{Q}}_\text{cl}^{(2)}+\hat{q}^{\mu}_{\perp}\tilde{\mathcal{Q}}_\text{cl}^{(3)}
+ \tilde {\mathcal{Q}}_\text{cl}^{(4)}\hat{q}_{\perp}^{\nu}\partial_{q_{\perp\mu}}\tilde{a}_{q\nu}
\nn
\\
&&
+ \tilde {\mathcal{Q}}_\text{cl}^{(5)}\hat{q}^{\nu}\partial_{q_{\perp}^{\nu}}\tilde{a}_{q}^{\mu}
+ \tilde {\mathcal{Q}}_\text{cl}^{(6)}\eta^{\nu\rho}\partial_{q^{\nu}_{\perp}}\partial_{q^{\rho}_{\perp}}\tilde{a}_q^{\mu}
+ \tilde {\mathcal{Q}}_\text{cl}^{(7)}\hat{q}_{\perp}^{\nu}\hat{q}_{\perp}^{\rho}\partial_{q^{\nu}_{\perp}}\partial_{q^{\rho}_{\perp}}\tilde{a}_q^{\mu}
\Big),
\label{eq:C2}
\end{eqnarray}
where 
\begin{subequations}
\begin{eqnarray}\nonumber
&&
\tilde{\mathcal{Q}}_\text{cl}^{(1)}=\frac{1}{|{\bf q}|q_0T}\Big[ 
|{\bf q}|q_0^2 \big\{ 
(1-2 f_{Vq}) (j_0^L-j_1^T+j_2^T)-2 |{\bf q}|(\hat{q}_{\perp}^{\nu}\partial_{q^{\nu}_{\perp}}f_{Vq}) (j_1^L+j_1^T)
\big\}
\\
&& \hspace{2cm}
-(2 f_{Vq}-1) |{\bf q}|^3 (j_0^T-j_1^L-j_1^T)+2 {(}\hat{q}_{\perp}^{\nu}\partial_{q^{\nu}_{\perp}}f_{Vq}) j_2^T q_0^4
\Big],
\\
&&
\tilde{\mathcal{Q}}_\text{cl}^{(2)}=-2\Big((j_1^L+j_1^T)|{\bf q}|^2-j_2^Tq_0^2\Big)\frac{f_{Vq}}{|{\bf q}|T}\hat{q}_{\perp}\cdot\tilde{a}_q,
\\
&&
\tilde{\mathcal{Q}}_\text{cl}^{(3)}=-2\Big((j_1^L+j_1^T)|{\bf q}|^2-j_2^Tq_0^2\Big)\frac{f_{Vq}}{q_0T}\hat{q}_{\perp}\cdot\tilde{a}_q,
\\
&&
\tilde{\mathcal{Q}}_\text{cl}^{(5)}=2\Bigg(\frac{q_0^2}{|{\bf q}|}j_2^T-|{\bf q}|(j_1^L+j_1^T)\Bigg)\frac{q_0f_{Vq}}{T},
\\
&&
\tilde {\mathcal{Q}}_\text{cl}^{(4)} = \tilde {\mathcal{Q}}_\text{cl}^{(6)} =\tilde {\mathcal{Q}}_\text{cl}^{(7)} = 0.
\end{eqnarray}
\end{subequations}

\subsection{Assembling all the pieces}

\label{sec:AKE-results}

Combining Eqs.~(\ref{int_Qcla_2}), (\ref{int_Qclb_2}), and (\ref{int_Qclc_2}) and carrying out the integration, we acquire
\begin{eqnarray}\nonumber
{\mathcal C}_1 &\equiv&
\delta(q^2-m^2)\int_p\big(Q_2^{\mu\nu}\tilde{a}_{q'\nu}-Q_1\tilde{a}^{\mu}_q\big)
\\\nonumber
&\approx&  \frac{m_D^2T\delta(q^2-m^2)}{8\pi E_q}\ln (1/g_c)\Big(\tilde{a}_q^{\mu}\mathcal{Q}_\text{cl}^{(1)}+u^{\mu}\mathcal{Q}_\text{cl}^{(2)}+\hat{q}^{\mu}_{\perp}\mathcal{Q}_\text{cl}^{(3)}
+\mathcal{Q}_\text{cl}^{(4)}\hat{q}_{\perp}^{\nu}\partial_{q_{\perp\mu}}\tilde{a}_{q\nu}
\\
&&\quad+\mathcal{Q}_\text{cl}^{(5)}\hat{q}^{\nu}\partial_{q_{\perp}^{\nu}}\tilde{a}_{q}^{\mu}+\mathcal{Q}_\text{cl}^{(6)}\eta^{\nu\rho}\partial_{q^{\nu}_{\perp}}\partial_{q^{\rho}_{\perp}}\tilde{a}_q^{\mu}+\mathcal{Q}_\text{cl}^{(7)}\hat{q}_{\perp}^{\nu}\hat{q}_{\perp}^{\rho}\partial_{q^{\nu}_{\perp}}\partial_{q^{\rho}_{\perp}}\tilde{a}_q^{\mu}
\Big),
\label{eq:C1}
\end{eqnarray}
where
\begin{subequations}
\begin{eqnarray}
&&
\mathcal{Q}_\text{cl}^{(1)}=\frac{1}{2}\Bigg(j_1^L+3j_1^T-j_0^L-3j_0^T+\frac{q_0^2}{|{\bf q}|^2} (j_1^T-j_2^T)\Bigg),
\\
&&
\nonumber
\mathcal{Q}_\text{cl}^{(2)}=-\frac{1}{2}\Bigg[2q_0(\partial_{q^{\nu}_{\perp}}\tilde{a}_q^{\nu}) \left(j^L_0-(j_1^L+j_1^T)+j_2^T \frac{q_0^2}{|{\bf q}|^2}\right)
+2q_0\hat{q}^{\nu}_{\perp}\hat{q}^{\rho}_{\perp}(\partial_{q^{\rho}_{\perp}}\tilde{a}_{q\nu}) \Big(j_0^L -3(j_1^L+j_1^T)
+3j_2^T\frac{q_0^2}{|{\bf q}|^2}\Big)
\\\nonumber
&& \hspace{2cm}
-\hat{q}^{\nu}_{\perp}\tilde{a}_{q\nu}
\frac{|{\bf q}|}{q_0}\Bigg(j_1^L\frac{(2 q_0-T)}{T}-3 j_0^L-j_0^T
-2j_1^T \frac{q_0(m^2-|{\bf q}|^2)}{|{\bf q}|^2T}
+j_1^T \frac{m^2}{|{\bf q}|^2} \frac{(2 q_0+3 T)}{T}
\\
&& \hspace{2cm}
+j_2^T \frac{q_0^2}{|{\bf q}|^2} \frac{(T-2 q_0)}{T}\Bigg)
\Bigg],
\\
&&
\nonumber
\mathcal{Q}_\text{cl}^{(3)}=-\frac{1}{2}\Bigg[\frac{2}{|{\bf q}|}
\Bigg(\big(q_0^2j^T_1-|{\bf q}|^2j_0^T\big)(\partial_{q^{\nu}_{\perp}}\tilde{a}_q^{\nu})
+\hat{q}^{\nu}_{\perp}\hat{q}^{\rho}_{\perp}(\partial_{q^{\rho}_{\perp}}\tilde{a}_{q\nu})  \Big(j_0^L |{\bf q}|^2+3j_2^T q_0^2-3|{\bf q}|^2(j_1^L+j_1^T)\Big)\Bigg)
\\\nonumber
&& \hspace{2cm}
-\hat{q}^{\nu}_{\perp}\tilde{a}_{q\nu}
\Bigg(\frac{|{\bf q}|^2}{q_0^2} (j_0^T-j_0^L)-\frac{m^2}{q_0^2}(j^L_0+j_0^T)
+3 \frac{m^2}{q_0^2}j_1^L+j_1^L \frac{|{\bf q}|^2}{q_0^2T} (2 q_0-3 T)-2 j_1^T \frac{m^4}{q_0^2|{\bf q}|^2}
\\
&& \hspace{2cm}
+5 j_1^T \frac{m^2}{|{\bf q}|^2}+2 j_1^T \frac{|{\bf q}|^2}{q_0^2T} (q_0-2 T)+2 j_2^T \frac{m^2 q_0}{|{\bf q}|^2T}-6 j_2^T \frac{m^2}{|{\bf q}|^2}-2 j_2^T \frac{q_0^3}{|{\bf q}|^2T}+3 j_2^T \frac{q_0^2}{|{\bf q}|^2}\Bigg)
\Bigg],
\\
&&
\mathcal{Q}_\text{cl}^{(4)}=|{\bf q}|\big(j_1^L+j_1^T-j_0^L-j_0^T\big)+\frac{q_0^2}{|{\bf q}|}\big(j_1^T-j_2^T\big),
\\
&&
\mathcal{Q}_\text{cl}^{(5)}=\frac{q_0}{|{\bf q}|T}\Big(\big(j_1^L+j_1^T\big)|{\bf q}|^2-j_2^Tq_0^2\Big),
\\
&&
\mathcal{Q}_\text{cl}^{(6)}={-}\frac{1}{2}\Bigg(j_0^T|{\bf q}|^2+\big(j_0^L-j_1^L-2j_1^T\big)q_0^2+j_2^T\frac{q_0^4}{|{\bf q}|^2}\Bigg),
\\
&&
\mathcal{Q}_\text{cl}^{(7)}=-\frac{1}{2}\Bigg(j_0^T|{\bf q}|^2+\big(j_0^L-3j_1^L-4j_1^T\big)q_0^2+3j_2^T\frac{q_0^4}{|{\bf q}|^2}\Bigg).
\end{eqnarray}
\end{subequations}
Note that we have implemented the following relations in computations,
\begin{eqnarray}\nonumber
&&\Theta^{\mu\nu}\tilde{a}_{q\nu}=\tilde{a}_q^{\mu}-(\tilde{a}_q\cdot u)u^{\mu}=\tilde{a}_q^{\mu}+u^{\mu}\frac{|{\bf q}|}{E_q}\hat{q}_{\perp}\cdot\tilde{a}_q,
\\\nonumber
&&\partial_{q_{\perp\rho}}(\tilde{a}_q\cdot u)=-\partial_{q_{\perp\rho}}\left(\frac{q_{\perp}\cdot \tilde{a}_q}{E_q}\right)
=-\frac{1}{E_q}\Bigg(\Theta^{\rho\nu}\tilde{a}_{q\nu}+\frac{|{\bf q}|^2}{E_q^2}\hat{q}^{\rho}_{\perp}\hat{q}_{\perp}\cdot \tilde{a}_{q}+q_{\perp}^{\nu}\partial_{q_{\perp\rho}}\tilde{a}_{q\nu}\Bigg),
\\
&&\Theta^{\mu\nu}q^{\beta}_{\perp}\partial_{q^{\beta}_{\perp}}\tilde{a}_{q\nu}
=q^{\beta}_{\perp}\partial_{q^{\beta}_{\perp}}\tilde{a}_{q}^{\mu}-u^{\mu}q^{\beta}_{\perp}\partial_{q^{\beta}_{\perp}}(\tilde{a}_{q}\cdot u)
=q^{\beta}_{\perp}\partial_{q^{\beta}_{\perp}}\tilde{a}_{q}^{\mu}
+\frac{u^{\mu}}{E_q}\Big(\frac{m^2}{E_q^2}q_{\perp}^{\beta}+q_{\perp}^{\alpha}q_{\perp}^{\beta}\partial_{q^{\alpha}_{\perp}}\Big)\tilde{a}_{q\beta}.
\nn
\end{eqnarray}
By inputting exact expressions of $j^{T/L}_i$ from Eqs.~(\ref{jT}) and (\ref{jL}), it is found
\begin{subequations}
\begin{eqnarray}
&&
\mathcal{Q}_\text{cl}^{(1)}=-2,
\\
&&
\mathcal{Q}_\text{cl}^{(2)} =\Bigg(\frac{q_0^2}{|{\bf q}|T}-\frac{2|{\bf q}|}{q_0}\Big(2+\frac{m^2}{|{\bf q}|^2}\Big)-\frac{\eta_qm^2}{|{\bf q}|^2T}(q_0-2T)\Bigg)\hat{q}_{\perp}\cdot\tilde{a}_q
\\
&& \hspace{2cm}
+\Bigg(\frac{q_0(m^2-|{\bf q}|^2)}{|{\bf q}|^2}-\frac{\eta_qm^2q_0^2}{|{\bf q}|^3}\Bigg)\partial_{q_{\perp}^{\nu}}\tilde{a}_q^{\nu}
+\Bigg(\frac{q_0(3m^2+|{\bf q}|^2)}{|{\bf q}|^2}-\frac{3\eta_qm^2q_0^2}{|{\bf q}|^3}\Bigg)\hat{q}_{\perp}^{\nu}\hat{q}_{\perp}^{\rho}\partial_{q_{\perp}^{\rho}}\tilde{a}_q^{\nu},
\nn
\\
&&
\nonumber
\mathcal{Q}_\text{cl}^{(3)}=\Bigg(\frac{q_0}{T}-2-\frac{\eta_qm^2}{|{\bf q}|q_0T}(q_0-2T)\Bigg)\hat{q}_{\perp}\cdot\tilde{a}_q
+\Bigg(\frac{q_0^2}{|{\bf q}|}-\frac{\eta_qm^2q_0}{|{\bf q}|^2}\Bigg)\partial_{q_{\perp}^{\nu}}\tilde{a}_q^{\nu}
\\
&& \hspace{2cm}
+\Bigg(\frac{(3m^2+|{\bf q}|^2)}{|{\bf q}|}-\frac{3\eta_qm^2q_0}{|{\bf q}|^2}\Bigg)\hat{q}_{\perp}^{\nu}\hat{q}_{\perp}^{\rho}\partial_{q_{\perp}^{\rho}}\tilde{a}_q^{\nu},
\\
&&
\mathcal{Q}_\text{cl}^{(4)}=-2|{\bf q}|,
\\
&&
\mathcal{Q}_\text{cl}^{(5)}=\frac{q_0^3}{|{\bf q}|T}\left(1-\frac{\eta_qm^2}{q_0|{\bf q}|}\right),
\\
&&
\mathcal{Q}_\text{cl}^{(6)}=-\frac{q_0^2}{2}\left(3-\frac{q_0^2}{|{\bf q}|^2}+\frac{\eta_qm^4}{q_0|{\bf q}|^3}\right),
\\
&&
\mathcal{Q}_\text{cl}^{(7)}=\frac{m^2q_0}{2|{\bf q}|^3}\big(3|{\bf q}|q_0+\eta_q(|{\bf q}|^2-3q_0^2)\big),
\end{eqnarray} 
\end{subequations}
and
\begin{subequations}
\begin{eqnarray}
&&
\tilde{\mathcal{Q}}_\text{cl}^{(1)}=\frac{2q_0}{|{\bf q}|T}\Big[ 
|{\bf q}| (1-2 f_{Vq})
+ (\hat{q}_{\perp}^{\nu}\partial_{q^{\nu}_{\perp}}f_{Vq}) \Bigl(\frac{m^2q_0}{|{\bf q}|}\eta_q-q_0^2\Bigr)\Big],
\\
&&
\tilde{\mathcal{Q}}_\text{cl}^{(2)}=2\Big(
\frac{m^{2}q_0}{|{\bf q}|}\eta_q-q_0^2\Big)\frac{f_{Vq}}{|{\bf q}|T}\hat{q}_{\perp}\cdot\tilde{a}_q,
\\
&&
\tilde{\mathcal{Q}}_\text{cl}^{(3)}=2\Big(
\frac{m^2q_0}{|{\bf q}|}\eta_q-q_0^2\Big)\frac{f_{Vq}}{q_0T}\hat{q}_{\perp}\cdot\tilde{a}_q,
\\
&&
\tilde{\mathcal{Q}}_\text{cl}^{(5)}=2\Bigg(\frac{m^2q_0}{|{\bf q}|}\eta_q-q_0^2
\Bigg)\frac{q_0f_{Vq}}{|{\bf q}|T},
\\
&&
\tilde {\mathcal{Q}}_\text{cl}^{(4)} = \tilde {\mathcal{Q}}_\text{cl}^{(6)} =\tilde {\mathcal{Q}}_\text{cl}^{(7)} = 0.
\end{eqnarray}
\end{subequations}

Finally, the sum of all the pieces ${\mathcal C}_1 + {\mathcal C}_2  $ from Eqs.~(\ref{eq:C2}) and (\ref{eq:C1}) 
provide the collision term of the AKE: 
\begin{eqnarray}\nonumber
\lambda^{-1}_c \delta(q^2-m^2)  \hat{\mathcal{C}}^{\mu}_{\text{cl}}
&\approx&  \frac{m_D^2\delta(q^2-m^2)T}{8\pi E_q}\ln (1/g_c)\Big(\tilde{a}_q^{\mu}\grave{\mathcal{Q}}_\text{cl}^{(1)}+u^{\mu}\grave{\mathcal{Q}}_\text{cl}^{(2)}+\hat{q}^{\mu}_{\perp}\grave{\mathcal{Q}}_\text{cl}^{(3)}+\grave{\mathcal{Q}}_\text{cl}^{(4)}\hat{q}_{\perp}^{\nu}\partial_{q_{\perp\mu}}\tilde{a}_{q\nu}
\\
&&
+\grave{\mathcal{Q}}_\text{cl}^{(5)}\hat{q}^{\nu}\partial_{q_{\perp}^{\nu}}\tilde{a}_{q}^{\mu}
+\grave{\mathcal{Q}}_\text{cl}^{(6)}\eta^{\nu\rho}\partial_{q^{\nu}_{\perp}}\partial_{q^{\rho}_{\perp}}\tilde{a}_q^{\mu}+\grave{\mathcal{Q}}_\text{cl}^{(7)}\hat{q}_{\perp}^{\nu}\hat{q}_{\perp}^{\rho}\partial_{q^{\nu}_{\perp}}\partial_{q^{\rho}_{\perp}}\tilde{a}_q^{\mu}
\Big).
\end{eqnarray}
The coefficients $\grave{\mathcal{Q}}_\text{cl}^{(i)} 
 \equiv  \tilde {\mathcal{Q}}_\text{cl}^{(i)} +  {\mathcal{Q}}_\text{cl}^{(i)}  \ (i=1, \dots, 7)$ 
are shown in Eqs.~(\ref{grave_Qcl_1})-(\ref{grave_Qcl_7}).

%\bibliography{polarization_ref}
\bibliography{spin_transport_collisions_manuscript_revised.bbl}
\end{document}